\documentclass[aps,pra,reprint,superscriptaddress,nofootinbib]{revtex4-2}

\usepackage{tikz}
\usetikzlibrary{positioning,arrows.meta,calc}
\usepackage[utf8]{inputenc}
\usepackage[T1]{fontenc}
\usepackage[english]{babel}
\usepackage{microtype}

\usepackage{graphicx}
\usepackage{float}
\usepackage{adjustbox}
\usepackage{placeins}
\usepackage[caption=false]{subfig}
\usepackage{tabularx,booktabs}
\usepackage{multirow}

\makeatletter

\newcommand{\supplementtoc}{%
  \@starttoc{stc}%
}

\newcommand{\startsupplementtoc}{%
  \let\originaladdcontentsline\addcontentsline
  \renewcommand{\addcontentsline}[3]{%
    \def\tempa{##1}%
    \def\tempb{toc}%
    \ifx\tempa\tempb
      \originaladdcontentsline{stc}{##2}{##3}%
    \else
      \originaladdcontentsline{##1}{##2}{##3}%
    \fi
  }%
}

\makeatother

\usepackage{amsmath}
\usepackage{amsthm}
\usepackage{amsfonts}
\usepackage{amssymb}
\usepackage{mathtools}
\usepackage{bm}
\usepackage{bbm}
\usepackage{dsfont}
\usepackage{braket}
\usepackage{nccmath}
\usepackage{chngcntr}

\usepackage[normalem]{ulem}
\usepackage[shortlabels]{enumitem}
\usepackage{csquotes}
\usepackage{comment}
\usepackage{verbatim}
\usepackage[toc,page]{appendix}
\usepackage{natbib}
\usepackage{xcolor}
\usepackage[colorlinks=true,citecolor=blue,linkcolor=magenta]{hyperref}

\newcommand{\Tr}{\operatorname{Tr}}
\newcommand{\rbar}{\bar{\rho}}
\newcommand{\Holevo}{\chi}
\newcommand{\abs}[1]{\left|#1\right|}
\newcommand{\norm}[1]{\left\lVert#1\right\rVert}
\newcommand{\eG}{e_G^{\xi}}
\newcommand{\qF}{F_{\rm Q}}
\newcommand{\cF}{F_{\rm C}}
\newcommand{\etaone}{\eta_{\rm 1P}^{\Delta}}
\newcommand{\etaD}{\eta_{\rm prodP}^{\Delta}}
\newcommand{\etaX}{\eta_{\rm prodP}^{\Holevo}}
\newcommand{\etaF}{\eta_{\rm prodP}^{F}}

\begin{document}

\title{Dynamical quantum simulations as laboratories for quantum information}

\author{Su Yeon Chang}
\affiliation{Theoretical Division, Los Alamos National Laboratory, Los Alamos, NM 87545, USA}

\author{Hailey S. Murray}
\affiliation{School of Applied and Engineering Physics, Cornell University, Ithaca, NY 14850, USA}

\author{Brian Barch}
\affiliation{Theoretical Division, Los Alamos National Laboratory, Los Alamos, NM 87545, USA}

\author{Greeshma Oruganti}
\affiliation{School of Electrical and Computer Engineering, Cornell University, Ithaca, New York 14850, USA}

\author{Luis~Pedro~Garc\'ia-Pintos}
\affiliation{Theoretical Division, Los Alamos National Laboratory, Los Alamos, NM 87545, USA}

\author{Martin Larocca}
\affiliation{Theoretical Division, Los Alamos National Laboratory, Los Alamos, NM 87545, USA}

\author{Mark M.~Wilde}
\affiliation{School of Electrical and Computer Engineering, Cornell University, Ithaca, New York 14850, USA}

\author{M. Cerezo}
\email{cerezo@lanl.gov}
\affiliation{Information Sciences, Los Alamos National Laboratory, Los Alamos, NM 87545, USA}

\begin{abstract}
Dynamical quantum simulations are commonly designed to access a particular response function or correlation function. Yet the underlying protocol prepares a family of many-body states that contains information well beyond the observable for which the simulation was originally designed. In this work, we explore the idea of using these states as laboratories for quantum information. As a case study, we consider the local-pulse protocols used to compute Green functions in finite XX and XXX spin chains, and we interpret opposite source rotations as encoding a classical bit into the many-body system. This allows us to follow how the resulting record propagates, how it is encoded across local subsystems, and how readily it can be accessed with different measurements. We also find a clear separation between representation and readout. While fermionic covariance data exactly reconstruct the signed response in the XX chain, simple spin measurements recover a larger fraction of the locally available information in the XXX chain. This reversed ordering persists when we move from finite-pulse discrimination to local source-angle estimation. Going beyond the source record, the same reduced states also allow us to reinterpret results in terms of Bell state recovery and quantum communication rates, as well as providing direct access to the evolving bipartite and genuine multipartite entanglement of the system. Our results show that dynamical quantum simulations developed for response calculations can naturally be used to ask a much broader set of quantum-information questions about the many-body states they already prepare. 
\end{abstract}
\maketitle

\small
\tableofcontents
\normalsize
\clearpage
\section{Introduction}
\label{sec:introduction}

Quantum simulations constitute one of the most natural applications of quantum computers~\cite{feynman1982simulating,lloyd1996universal,georgescu2014quantum,daley2022practical}, as they provide a direct route to studying the properties of quantum systems that can become difficult to access with classical numerical methods. While much of the early effort in this direction focused on equilibrium properties, programmable quantum computers can now be used to study dynamical quantities~\cite{bernien2017probing,zhang2017observation,kokail2019self,lee2026benchmarking,granet2026dynamical}. For instance, currently available hardware can estimate quantities such as dynamical structure factors~\cite{knap2013probing,baez2020dynamical,lee2026digital,granet2026dynamical}, which encode the excitation spectrum of a material and can be directly compared with neutron scattering data from experiments on quantum materials~\cite{lee2026benchmarking}. These developments suggest that quantum computers are beginning to provide a useful platform not only for simulating idealized models, but also for connecting microscopic many-body calculations with experimentally measurable properties of real materials.

However, a quantum simulation designed to compute a response function necessarily prepares much more than the final quantity that is reported. Before measuring a Green function or a particular correlation function, one first prepares an initial many-body state, locally perturbs it, and evolves the resulting state under the Hamiltonian of interest. By the time the final measurement is performed, the quantum computer has therefore prepared a nontrivial family of dynamical many-body states~\cite{knap2013probing,baez2020dynamical,lee2026benchmarking}. From this perspective, the preparation of the dynamical state and the observable used to probe it are two distinct parts of the experiment. This simple observation raises the possibility that simulations developed for one physical quantity could be reused to learn substantially more about the states that they already prepare.

Viewing many-body dynamics through an information-theoretic lens is itself not novel. Spin chains have long been studied as quantum communication channels, with capacities and information-transfer rates used to characterize correlations and transport \cite{giovannetti2005information,rossini2007information}. There are also important examples showing that many-body experiments contain information of direct interest to quantum information science. Entanglement entropies have been measured directly in quantum simulators~\cite{islam2015measuring,brydges2019probing}, while quantum Fisher information can be related to dynamical susceptibilities and used to witness multipartite entanglement~\cite{hauke2016measuring}. Indeed, this connection has enabled entanglement to be inferred from neutron scattering spectra~\cite{scheie2021witnessing}. These results provide a remarkable bridge between quantities traditionally studied in condensed-matter physics and concepts originating in quantum information. At the same time, they naturally lead to a broader question: \begin{quote}
    \textit{If a programmable quantum simulator gives us access to the many-body states from which the response is obtained, rather than only to the response function itself, what additional quantum-information properties can be extracted from the same simulation?}
\end{quote}

\begin{figure}[t]
 \centering
 \includegraphics[width=1\columnwidth]{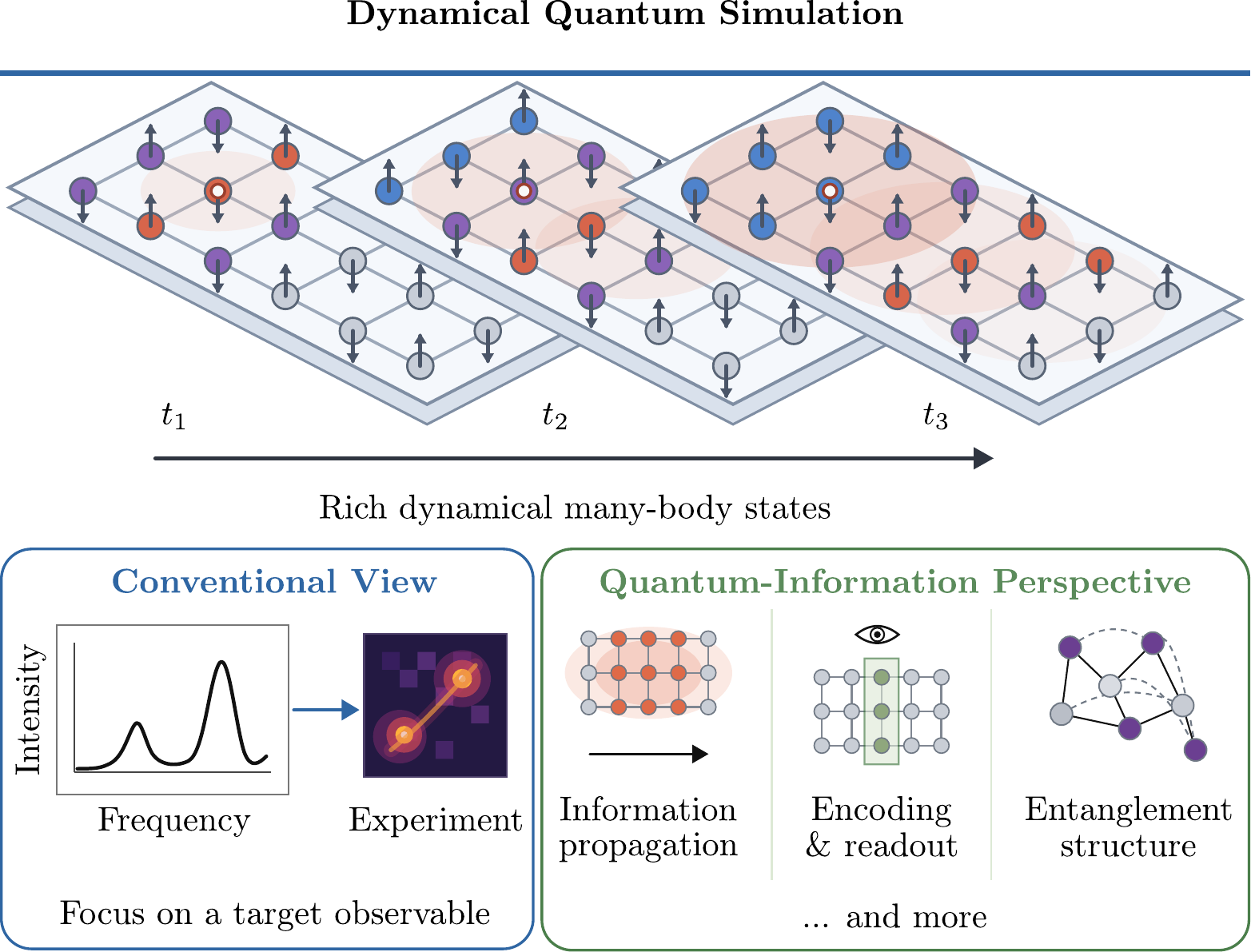}
\caption{\textbf{Schematic representation of our framework.} A dynamical quantum simulation prepares a family of rich many-body states as the system evolves. In the conventional view, these states are used to estimate a target dynamical quantity, such as a response function or spectrum that can be compared with experiment. Here we instead take the prepared states themselves as objects of study and use them to investigate quantum-information properties such as information propagation, encoding and readout, and entanglement structure.}
 \label{fig:schematic}
\end{figure}

Motivated by this question, here we propose to view dynamical quantum simulations as laboratories for quantum information. The basic idea is to take the family of states already generated during a dynamical simulation and interrogate them through a broader set of measurements and quantum-information diagnostics. As shown in Fig.~\ref{fig:schematic}, such a viewpoint allows one to ask, for instance, how information introduced through a local perturbation spreads through the system, how large a subsystem must be observed before that information can be recovered, whether it is encoded independently in local degrees of freedom or jointly across several sites, and how readily it can be accessed with experimentally simple measurements. One can similarly ask whether a compact representation of the state preserves the information relevant for a given task, or if one can use the same reduced states to study the evolution of bipartite and genuine multipartite entanglement. Hence, rather than associating a dynamical simulation with a single target observable, we seek to use the prepared states themselves as the objects of study.

In this work, we explore this perspective through the local-pulse protocol used to compute Green functions in finite XX and XXX spin chains. By considering opposite local rotations, we can interpret the pulse sign as a classical bit encoded into the many-body system and follow how this information is carried by the ensuing dynamical states. We find that the information collected by a local observer grows as its spatial aperture is increased, and that the record changes from a simple one-site encoding to a noncommuting and jointly distributed multisite one. More interestingly, our results reveal a clear separation between how simply this information can be represented and how easily it can be read. In the XX chain, fermionic covariance matrices exactly determine the signed response, whereas the same description is no longer complete for the XXX chain. Nevertheless, simple spin measurements recover a larger fraction of the locally available information in the XXX model. This reversed ordering persists when we change the task from discriminating two finite pulses to estimating a continuously varying source angle. Finally,  we show that the same results can be used to reinterpret these results in terms of Bell state recovery and quantum communication rates; also, the same reduced states can be used to track the reorganization of bipartite entanglement and to certify genuine multipartite entanglement. Altogether, these results provide an example of the broader opportunities that arise when dynamical quantum simulations are viewed not only as tools for computing response functions, but also as laboratories in which to explore the quantum-information structure of many-body states.

\section{From a response protocol to a local information problem}
\label{sec:protocol}

In this section, we introduce the basic ingredients of the framework used throughout this work.  Our starting point is to reinterpret a standard quantum dynamical simulation experiment as a classical information source encoding binary labels into quantum states. Here we also introduce some basic quantum-information quantities that will be used to answer these questions.

\subsection{Many-body setting and local source pulse}

To illustrate our framework, we use the XX and XXX spin chains as a concrete test bed for the framework developed below. This choice is motivated in part by the recent neutron-scattering benchmarking experiment of Ref.~\cite{lee2026benchmarking}, where closely related local-pulse Green-function simulations were implemented on superconducting quantum hardware. In that work, the protocol was used for both an XX chain and for KCuF$_3$, whose low-energy spin dynamics are well described by the one-dimensional isotropic Heisenberg model. The present setting therefore lets us ask what additional information could, in principle, be extracted from states prepared by a dynamical simulation that has already been realized experimentally.

The two models also provide a useful theoretical contrast. Under the Jordan--Wigner transformation, the XX chain maps to free spinless fermions and preserves Gaussianity, whereas the XXX interaction generates density--density interactions and breaks this Gaussian structure. This allows us to compare how the same source record is represented and read out in a free and an interacting many-body system.

In particular, we emphasize that the broader program presented in this work is not tied to either model. The same questions can be asked whenever a dynamical quantum simulation prepares a family of many-body states before the final observable is measured. We focus on the XX and XXX chains, which are experimentally relevant and provide a well-controlled setting for developing the key ideas.

Concretely, we consider an antiferromagnetic spin-$1/2$ chain with open boundary conditions. We take $N=50$ spins governed by
\begin{equation}
 H_\epsilon
 :=
 2J\sum_{j=1}^{N-1}
 \left(
 S_j^xS_{j+1}^x
 +
 S_j^yS_{j+1}^y
 +
 \epsilon S_j^zS_{j+1}^z
 \right),
 \label{eq:hamiltonian}
\end{equation}
where $S_j^\alpha=\sigma_j^\alpha/2$, $J>0$, and $\epsilon=0$ and $1$ correspond to the XX and XXX models, respectively. For each model, the initial state
\begin{equation}
 \rho_{0,\epsilon}
 =
 \ket{g_\epsilon}\!\bra{g_\epsilon}
\end{equation}
is the finite-chain ground state in the zero-total-$S^z$ sector. This is the natural initial state for the zero-temperature response protocol considered here and matches the setting in which the corresponding dynamical structure factors are computed. All details of our numerical simulations are presented in Supplemental Sec.~\ref{supp:details}.

We now follow the local-pulse response protocol used in Refs.~\cite{baez2020dynamical,lee2026benchmarking} to compute dynamical structure factor from quantum simulations of inelastic neutron scattering. We refer the reader to Supplemental Sec.~\ref{supp:Green function-dsf} for additional details. In such experiments, the relevant dynamical quantity is the retarded spin Green function
\begin{equation}
 G^R_{\alpha\beta}(i,j,t)
 \coloneqq
 -i\Theta(t)
 \left\langle
 \left[
 S_i^\alpha(t),S_j^\beta
 \right]
 \right\rangle\,,
 \label{eq:retarded-green}
\end{equation}
where $\Theta(t)$ denotes the Heaviside step function
\begin{equation}
\Theta(t)\coloneqq
\begin{cases}
1, & t>0,\\
\frac{1}{2}, & t=0,\\
0, & t<0.
\end{cases}
\end{equation}
The Fourier transform of the retarded spin Green function $G^R_{\alpha\beta}(i,j,t)$ determines the dynamical structure factor measured in neutron scattering.

On a quantum computer, the commutator entering this response can be accessed by applying a local spin rotation and evolving under the Hamiltonian of the material. Let~$j_0$ denote the source site, and take the local generator to be $P\coloneqq \sigma_{j_0}^z$. We consider the two opposite quarter turns
\begin{equation}
 V_s\coloneqq \exp(-is\pi P/4)=\frac{I-isP}{\sqrt{2}},
 \qquad s\in\{+1,-1\},
 \label{eq:pulse}
\end{equation}
where $s$ labels the direction of the pulse. After the perturbation, the system evolves under $U_\epsilon(t)\coloneqq \exp(-iH_\epsilon t)$, producing the two global states
\begin{equation}
 \ket{\psi_{s,\epsilon}(t)}
 \coloneqq 
 U_\epsilon(t)V_s\ket{g_\epsilon}.
 \label{eq:global-branches}
\end{equation}

For an observable $O$ measured after the evolution, let
\begin{equation}
 \left\langle O\right\rangle_s
 \coloneqq 
 \bra{\psi_{s,\epsilon}(t)}
 O
 \ket{\psi_{s,\epsilon}(t)} .
\end{equation}
Taking the difference between the two pulse directions isolates the commutator,
\begin{equation}
 \frac{\left\langle O\right\rangle_+-\left\langle O\right \rangle_-}{2}
 =
 -\frac{i}{2}
 \left\langle
 \left[
 O(t),P
 \right]
 \right\rangle ,
 \label{eq:phase-cycle}
\end{equation}
where $O(t) \coloneqq U_{\epsilon }^\dag(t)  O U_{\epsilon }(t)$ and the expectation value on the right-hand side is taken in the ground state. Thus, for appropriate choices of $O$ and $P$, the usual retarded response is obtained directly from the two pulse experiments.

For the systems considered here, the two global states are orthogonal. Global spin-flip symmetry gives $\langle g_\epsilon|P|g_\epsilon\rangle=0$. Indeed, because \(H_\epsilon\) is invariant under the global spin-flip transformation \(F\coloneqq\prod_{j=1}^N \sigma_j^x\), while \(F\sigma_{j_0}^zF^\dagger=-\sigma_{j_0}^z\), a ground state chosen with definite \(F\)-parity satisfies \(\langle g_\epsilon|\sigma_{j_0}^z|g_\epsilon\rangle=0\). Furthermore, $V_+^\dagger V_-=iP$. Therefore
\begin{equation*}
 \langle\psi_{+,\epsilon}(t)|\psi_{-,\epsilon}(t)\rangle
 =
 i\langle g_\epsilon|P|g_\epsilon\rangle
 =
 0 .
\end{equation*}
The source sign is therefore perfectly distinguishable when the full chain is available. The question below is how much of the recorded bit can still be recovered when the observer has access only to a finite region.

In the usual Green-function protocol, selected local spin observables are measured on the states in Eq.~\eqref{eq:global-branches} to recover the retarded response. We instead consider the reduced state on a contiguous region $R$. In all numerical results below we place the source at $j_0=25$, one of the two central sites of the $N=50$ open chain, and label the center of $R$ by its position $d$ relative to the source. We consider regions containing $k\in\{1,3,5,7\}$ spins, with $k$ playing the role of a tunable spatial aperture. An observer at $(R,t)$ then receives one of the two reduced states
\begin{equation}
 \rho_{s,\epsilon}(R,t)
 \coloneqq 
 \Tr_{\bar R}
 \left[
 \ket{\psi_{s,\epsilon}(t)}
 \!\bra{\psi_{s,\epsilon}(t)}
 \right]\,,
 \label{eq:branch-state}
\end{equation}
where $ \Tr_{\bar R}$ denotes the partial trace over all qubits not in~$R$. Unless needed explicitly, we henceforth suppress the dependence on $\epsilon$, $R$, and $t$, and write the two local states simply as $\rho_+$ and $\rho_-$.

In the following sections, we will make the reduced density matrices $\rho_\pm$ the central object of our study, and ask how much information about the applied pulse can be recovered from different parts of the chain. As such, a single-site observer asks how much information about the source can be recovered locally, whereas increasing $k$ allows the observer to collect information stored over a larger portion of the evolving state.

\begin{figure*}[t]
 \centering
 \includegraphics[width=0.98\textwidth]{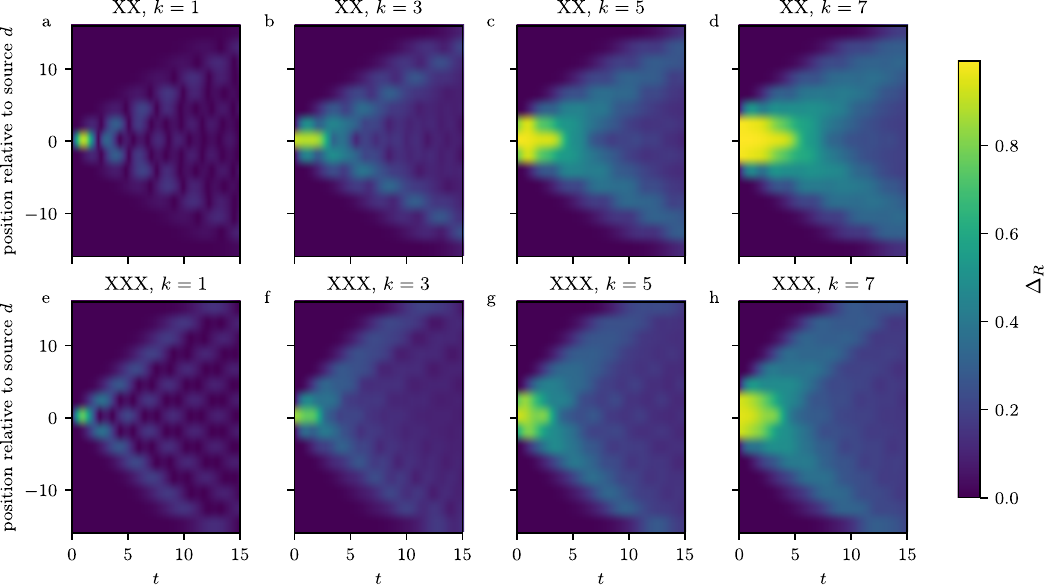}
 \caption{\textbf{Trace distance wakes for all observation apertures.} Trace distance maps for XX (a-d) and XXX (e-h) at widths  $k=1,3,5,7$. Every panel uses the same absolute scale. At fixed time and center, the regions are nested, so enlarging the aperture cannot reduce the distinguishability. The maps locate the record over the space-time window. }
 \label{fig:D-wakes}
\end{figure*}

\subsection{The response as one projection of the local ensemble}

The two states define the equal-prior ensemble
\begin{equation}
 \mathcal E_{R,t}\coloneqq \left\{\left(\frac12,\rho_+\right),
 \left(\frac12,\rho_-\right)\right\}.
 \label{eq:source-ensemble}
\end{equation}
From here, it is useful to separate the local state into two contributions: one that remains unchanged under pulse reversal and another that changes sign. We therefore define the average local state and the signed response respectively as
\begin{equation}
 \rbar\coloneqq \frac{\rho_++\rho_-}{2},\qquad \mathrm{and} \qquad 
 \xi\coloneqq \frac{\rho_+-\rho_-}{2}\,,
 \label{eq:bar-chi}
\end{equation}
so that 
\begin{equation}\label{eq:recover}
    \rho_\pm= \rbar\pm  \xi\,.
\end{equation}
The state $\rbar$ is the density matrix assigned by an observer who knows that a pulse was applied but does not know its sign. By contrast, $\xi$ is not a density matrix but a Hermitian, traceless operator that encodes the dependence of the local state on the pulse sign. It contains the complete odd component of the local dynamics. 

Moreover, the signed response $\xi$ can be directly related to the basic framework of response theory. For any observable $O_R$ supported in $R$, direct expansion of the two pulses gives
\begin{equation}
 \Tr_R(O_R\xi)=-\frac{i}{2}\Tr\!\left([O_R(t),P]\rho_{0,\epsilon}\right).
 \label{eq:response-bridge}
\end{equation}
Thus, the retarded response measured in the Green function protocol is one particular projection of $\xi$. More generally, choosing different observables $O_R$ gives access to different components of the local response.

\subsection{Two operational questions about the source record}

The local ensemble picture of the Green function experiment naturally leads to two questions. We first ask how well an observer who receives one copy of the local state can guess whether the positive or negative pulse was applied. For equal prior probabilities, the optimal one-copy success probability is determined by the trace distance~\cite{wilde2013quantum},
\begin{equation}
 \Delta_R(t) \coloneqq  \norm{\xi}_1 = \frac12\norm{\rho_+-\rho_-}_1,
 \quad
 P_{\rm succ}\coloneqq \frac{1+\Delta_R}{2}.
 \label{eq:Trace distance}
\end{equation}
Hence $\Delta_R=0$ corresponds to a random guess, whereas $\Delta_R=1$ means that the two states can be distinguished perfectly. The dual expression
\begin{equation}
 \Delta_R=\max_{\substack{O=O^\dagger,\\ \norm{O}_\infty\leq1}}
 \abs{\Tr(O\xi)}
 \label{eq:Trace distance-dual}
\end{equation}
shows that every bounded response observable provides a witness to the source sign. The maximization is achieved by the Helstrom observable $O_{\rm H}\coloneqq \operatorname{sign}(\xi)=\Pi_{+}-\Pi_{-}$, where $\Pi_+$ and $\Pi_-$ project onto the positive and negative eigenspaces of $\xi$, respectively. Measuring in this basis gives the optimal decision between the two pulse directions.

We next ask how much information about the source label $s$ is present in the qubits within the region $R$. The Holevo information of the local ensemble quantifies this information and is defined as
\begin{equation}
 \Holevo_R\coloneqq S(\rbar)-\frac{S(\rho_+)+S(\rho_-)}{2} \label{eq:holevo}
 \end{equation}
 where $S(\rho)$ is the von Neumann entropy:
 \begin{equation}
     S(\rho)\coloneqq -\Tr(\rho\log_2\rho).
\end{equation}
Moreover, from Eq.~\eqref{eq:recover}, the Holevo information can be written as the equal-weight quantum Jensen--Shannon divergence~\cite{Schumacher2001Optimal,majtey2005jensen},
\begin{equation}\label{eq:holevo-relative}
\Holevo_R=
\frac{1}{2}D(\rho_+\Vert\bar\rho)
+\frac{1}{2}D(\rho_-\Vert\bar\rho)\,,
\end{equation}
where the quantum relative entropy of states $\rho$ and $\sigma$ is defined as \cite{Umegaki1962Conditional}
\begin{equation}
    D(\rho\Vert\sigma)\coloneqq \Tr\!\left[\rho\left(\log_2\rho-\log_2\sigma\right)\right].
\end{equation}

To see the information-theoretic meaning of $\Holevo_R$, we introduce a classical register $C$ that records which pulse was applied. The joint classical--quantum state is
\begin{equation}
 \rho_{CR}
 \coloneqq
 \frac{1}{2}\sum_{s\in\{\pm1\}}
 \ket{s}\!\bra{s}_C\otimes\rho_s .
 \label{eq:cq-state}
\end{equation}
For this state, the quantum mutual information, defined for a bipartite quantum states $\rho_{AB}$ as $I(A:B)=S(\rho_A)+S(\rho_B)-S(\rho_{AB})$, between the pulse label and the quantum region is exactly
\begin{equation}
 I(C{:}R)=\Holevo_R.
 \label{eq:Holevo-mutual}
\end{equation}

Operationally, we can imagine repeating the experiment with the two pulse directions chosen with equal probability. An observer receives the corresponding reduced state on $R$ and performs some measurement with classical outcome $Y$. The mutual information $I(C{:}Y)$ quantifies how many bits about the pulse label are obtained from that measurement, and the Holevo bound gives
\begin{equation}
 I(C{:}Y)\leq \Holevo_R
\end{equation}
for any measurement on a single copy of the local state. Thus, $\Holevo_R$ is the measurement-independent information budget about the source sign available in the region $R$. Since the source label is one equiprobable bit, $0\leq\Holevo_R\leq1$: $\Holevo_R=0$ means that the two local states are identical and thus indistinguishable, while $\Holevo_R=1$ means that the source sign can in principle be recovered perfectly from $R$. More fundamentally, the Holevo information $\chi_R$ is equal to an achievable rate for classical communication over the channel $s\mapsto \rho_s$, and we argue later that it is in fact equal to the classical capacity of this channel.

We focus primarily on $\Delta_R$ and $\Holevo_R$ because they answer the two most direct operational questions associated with the equal-prior binary source ensemble considered here. The trace distance completely determines the optimal single-copy probability of deciding which pulse was applied, whereas $\Holevo_R$ quantifies the amount of source-label information present in the quantum region before a measurement is chosen. They therefore distinguish between the optimal binary decision and the information content of the local ensemble. Both quantities also satisfy data processing under partial trace, which makes them particularly natural for studying how much of the source record remains accessible as the observation region is changed. We will consider other tasks later in the paper, but $\Delta_R$ and $\Holevo_R$ provide a common reference for following the record through the many-body dynamics.

Moreover, although they describe the same binary ensemble, the two quantities need not assign the same significance to a given local state. A region can provide a noticeable advantage for deciding which pulse was applied while carrying only a small amount of source information in bits. Conversely, a measurement that saturates the Holevo bound extracts all of the source-label information available in the region, but this does not imply that the pulse sign can be inferred with high probability. The two quantities are nevertheless constrained by the dimension-independent bounds
\begin{equation}
1-h_2\!\left(
\frac{1-\Delta_R(t)}{2}
\right)
\leq
\Holevo_R(t)
\leq
\Delta_R(t),
\end{equation}
where \( h_2(x) \coloneqq -x\log_2 x-(1-x)\log_2(1-x) \) is the binary entropy function. The lower bound follows by considering the optimal Helstrom measurement and applying the corresponding classical Jensen--Shannon divergence bound~\cite{Liu2025QuantumStateTesting}, while the upper bound follows from a continuity bound for the quantum Jensen--Shannon divergence \cite[Theorem 14]{Audenaert2014}.

\section{A propagating, task-dependent record}
\label{sec:propagation}

Having introduced the framework, we now turn to the results obtained by applying it to the XX and XXX dynamics. We begin with the most basic question, namely where the information written by the local pulse propagates and how much of it can be recovered as the observer is given access to increasingly larger regions of the chain.

\subsection{Watching the observer aperture grow}

Here, we first ask where the source record goes after it has been encoded. Figure~\ref{fig:D-wakes} shows $\Delta_R$ for all considered apertures in both models.  Time $t$ runs horizontally and the position $d$ relative to the source runs vertically. All eight panels share the same absolute color scale, allowing us to directly compare how much of the source record becomes visible as the observer grows from one to seven sites.

The connection with the usual condensed-matter picture is most direct for $k=1$. For an observer centered on the source, symmetry fixes the optimal single-spin measurement to the $z$ direction, and the trace distance reduces to
\begin{equation}
    \Delta_1(t)=2\left|G^R_{zz}(j_0,j_0;t)\right|\,.
    \label{eq:D1-green}
\end{equation}
Thus, the one-site trace-distance wake is simply the magnitude of the local retarded response generated by the perturbation. Its outward-moving fronts therefore have the same origin as the familiar light-cone structure of the spin dynamics, while the oscillations behind the front reflect the underlying many-body propagation and interference captured by the Green function. In the Heisenberg chain, such real-space light cones are naturally understood in terms of propagating spinons, while persistent oscillations inside the cone have been associated with the quantum-wake dynamics of the isotropic model~\cite{lee2026benchmarking}.

As the aperture grows, so does what an observer is allowed to measure. At $k=1$, the signal is narrow and strongly oscillatory. Once neighboring spins are included, a much broader wake becomes visible. In particular, for $k=5$ and $k=7$ the source record remains appreciable throughout much of the region behind the propagating front, even when the one-site response has become small. This phenomenon is particularly visible for the XX chain, which exhibits a broad and long-lived trace-distance wake for the larger apertures. The XXX chain also retains a sizeable record, although with more pronounced oscillatory structure. The qualitative difference is suggestive of the distinct dynamical regimes of the two models, with ballistic spin transport in XX and superdiffusive spin transport together with interacting spinon dynamics in the isotropic Heisenberg chain~\cite{lee2026benchmarking}.

\begin{figure}
    \centering
    \includegraphics[width=1\linewidth]{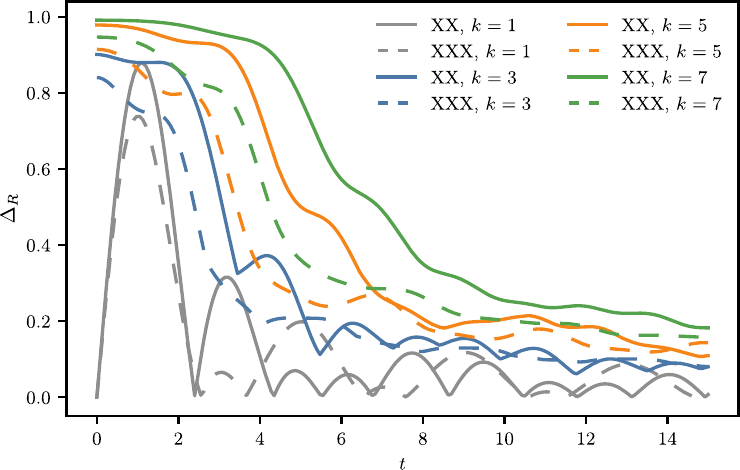}
    \caption{\textbf{Trace distance at the source as a function of time for the four observation apertures $k=1,3,5,7$.} Solid (dashed) lines correspond to the XX (XXX) chain, with colors indicating the aperture size. For $k=1$, the trace distance is directly related to the conventional local retarded response through $\Delta_1(t)=2|G^R_{zz}(j_0,j_0;t)|$. Increasing the aperture reveals a substantially larger and longer-lived record of the same perturbation. The more persistent oscillatory structure in XXX follows that of the quantum-wake behavior of the isotropic Heisenberg chain.}
    \label{fig:D-source}
\end{figure}

This distinction is particularly clear if we sit at the source and follow the dynamics in time, as shown in Fig.~\ref{fig:D-source}. A single spin initially contains essentially no information about the pulse direction, whereas already a three-site region makes the two states strongly distinguishable. At later times, $\Delta_1$ repeatedly becomes small as the local response oscillates, while the larger regions can still retain substantial distinguishability. In this sense, the vanishing of a local response does not imply that the system has lost the perturbation. Rather, the corresponding record has moved into neighboring spins and their correlations, where it becomes visible only to a larger observer.

\begin{figure*}[t]
 \centering
 \includegraphics[width=0.98\textwidth]{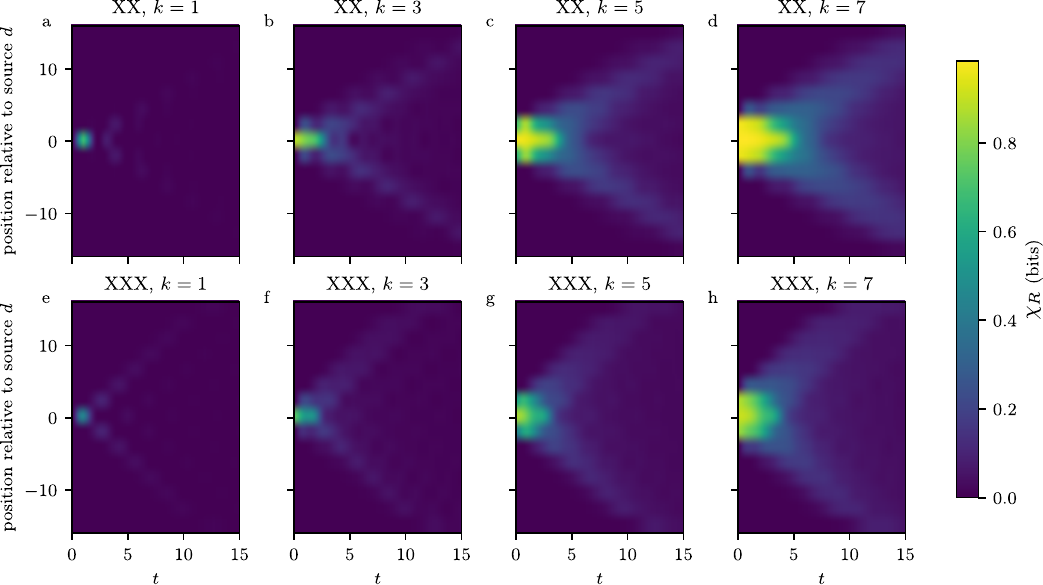}
\caption{\textbf{Holevo-information wakes for all observation apertures.} The layout and absolute spatial and temporal scales are the same as in Fig.~\ref{fig:D-wakes}. The same propagating structure is visible, but regions of small distinguishability are strongly suppressed because $\Holevo_R$ is quadratic in $\Delta_R$ when the two local branches are close. For $k=1$, where $\Delta_1$ is proportional to the magnitude of the local retarded Green function, this implies $\Holevo_1\propto |G^R_{zz}|^2$ in this regime.}
 \label{fig:holevo}
\end{figure*}

Importantly, we note that the growth in the trace distance with aperture is guaranteed by the data-processing inequality. Since the four observation windows are exactly nested, the reduced density matrix for $k=1$ can be obtained by tracing out qubits from that of $k=3$, and so on, meaning that
\begin{equation}
 \Delta_{1}\leq \Delta_{3}\leq \Delta_{5}\leq \Delta_{7},\qquad
 \Holevo_{1}\leq\Holevo_{3}\leq\Holevo_{5}\leq\Holevo_{7}.
 \label{eq:nested-data-processing}
\end{equation}
A larger observer can therefore never lose information available to a smaller one. What is nontrivial is how quickly this additional information accumulates and how long it remains locally accessible.

This aperture dependence is also different in the two models. Across the resolved part of the wake, as defined by the activity masks in Supplemental Sec.~\ref{supp:matching-domain}, the median XX trace distance grows from $0.0371$ at $k=1$ to $0.2936$ at $k=7$, while the corresponding XXX median grows from $0.0543$ to $0.2300$. A one-site observer therefore sees a larger typical signal in XXX, whereas a seven-site observer collects a larger one in XX.

\subsection{Decision and information paint different wakes}

The trace distance and Holevo information quantify different aspects of the same propagating record. For one, the trace distance is dimensionless and determines how well an observer can decide which pulse was applied, whereas $\Holevo_R$ is measured in bits and asks how much information about the classical source label is present in the reduced subsystem's qubits. Figure~\ref{fig:holevo} shows the Holevo wake for the same four apertures used in Fig.~\ref{fig:D-wakes}. The overall propagation structure is clearly the same, but the two quantities paint rather different pictures of the wake. In particular, the faint regions visible in $\Delta_R$ are much more strongly suppressed in $\Holevo_R$.

When the two reduced states $\rho_+$ and $\rho_-$ are nearly indistinguishable, so that $\Delta_R\ll1$, the Holevo information scales quadratically with their trace distance. This stronger suppression can be understood by expanding Eq.~\eqref{eq:holevo-relative} around the average local state. For a full-rank $\bar\rho$, the linear term in $\xi$ vanishes, while the two branches cancel all odd orders (see the Supplemental Sec.~\ref{supp:holevo-quadratic} for more details). Consequently,
\begin{equation}
\Holevo_R=C_R \Delta_R^2+O(\Delta_R^4),
\label{eq:holevo-quadratic}
\end{equation}
where the coefficient $C_R$ depends on the local average state and on the direction in which the two states separate. Moreover, applying the quantum Pinsker inequality~\cite{hiai1981sufficiency} gives
\begin{equation}
 \Holevo_R\geq \frac{\Delta_R^2}{2\ln2}\,.
 \label{eq:pinsker-holevo}
\end{equation}
Hence, quantum Pinsker's inequality further requires $C_R\geq1/(2\ln2)$. For the one-site states considered here, symmetry gives the exact relation $\Holevo_1=1-h_2[(1+\Delta_1)/2]$, and hence $\Holevo_1=\Delta_1^2/(2\ln2)+O(\Delta_1^4)$ (see the Supplemental Information).

\begin{figure}
    \centering
    \includegraphics[width=1\linewidth]{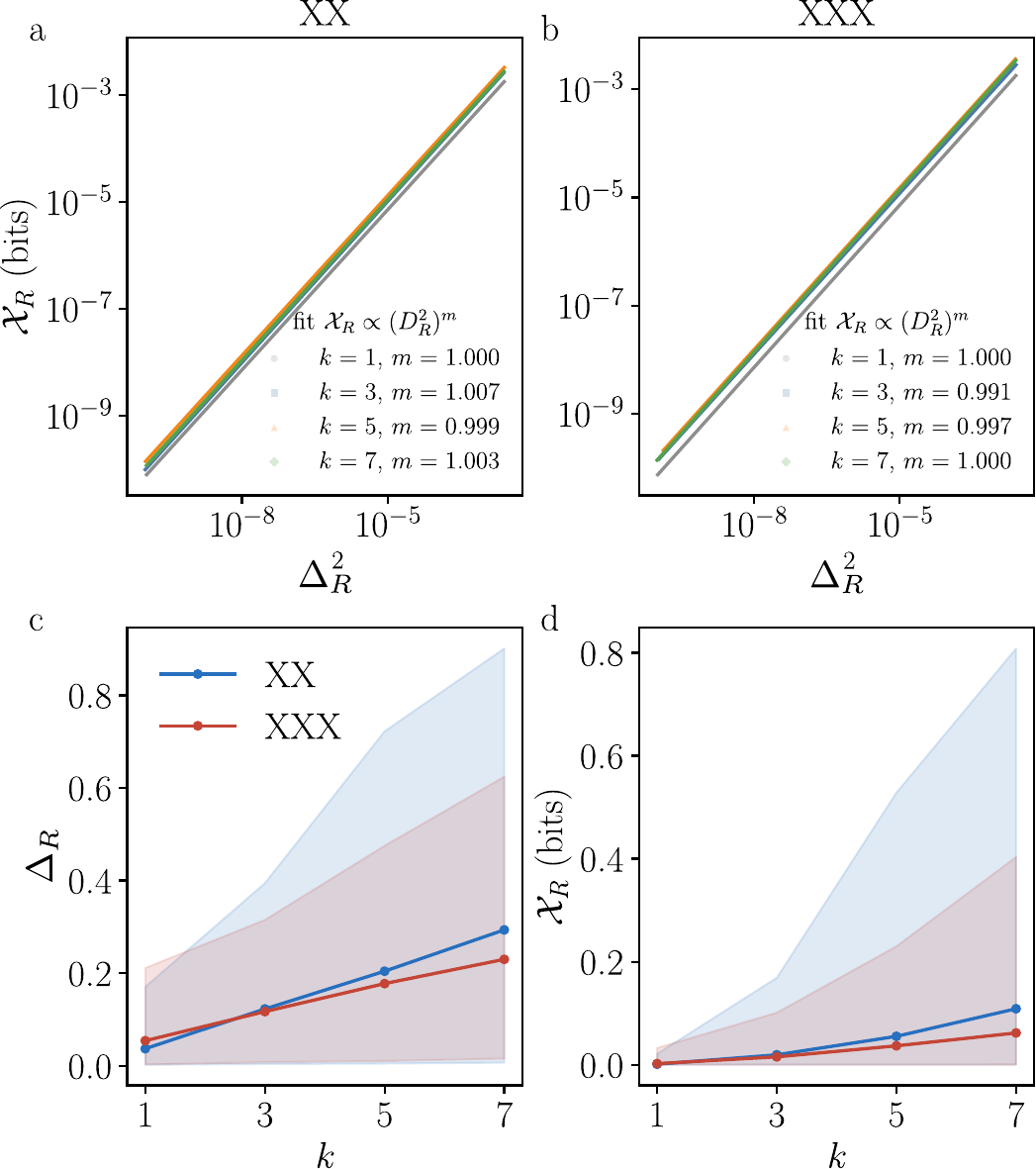}
\caption{\textbf{Weak-signal scaling and aperture dependence of the trace distance and Holevo information.} Panels (a) and (b) show $\mathcal{X}_R$ as a function of $\Delta_R^2$ for the XX and XXX chains, respectively, restricted to $10^{-5}\leq \Delta_R\leq0.05$. Colors denote the aperture width $k$, and solid lines show independent power-law fits $\mathcal{X}_R\propto(\Delta_R^2)^m$. The fitted exponents remain close to $m=1$ for every model and aperture, confirming the quadratic scaling $\mathcal{X}_R\propto \Delta_R^2$ when the two local branches are nearly indistinguishable. Panels (c) and (d) show the median $\Delta_R$ and $\mathcal{X}_R$, respectively, as a function of $k$, with shaded regions indicating the 5th--95th percentile range over the resolved space-time points. Both quantities increase with aperture, while the separation between XX and XXX becomes more pronounced for the larger observation regions.}
    \label{fig:aperture-summary}
\end{figure}

Indeed, as shown in Fig.~\ref{fig:aperture-summary} (a,b) we can numerically verify the scaling $\Holevo_R\propto \Delta_R^2$ throughout the small-$\Delta_R$ regime for every model and aperture considered. At one site the coefficient approaches $1/(2\ln2)$, so that
\begin{equation}
 \Holevo_1\simeq
 \frac{2}{\ln2}\left|G^R_{zz}\right|^2
 \qquad (\Delta_1\ll1).
 \label{eq:holevo-green-small}
\end{equation}
Thus the one-site result saturates the quadratic Pinsker coefficient in the limit $\Delta_1\rightarrow0$. In addition, here we find that  the light cone and oscillatory structure inherited from the many-body response remain present, but low-amplitude portions of that response contribute much less information about the source label.

Next, we compare how both the trace distance and the Holevo information depend on the observer aperture in Fig.~\ref{fig:aperture-summary} (c,d). Both quantities increase with $k$, as expected from data processing, but their growth differs noticeably between the two models. At $k=1$, XXX has the larger median trace distance, whereas for wider apertures XX overtakes it and the separation grows with $k$. The same behavior is even more pronounced for the Holevo information. At $k=7$, the median trace distances are $0.2936$ and $0.2300$ for XX and XXX, respectively, while the corresponding Holevo informations are $0.1087$ and $0.0618$ bits. Thus, the broader trace-distance wake observed in XX is accompanied by a larger amount of source information becoming accessible as the observation region grows.

This difference is especially interesting in light of the distinct condensed-matter dynamics of the two models. The XX and XXX chains respectively realize ballistic and superdiffusive spin-transport regimes, while the isotropic chain also exhibits the persistent oscillatory quantum wake associated with its spinon continuum~\cite{lee2026benchmarking}. Our results suggest that these differences are reflected not only in conventional response functions, but also in how information about the local perturbation is distributed across finite regions of the evolving many-body state.

Finally, $\Delta_R$ and $\Holevo_R$ remain distinct away from the small-$\Delta_R$ regime. This becomes particularly clear when comparing regions with similar trace distance. Fixing $\Delta_R$ fixes the optimal binary guessing probability, but does not fix how much information about the source label is contained in the reduced state. Indeed, $\Delta_R$ depends only on the trace norm of the state difference $\xi$, whereas $\Holevo_R$ also depends on the average state $\bar\rho$ and on how the two states are embedded within it. Two regions can therefore be equally useful for deciding which pulse was applied while carrying different amounts of source information.

Supplemental Sec.~\ref{supp:DX} makes this comparison directly. After conditioning on $\Delta_R$, the residual Holevo information retains a structured dependence on space and time in both models. Thus, regions with comparable binary distinguishability can still carry different amounts of information about the source label. Supplemental Sec.~\ref{supp:numerical-domain} then separates this effect from the location of the observer. Allowing a smaller contiguous window to move recovers much of the signal missed by a fixed centered observer, but even the best one-site observer typically retains only a fraction of the distinguishability available in the seven-site region. Allowing noncontiguous subsets does not further reduce the number of spins needed to retain $90\%$ or $95\%$ of the seven-site Holevo information relative to the best contiguous block.

\section{How the record is encoded}
\label{sec:encoding}

\subsection{A common-basis one-site record}

The aperture results tell us how much of the source record is available, but not how a concrete measurement can extract that information. We begin with the simplest case of a one-site observer. Any rank-one projective measurement on a qubit is specified by a unit vector $\bm n \in \mathbb{R}^3$ that defines a measurement direction on the Bloch sphere:
\begin{equation}
 \Pi_y^{(\bm n)}
 =
 \frac12\left(I+y\bm n\!\cdot\!\bm\sigma\right),
 \quad
 p(y|s,\bm n)
 =
 \Tr\!\left[\Pi_y^{(\bm n)}\rho_s\right],
 \label{eq:projective-measurement}
\end{equation}
where $y=\pm1$ is the measurement outcome and $\bm \sigma = (\sigma^x,\sigma^y,\sigma^z)$ denotes the vector of Pauli operators. For a fixed direction $\bm n$, these conditional probabilities define a classical channel from the source label $C$ to the measurement outcome $Y_{\bm n}$.

The amount of information about the source obtained from the measurement is quantified by the classical mutual information~\cite{cover1991elements,wilde2013quantum},
\begin{align}
 I(C{:}Y_{\bm n})
 &=
 \sum_{s,y}
 p(s,y)
 \log_2
 \frac{p(s,y)}{p(s)p(y)}\nonumber\\
 &=
 H(C)-H(C|Y_{\bm n}),
 \label{eq:classical-mutual-information}
\end{align}
where
\begin{equation}
 H(C)=-\sum_s p(s)\log_2 p(s),
\end{equation}
and
\begin{equation}
 H(C|Y_{\bm n})
 =
 -\sum_{s,y}p(s,y)\log_2 p(s|y)
\end{equation}
are the Shannon entropy of the source and the conditional entropy after the measurement, respectively. Since the two pulse directions are equiprobable, $H(C)=1$ bit. Thus, $I(C{:}Y_{\bm n})$ measures how much, on average, observing the measurement outcome reduces our uncertainty about which pulse was applied.

We then define the largest mutual information that can be obtained from a one-qubit projective measurement as
\begin{equation}
 I_{\rm proj}^{(1)}
 \coloneqq 
 \max_{\norm{\bm n}=1}
 I(C{:}Y_{\bm n}).
 \label{eq:projective-info}
\end{equation}
Of course, a projective measurement is only one possible way of interrogating the qubit. Thus, allowing an arbitrary POVM $\{M_y\}$ gives the one-copy accessible information
\begin{equation}
 I_{\rm acc}^{(1)}
 \coloneqq 
 \max_{\{M_y\}}
 I(C{:}Y).
 \label{eq:accessible-info}
\end{equation}
Since projective measurements form a restricted measurement family, and any measurement can only decrease the mutual information $I(C{:}R)=\Holevo_1$ present before measurement, we have
\begin{equation}
 I_{\rm proj}^{(1)}
 \leq
 I_{\rm acc}^{(1)}
 \leq
 \Holevo_1.
 \label{eq:accessible-chain}
\end{equation}
The last inequality is the Holevo bound~\cite{holevo1973bounds}.

\begin{figure*}[t]
 \centering
 \includegraphics[width=0.98\textwidth]{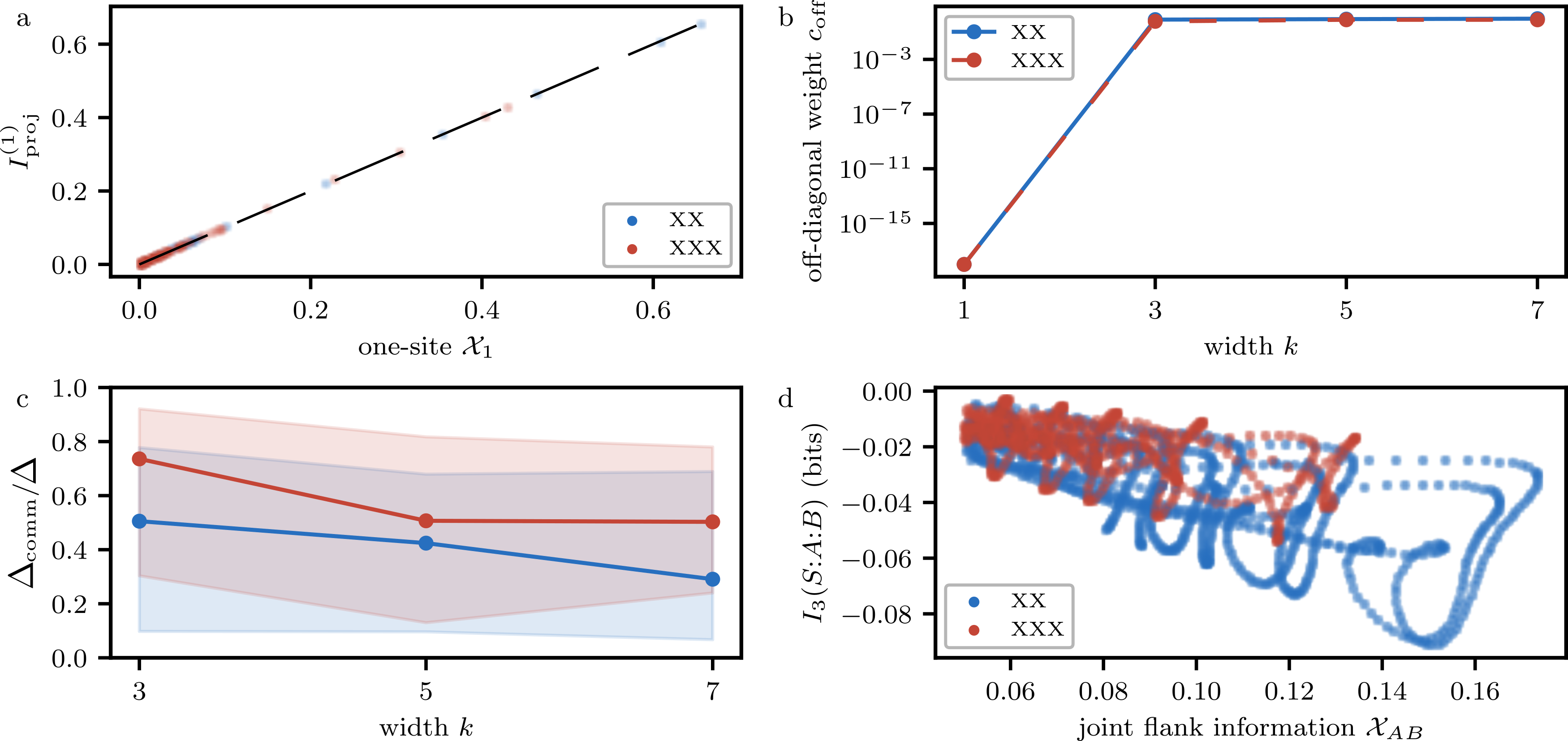}
 \caption{\textbf{From a common-basis one-site record to joint multisite encoding.} (a) The optimized one-qubit projective information equals the one-site Holevo information. (b) The off-diagonal weight $c_{\rm off}$ of $\xi$ in the eigenbasis of $\rbar$ becomes nonzero on wider blocks. (c) Median $\Delta_{\rm comm}/\Delta$ with 5th to 95th percentile ranges. Less of the XX decision bias survives the commuting restriction. (d) For the nonlocal width-seven geometry $A|\mathrm{center}|B$, the classical source-label tripartite information is negative across the nonlocal region shown.}
 \label{fig:encoding}
\end{figure*}

For a one-site observer, the encoding is particularly simple because the $U(1)$ symmetry of the dynamics constrains the structure of the reduced state. Let
\begin{equation}
    Q\coloneqq \sum_j S_j^z
\end{equation}
denote the total magnetization. Since $[H_\epsilon,Q]=[P,Q]=0$, both the source pulse and subsequent evolution preserve the zero-magnetization sector of the initial ground state. Consequently, the global states are invariant under rotations generated by $Q$, with 
\begin{equation}
    e^{-i\theta Q}=\bigotimes_j e^{-i\theta S_j^z}.
\end{equation}
Since the global rotation factorizes into local rotations, this invariance is preserved under the partial trace over all spins except site $i$. It follows that the reduced state of any individual spin, denoted by $\rho_{s,i}$ for pulse direction $s=\pm1$, is invariant under rotations about its local $z$ axis,
\begin{equation}
    e^{-i\theta S_i^z}\rho_{s,i}e^{i\theta S_i^z}=\rho_{s,i},
\end{equation}
and hence
\begin{equation}
    [\rho_{s,i},S_i^z]=0.
\end{equation}
The one-site reduced states therefore take the form
\begin{equation}
    \rho_{s,i}
    =
    \frac{1}{2}\left(I+m_{s,i}\sigma_i^z\right)\,,
\end{equation}
for some $-1\leq m_{s,i}\leq 1$. In particular, $\rho_{+,i}$ and $\rho_{-,i}$ are diagonal in the same local $z$ basis, so that
\begin{equation}
    [\rho_{+,i},\rho_{-,i}]=0.
\end{equation}

Both Hamiltonians also possess a global spin-flip symmetry. Let $F=\prod_j\sigma_j^x$, then $[F,H_\epsilon]=0$ and $FPF=-P$. Since the initial ground state has a definite $F$-parity, the two pulse branches are exchanged by $F$. Combining this with the $z$-diagonal one-site form gives
\begin{equation*}
m_{-,i}=-m_{+,i},\qquad \bar\rho_i=\frac{I}{2}.
\end{equation*}
This additional symmetry is what fixes the one-site average state to be maximally mixed; $U(1)$ symmetry alone fixes only the common $z$ basis.

This common eigenbasis has a direct operational consequence. Measuring $\sigma^z$ maps the two quantum states to their corresponding classical probability distributions without losing any information about the source label. The resulting classical mutual information therefore attains the Holevo information, and we have
\begin{equation}
 I_{\rm proj}^{(1)}
 =
 I_{\rm acc}^{(1)}
 =
 \Holevo_1.
 \label{eq:one-site-saturation}
\end{equation}
Figure~\ref{fig:encoding}(a) shows this equality throughout the one-site data.

Equation~\eqref{eq:one-site-saturation} means that all of the source information contained in a single spin can be extracted with a simple measurement of $\sigma^z$. It does not mean that a single spin necessarily contains a large amount of information. When $\Holevo_1$ is small, the measurement can saturate the Holevo bound while revealing very little about which pulse was applied. Similarly, when $\Delta_1$ is small, even the optimal probability of identifying the pulse direction remains close to one half.

\subsection{Coherences used by an optimal decision}

The common-basis structure of the one-site problem does not extend to larger apertures. The $U(1)$ symmetry still implies that a reduced state on a region $R$ commutes with its total magnetization,
\begin{equation}
    \left[\rho_{s,R},Q_R\right]=0,
    \qquad
    Q_R=\sum_{j\in R}S_j^z .
\end{equation}
For a single spin, the eigenspaces of $Q_R$ are one dimensional, which forces both states to be diagonal in the same $z$ basis. For several spins, however, a fixed-magnetization sector generally contains more than one basis state. The symmetry therefore only requires $\rho_+$ and $\rho_-$ to be block diagonal in the same magnetization sectors, but the two states need not commute within those blocks.

To quantify this difference, we write the spectral decomposition of the sign-forgotten state as
\begin{equation}
    \rbar=\sum_\lambda\lambda\Pi_\lambda\,,
\end{equation}
where $\lambda$ denote the eigenvalues of $\rbar$ and $\Pi_\lambda$ the projector onto the respective eigenspace. Then, define the pinching map
\begin{equation}
 \mathcal P_{\rbar}(A)=\sum_\lambda\Pi_\lambda A\Pi_\lambda.
 \label{eq:pinching}
\end{equation}
This map removes the components of an operator that connect distinct eigenspaces of $\rbar$. We therefore use
\begin{equation}
 c_{\rm off}
 \coloneqq
 \frac{\norm{\xi-\mathcal P_{\rbar}(\xi)}_2}
      {\norm{\xi}_2}
 \label{eq:coff}
\end{equation}
to measure how much of the signed response lies off diagonal in this basis. At one site, $c_{\rm off}=0$ exactly (see  above). As shown in Fig.~\ref{fig:encoding}(b), once the observer contains several spins, the off-diagonal contribution becomes sizeable, showing that the difference between the two states is no longer contained only in populations of a common basis.

Importantly, this distinction matters directly for the binary decision problem. Suppose that the observer is restricted to observables that commute with $\rbar$. The largest decision bias available under this restriction is
\begin{align}
 \Delta_{\rm comm}
 &\coloneqq 
 \norm{\mathcal P_{\rbar}(\xi)}_1
 \nonumber\\
 &=
 \max_{\substack{O=O^\dagger,\ \norm{O}_\infty\leq1,\\
                  [O,\rbar]=0}}
 \abs{\Tr(O\xi)}.
 \label{eq:Dcomm}
\end{align}
We define
\begin{equation}
 g_{\rm coh}
 \coloneqq 
 1-\frac{\Delta_{\rm comm}}{\Delta}
 \label{eq:gcoh}
\end{equation}
as the fraction of the optimal decision bias that is lost under this commuting-readout restriction. Equivalently, $\Delta_{\rm comm}/\Delta$ tells us how much of the optimal binary decision can still be recovered after the off-diagonal part of $\xi$ has been discarded.

Figure~\ref{fig:encoding}(c) shows that this effect differs noticeably between the two models. At width seven, the median ratio $\Delta_{\rm comm}/\Delta$ is $0.2908$ in XX and $0.5030$ in XXX. Thus, a readout that ignores these coherences retains only about $29\%$ of the optimal XX decision bias, compared with about $50\%$ in XXX. The optimal decision in XX therefore makes substantially greater use of the off-diagonal part of the signed response.

From the many-body point of view, the change from one to several spins is simple to interpret. For a one-site observer, reversing the pulse changes only the probabilities of the local $z$ states. For a larger region, the pulse also changes coherences between many-body states within the same local magnetization sector. The source record seen by the larger observer is therefore carried by more than population changes in a single common basis.

\subsection{Joint source information across separated flanks}

The previous subsection shows that the two states no longer share a common basis once several spins are observed. This still does not tell us how the source information is distributed within the block. In particular, as the perturbation propagates away from the source, we can ask whether the left and right parts of the wake carry information independently, or whether they become more informative when they are observed together.

For the width-seven block, we use the partition
\begin{equation}
 A\,|\,\mathrm{center}\,|\,B,
 \label{eq:flank-geometry}
\end{equation}
where $A$ and $B$ are the left and right three-site flanks and the central site is omitted. Using $C$ for the classical source label as in Eq.~\eqref{eq:cq-state}, $I(C{:}A)$ and $I(C{:}B)$ quantify how much each flank reveals about the pulse direction on its own, while $I(C{:}AB)$ gives the information available when the two flanks are observed jointly.

We compare these quantities through
\begin{align}
 I_3(C{:}A{:}B)
 &\coloneqq 
 I(C{:}A)+I(C{:}B)-I(C{:}AB)
 \nonumber\\
 &=
 \Holevo_A+\Holevo_B-\Holevo_{AB}\,,
 \label{eq:i3}
\end{align}
where in the last equality we used Eq.~\eqref{eq:Holevo-mutual}. Using the chain rule for conditional mutual information, the same quantity can be written as
\begin{equation}
 I_3(C{:}A{:}B)
 =
 I(C{:}A)-I(C{:}A|B).
 \label{eq:i3-conditional}
\end{equation}
Thus, $I_3<0$ means that $A$ carries more information about the source once $B$ is also known than it does on its own. Equivalently, the two flanks together contain more source information than the sum of their separate contributions. We use negative $I_3$ as a signature of joint, or synergy-like, source encoding.

Figure~\ref{fig:encoding}(d) shows this quantity for the nonlocal width-seven data with $t\geq1$, $|d|\geq6$, and $\Holevo_{AB}\geq0.05$. Although $I_3$ is negative in both models, the two distributions are strikingly different. In XX, increasing $\Holevo_{AB}$ is accompanied by several pronounced branches extending to increasingly negative $I_3$, with values reaching below $-0.08$ bits. By contrast, the XXX data remain much more tightly concentrated near zero, with most observations lying between roughly $-0.01$ and $-0.03$ bits even over a comparable range of $\Holevo_{AB}$.

This separation is especially clear when the two models are compared at similar values of $\Holevo_{AB}$. The XX flanks can carry substantially more information jointly than is visible from the two flanks separately, whereas in XXX a larger fraction of the same total flank information is already present in the individual marginals. The way in which the propagating record is distributed between the two sides of the block is therefore qualitatively different in the two models.

This behavior is not tied to a single choice of partition. Supplemental Sec.~\ref{supp:i3} and Fig.~\ref{fig:s5} repeat the calculation for several choices of $A$ and $B$. The magnitude of $I_3$ depends on how the block is divided and becomes small when the chosen regions themselves contain little source information. The joint contribution therefore follows the spatial structure of the propagating record rather than being a uniform property of the seven-site block.

A natural question is whether this joint source information requires entanglement between the two observed regions. We can test this directly using a width-three block. After tracing out the central spin, the two flanks are single qubits, so positivity under partial transpose is equivalent to separability. In the nonlocal weak-signal regime, the positive-pulse, negative-pulse, and sign-forgotten two-qubit flank states are all separable. Nevertheless, as shown in Supplemental Sec.~\ref{supp:i3} and Fig.~\ref{fig:s5}(b), the XX data retain a pronounced negative $I_3$, whereas the XXX data remain concentrated much closer to zero. Thus, the two separated spins can carry the source information jointly even when they are not entangled. Moreover, the stronger joint encoding found in XX survives under this restriction, showing that the XX--XXX separation cannot be attributed simply to entanglement between the two flanks.

\section{Representation and physical readout}
\label{sec:inversion}

\subsection{What covariance information reconstructs}

The XX and XXX chains provide a natural setting in which to ask how much of the source record can be retained by a compact representation of the local state. After a Jordan--Wigner transformation, the XX Hamiltonian is quadratic in fermionic operators, so its dynamical states remain fermionic Gaussian and are completely specified by their two-point correlations. The XXX interaction breaks this Gaussian structure. We can therefore use the same fermionic description in both models and ask how much information is lost when it ceases to be exact.

More concretely, a $k$-site spin block is represented by $2k$ Majorana operators $c_a$. Rather than retaining the full reduced density matrix, whose number of parameters grows exponentially with $k$, we keep its fermionic covariance matrix, defined as
\begin{equation}
 \Gamma_{ab}(\rho)
 \coloneqq 
 \frac{i}{2}\Tr\!\left(\rho[c_a,c_b]\right).
 \label{eq:covariance}
\end{equation}
This matrix contains all two-point Majorana correlations and requires only $O(k^2)$ parameters. For a fermionic Gaussian state it is a complete description of the state~\cite{bravyi2004lagrangian,gigena2015entanglement,gigena2016one, di2019fermionic,tarabunga2026computable}.

For each physical state $\rho_s$, we construct the unique fermionic Gaussian state $\rho_s^G$ whose covariance matrix agrees with that of $\rho_s$. Explicitly, we bring $\Gamma(\rho_s)$ to its canonical block-diagonal form, use the resulting normal-mode eigenvalues to fix the occupations of the corresponding Gaussian modes, and then transform back to the original Majorana basis. By construction,
\begin{equation}
    \Gamma(\rho_s^G)=\Gamma(\rho_s),
\end{equation}
although $\rho_s^G$ need not agree with $\rho_s$ beyond its two-point fermionic correlations. We then define
\begin{equation}
 \rbar^G\coloneqq\frac{\rho_+^G+\rho_-^G}{2},
 \qquad
 \xi^G\coloneqq\frac{\rho_+^G-\rho_-^G}{2}.
 \label{eq:gaussian-branches}
\end{equation}

For XX, this reconstruction is exact. The ground state is fermionic Gaussian, the local $z$ rotation is a Gaussian unitary, the time evolution is generated by a quadratic Hamiltonian, and partial trace preserves Gaussianity. Hence
\begin{equation}
 \rho_s^G=\rho_s,
 \qquad
 \xi^G=\xi
\end{equation}
for every time and observation region. The XXX chain provides a nontrivial test. The $Z_iZ_{i+1}$ interaction breaks Gaussianity, and we can ask how much of the physical source record is nevertheless retained by covariance information alone.

\begin{figure}[t]
 \centering
 \includegraphics[width=0.98\columnwidth]{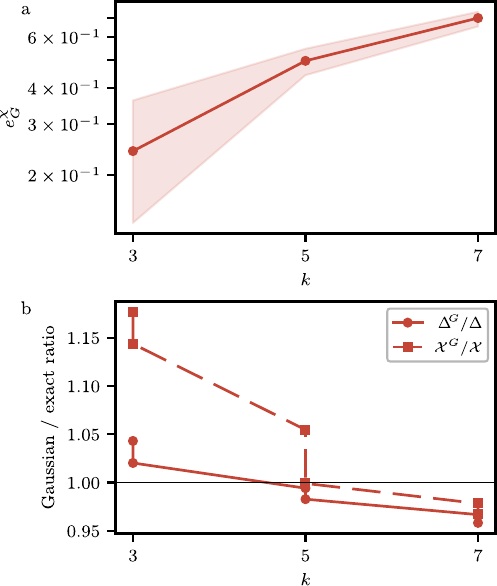}
\caption{\textbf{Gaussian reconstruction of the XXX source record at the operator and scalar levels.} (a) Median normalized signed-response error $\eG$ as a function of the aperture $k$, with shaded regions showing the 5th--95th percentile range. The covariance reconstruction becomes progressively less accurate at the operator level as the aperture grows. (b) Ratios of the Gaussian and physical trace distances, $\Delta^G/D$, and Holevo informations, $\Holevo^G/\Holevo$. The horizontal line marks exact agreement. Despite the growing operator-level error in panel (a), both scalar quantities remain close to their physical values.}
 \label{fig:gaussianity}
\end{figure}

We compare the covariance reconstruction at two levels. First, at the operator level, we use the signed-response error
\begin{equation}
 \eG\coloneqq\frac{\norm{\xi-\xi^G}_1}{\norm{\xi}_1}.
 \label{eq:eG}
\end{equation}
Second, we compare the trace distance and Holevo information of the Gaussian surrogate ensemble,
\begin{equation}
 \Delta^G\coloneqq \norm{\xi^G}_1,\qquad
 \Holevo^G\coloneqq S(\rbar^G)-\frac{S(\rho_+^G)+S(\rho_-^G)}{2}.
 \label{eq:gaussian-scalars}
\end{equation}
The distinction is important. Two operators can be far apart while having similar trace norms, and two different ensembles can have similar Holevo information. Thus, $\Delta^G/\Delta$ or $\Holevo^G/\Holevo$ near one does not imply that the signed operator or the physical states have been reconstructed accurately.

Figure~\ref{fig:gaussianity} shows the nontrivial XXX covariance reconstruction (we omit the XX reconstruction, as is exact by the argument above.) At the operator level, the reconstruction becomes steadily worse as the aperture grows. The median error $\eG$ increases from about $0.24$ at $k=3$ to almost $0.70$ at $k=7$, showing that the covariance matrix captures only a limited part of the full signed-response operator on the larger blocks.

The scalar quantities tell a rather different story. As shown in Fig.~\ref{fig:gaussianity}(b), both $\Delta^G/\Delta$ and $\Holevo^G/\Holevo$ remain comparatively close to one over the same range of apertures. At $k=3$ the Gaussian surrogate somewhat overestimates both quantities, particularly the Holevo information, while at $k=7$ both ratios lie only a few percent below one. Thus, even when $\xi^G$ is far from the physical $\xi$ as an operator, its trace norm and the Holevo information of the corresponding ensemble can still be reproduced rather well.

This distinction is important for what follows. Agreement of a few scalar quantities does not imply that covariance data provide an accurate representation of the physical state or response. In XXX, the covariance matrix can reproduce approximately the total distinguishability and source information while missing much of the operator structure that carries them. The same limitation appears for other properties of the local state. Supplemental Sec.~\ref{supp:gaussian-ent} and Fig.~\ref{fig:s6} show that the Gaussian surrogate also fails to reproduce the internal entanglement of the XXX states, with larger signed-response errors accompanied by larger errors in the logarithmic negativity.

\subsection{The model ordering reverses under spin readout}

\begin{figure*}[t]
 \centering
 \includegraphics[width=0.98\textwidth]{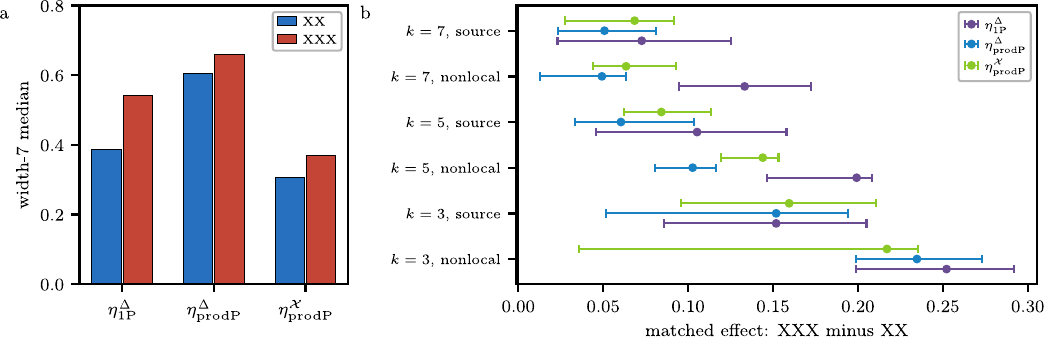}
\caption{\textbf{Restricted spin measurements recover a larger fraction of the source record in XXX.} (a) Width-seven median accessibility for the best single Pauli string, $\etaone$, and for product-Pauli measurements optimized for binary decision, $\etaD$, or source-label mutual information, $\etaX$. All three fractions are larger in XXX. (b) XXX-minus-XX differences after comparing observations with similar available information. Decision readouts are matched at comparable trace distance $\Delta$, while the information readout is matched at comparable $\Holevo$. The comparison is performed separately for each aperture and for source and nonlocal regions. All displayed differences remain positive.}
 \label{fig:inversion}
\end{figure*}

The covariance comparison gives a very simple ordering of the two models. In XX, the local branches are exactly Gaussian and are therefore completely determined by their covariance matrices. In XXX, this description is incomplete and the full signed-response operator is poorly reconstructed on the larger apertures. We now ask a different question: \textit{how much of the same source record can be accessed through simple measurements of the physical spins?}

We consider three readout tasks. The first restricts the observer to a single Pauli string,
\begin{equation}
 \etaone=\frac{\Delta_{\rm 1P}}{\Delta},\qquad
 \Delta_{\rm 1P}
\coloneqq 
 \max_{\alpha\neq I^{\otimes k}}
 \abs{\Tr(P_\alpha\xi)},
 \label{eq:eta1P}
\end{equation}
where $\alpha\in\{I,X,Y,Z\}^{\otimes k}$. Here a single string may act nontrivially on several spins, but it returns only one binary correlator.

We next consider measurements in which each spin is measured in a local Pauli basis. For a choice $\boldsymbol{\alpha}=(\alpha_1,\ldots,\alpha_k)$, with $\alpha_j\in\{X,Y,Z\}$, a single shot produces the full outcome string $\bm y=(y_1,\ldots,y_k)$, where $y_j=\pm1$. We retain this complete outcome string rather than compressing it to a single Pauli correlator. The measurement therefore defines the classical distributions $p(\bm y|s,\boldsymbol{\alpha})$ for the two source branches.

For a fixed product basis, the optimal classical decision bias is the total-variation distance between these two outcome distributions. We therefore define
\begin{equation}
\Delta_{\rm prodP}
 \coloneqq 
 \max_{\boldsymbol{\alpha}}
 \frac12\sum_{\bm y}
 \left|p(\bm y|+,\boldsymbol{\alpha})
       -p(\bm y|-,\boldsymbol{\alpha})\right|,
 \label{eq:DprodP}
\end{equation}
and, denoting the complete outcome string in basis $\boldsymbol{\alpha}$ by $Y_{\boldsymbol{\alpha}}$,
\begin{equation}
I_{\rm prodP}
 \coloneqq 
 \max_{\boldsymbol{\alpha}} I(C{:}Y_{\boldsymbol{\alpha}}).
 \label{eq:IprodP}
\end{equation}

We then optimize the choice of local Pauli axes for two different tasks. For binary discrimination, we use the full outcome distribution to make the optimal classical decision between the two source signs. For source information, we instead maximize the mutual information between the source label and the complete outcome string. This gives
\begin{equation}
 \etaD=\frac{\Delta_{\rm prodP}}{\Delta},\qquad
 \etaX=\frac{I_{\rm prodP}}{\Holevo}.
 \label{eq:product-accessibility}
\end{equation}
Thus, $\etaD$ asks what fraction of the optimal Helstrom decision bias can be recovered with product-Pauli measurements, while $\etaX$ asks what fraction of the source-label Holevo information can be extracted with the same class of measurements.

The result is opposite to the covariance ordering. In particular, Table~\ref{tab:inversion} and Fig.~\ref{fig:inversion}(a) summarize the width-seven readout comparison. All three restricted readout fractions are larger in XXX. The best single Pauli string recovers $0.3868$ of the optimal decision bias in XX and $0.5424$ in XXX. Product-Pauli measurements recover $0.6050$ and $0.6604$ of the decision bias, respectively, while the corresponding fractions of the Holevo information are $0.3066$ and $0.3697$. Thus, although the XX branches admit an exact Gaussian description, a smaller fraction of their local source record is accessible through these direct spin measurements.

\begin{table}[b]
 \caption{\textbf{Width-seven readout fractions.} Restricted spin measurements recover a larger fraction of the available source record in XXX, despite the exact Gaussian description of the XX branches.}
 \label{tab:inversion}
 \small
 \begin{ruledtabular}
 \begin{tabular}{@{}lcc@{}}
 Quantity & XX & XXX\\
 \hline
 Covariance reconstruction & exact & $\eG=0.6982$\\
 One-string decision fraction $\etaone$ & $0.3868$ & $0.5424$\\
 Product-Pauli $\Delta$ fraction $\etaD$ & $0.6050$ & $0.6604$\\
 Product-Pauli $\Holevo$ fraction $\etaX$ & $0.3066$ & $0.3697$\\
 \end{tabular}
 \end{ruledtabular}
\end{table}

A comparison of the raw medians is not quite sufficient, since XX and XXX do not generate identical wakes. In particular, one model may simply contain more distinguishability or more Holevo information at a given space-time point. We therefore compare the two models at similar values of the relevant information budget. For the two decision tasks we match on $\Delta$, while for the mutual-information task we match on $\Holevo$. The comparison is performed separately for each aperture and for source and nonlocal regions, with the two models matched in the relevant information budget and uncertainty obtained by moving-time-block resampling. The precise activity masks, matching rule, and resampling procedure are given in Supplemental Sec.~\ref{supp:matching-domain}.

Figure~\ref{fig:inversion}(b) shows the resulting XXX-minus-XX differences. They are positive for every readout task, aperture, and spatial region shown. The reversal is therefore not explained simply by different signal strengths in the two wakes. At comparable local distinguishability or source information, the restricted Pauli measurements still recover a larger fraction of the available record in XXX.

\begin{figure*}[t]
 \centering
 \includegraphics[width=0.98\textwidth]{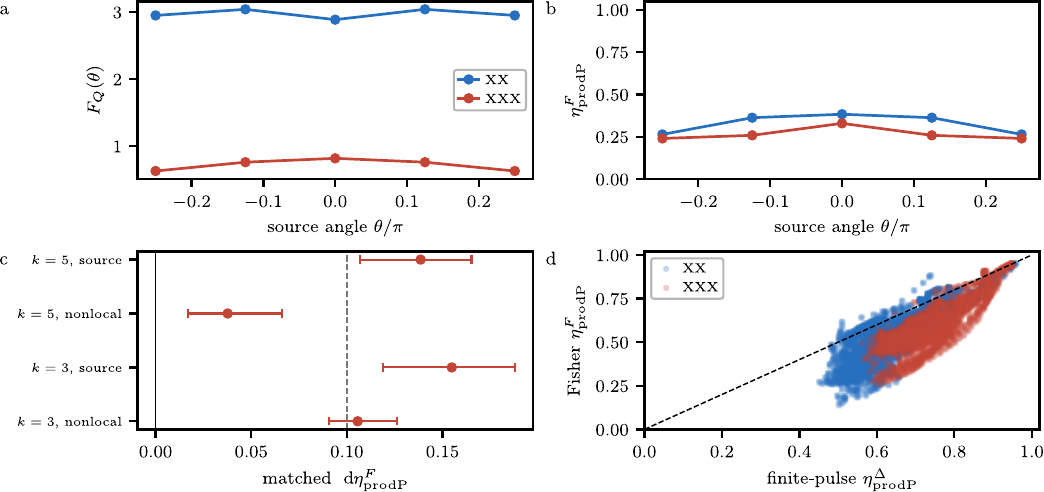}
\caption{\textbf{Source-angle sensitivity and its accessibility under product-Pauli readout.} (a) Quantum Fisher information $F_Q(\theta)$ for representative width-five source-region states as the pulse angle is varied. (b) Corresponding fraction $\eta_{\rm prodP}^F=F_C^{\rm prodP}/F_Q$ accessible to optimized product-Pauli measurements. For this representative trajectory, $\eta_{\rm prodP}^F$ is generally larger in XX, while the two models also have substantially different $F_Q$. (c) Matched XXX-minus-XX differences in $\eta_{\rm prodP}^F$ for widths three and five in source and nonlocal regions, after comparing observations with similar local QFI. All four matched differences are positive. (d) Product-Pauli accessibility for finite-pulse discrimination versus local angle estimation at the same physical observations. The dashed line denotes equal accessibility for the two tasks. Unless an angle dependence is shown explicitly, the Fisher quantities are evaluated at $\theta=0$.}
 \label{fig:qfi}
\end{figure*}

This gives a clear separation between representation and physical readout. The XX dynamics are simple in the fermionic representation: after the Jordan--Wigner transformation, covariance information is complete. However, that simplicity does not imply that the corresponding information is simple to access through local spin measurements. Indeed, the results of Sec.~\ref{sec:encoding} already point in the same direction: the wider XX blocks make greater use of coherences relative to the sign-forgotten state and show a stronger joint contribution between separated flanks. The XXX dynamics break Gaussian closure, but a larger fraction of the resulting source record is nevertheless visible to the spin-Pauli measurements considered here.

From a condensed-matter perspective, this distinction is useful because the notion of ``simple'' depends on which degrees of freedom are being used. The XX chain is free after the Jordan--Wigner transformation, so its dynamics are simple in terms of fermionic modes and their covariances. The measurements considered here, however, act directly on the physical spins. Our results show that simplicity in the quasiparticle representation need not translate into simplicity of the corresponding spin readout.

Supplemental Sec.~\ref{supp:inversion-controls} and Fig.~\ref{fig:s7} show that the same reversed model ordering persists under the commuting-readout restriction introduced in Sec.~\ref{sec:encoding}. Supplemental Sec.~\ref{supp:pauli-hierarchy} examines the Pauli-weight structure of the Helstrom observable and how much of the optimal decision bias is recovered by simple low-weight constructions.

\section{Local source-angle estimation}
\label{sec:fisher}

So far, the source has carried a binary label, i.e., the observer is asked to distinguish the two quarter turns $\theta=\pm\pi/4$. We now change the task and treat the pulse angle itself as a continuous parameter,
\begin{equation}
 V(\theta)=\exp(-i\theta P).
 \label{eq:continuous-pulse}
\end{equation}
This gives a family of reduced states $\rho_R(\theta,t)$ and allows us to ask how sensitively the state in $R$ responds to a small change of the source angle.

Indeed, this infinitesimal problem is directly connected to the signed response studied above. Differentiating the local state at $\theta=0$ gives
\begin{equation}
 \left.
 \partial_\theta\rho_R(\theta,t)
 \right|_{\theta=0}
 =
 -i\,\Tr_{\bar R}
 \left[
 U(t)[P,\rho_0]U^\dagger(t)
 \right]
 =
 2\xi_R(t).
 \label{eq:chi-tangent}
\end{equation}
Thus, the same operator $\xi$ that separates the two quarter-turn states also determines the infinitesimal response of the local state to the source angle. In particular, for any observable $O_R$,
\begin{equation}
 \left.
 \partial_\theta\langle O_R\rangle_\theta
 \right|_{\theta=0}
 =
 -i\langle[O_R(t),P]\rangle.
 \label{eq:observable-angle-response}
\end{equation}
Hence the usual retarded Green function probes one particular projection of the tangent $\partial_\theta\rho_R$.

To characterize the sensitivity of the full reduced state, rather than that of a chosen observable, we use the quantum Fisher information (QFI), denoted as $\qF(\theta)$~\cite{braunstein1994statistical,hyllus2012fisher, paris2009quantum,sone2020generalized} and defined as
\begin{equation}
    \qF(\theta)
 \coloneqq
 \Tr(\rho_RL_\theta^2),
 \label{eq:qfi}
\end{equation}
where the symmetric logarithmic derivative $L_\theta$ is defined through
\begin{equation}
 \partial_\theta\rho_R
 =
 \frac12\left(L_\theta\rho_R+\rho_RL_\theta\right).
\end{equation}
The QFI determines the largest sensitivity to an infinitesimal change of $\theta$ that can be obtained by optimizing over measurements on the region $R$. For $\nu$ independent preparations, the quantum Cram\'er--Rao bound gives
\begin{equation}
 \operatorname{Var}(\hat\theta)
 \geq
 \frac{1}{\nu\qF(\theta)}
\end{equation}
for every locally unbiased estimator. 

Equations~\eqref{eq:chi-tangent}--\eqref{eq:qfi} also make the connection with the conventional response picture explicit. A Green function asks how one selected spin observable responds to the local perturbation, whereas the QFI asks how strongly the entire reduced state changes under that same infinitesimal perturbation.

A particular measurement generally accesses only part of this sensitivity. If its outcomes have probabilities $p_y(\theta)$, the corresponding classical Fisher information is
\begin{equation}
 \cF(\theta)
 \coloneqq 
 \sum_{y:p_y>0}
 \frac{[\partial_\theta p_y(\theta)]^2}{p_y(\theta)}.
 \label{eq:cfi}
\end{equation}
We optimize $\cF(\theta)$ over the same product-Pauli measurements considered for the finite-pulse problem and define
\begin{equation}
 \etaF(\theta)
 \coloneqq
 \frac{\cF^{\rm prodP}(\theta)}{\qF(\theta)}.
 \label{eq:etaF}
\end{equation}
Here, $\cF^{\rm prodP}(\theta)$ denotes the maximum classical Fisher information obtained by optimizing over all product-Pauli measurement bases, while retaining the complete string of local measurement outcomes. The ratio $\etaF(\theta)$ is therefore the fraction of the available local source-angle sensitivity that can be extracted by measuring each spin in a local Pauli basis and retaining the complete outcome string. Equivalently, $1/\etaF$ is the sampling overhead relative to an optimal measurement attaining the QFI. Thus, for a scaling family of systems, $\etaF \geq 1/\operatorname{poly}(n)$ implies only polynomial overhead, whereas an exponentially small $\etaF$ would require exponentially more measurements.

Figure~\ref{fig:qfi}(a) shows $F_Q(\theta)$ for a representative width-five region near the source, while panel (b) shows the corresponding product-Pauli accessibility $\eta_{\rm prodP}^F(\theta)$. For this particular trajectory, $\eta_{\rm prodP}^F$ is generally larger in XX. This raw comparison, however, is not a comparison at fixed available sensitivity. As seen in panel (a), the two models have substantially different values of $F_Q$, and the fraction $\eta_{\rm prodP}^F$ itself depends on the state and on the amount of QFI present.

To compare the readout efficiency of the two models on equal footing, we therefore compare XX and XXX observations with similar values of $F_Q$, separately for each aperture and spatial region. The matching is performed in $\log_{10}F_Q$ and, unless otherwise stated, at $\theta=0$. The resulting XXX-minus-XX differences are shown in Fig.~\ref{fig:qfi}(c). All four are positive. Hence, although the raw representative profile in panel (b) can favor XX, at comparable local QFI the product-Pauli measurements recover a larger fraction of the available angle sensitivity in XXX. Since the comparison in Fig.~\ref{fig:qfi}(c) is performed at matched $F_Q$, the larger $\eta_{\rm prodP}^F$ in XXX also corresponds to a larger absolute product-Pauli Fisher information $\cF^{\rm prodP}$ within the matched pairs. This should be distinguished from the raw comparison in Fig.~\ref{fig:qfi}(b), where XX can have both a larger $F_Q$ and a larger $\cF^{\rm prodP}$ for the particular trajectory shown.

Figure~\ref{fig:qfi}(d) compares $\etaD$ and $\etaF$ at the same space-time points. The two tasks are correlated but clearly not equivalent. The XX data form a relatively compact cloud at moderate $\etaD$ and lower $\etaF$, whereas the XXX data extend to larger values of both quantities and follow a visibly different curved structure. Most points in both models lie below the diagonal, so product-Pauli measurements generally recover a smaller fraction of the infinitesimal angle sensitivity than of the finite-pulse distinguishability. The different clustering further shows that the relation between the two readout tasks is itself model dependent.

Supplemental Sec.~\ref{supp:fisher-controls} and Fig.~\ref{fig:s8} ask whether the product basis optimized for the finite-pulse problem can be reused for source-angle estimation. For widths three and five, a basis optimized for the finite-pulse Holevo information retains nearly all of the Fisher information available within the product-Pauli family. By contrast, a basis optimized for binary discrimination can lose a noticeably larger fraction, particularly in the XX width-five source region.

\section{Coherent source encoding and quantum-information recovery}
\label{sec:qubit-encoding}

The preceding sections concern classical information encoded in quantum states: the direction of a finite pulse, or the value of a continuously varying source angle. We now retain the source coherently and ask two fully quantum questions. \textit{Can an observer recover a Bell state whose entanglement has been distributed between a control qubit and the chain?} \textit{At what rate can the resulting channel transmit qubits?} The same two many-body states introduced in Sec.~\ref{sec:protocol} answer both questions, but the relevant information is now the source record in the inaccessible region. We obtain an exact expression for the optimal single-copy Bell-state fidelity and prove that the coherent information is equal to one minus the complementary-region Holevo information. For the spin-flip-symmetric setting considered here, the latter expression is also the exact quantum capacity. We also connect the earlier Holevo information to classical capacity, explicitly distinguishing whether the receiver retains the control. Throughout, the preparation is pure and the chain evolution is unitary. All proofs, including an explicit optimal recovery operation, are given in Supplemental Sec.~\ref{supp:qubit-proofs}.

\subsection{Replacing the classical source by a coherent control}

Let $A_{\mathrm{in}}$ denote the control qubit before encoding and $A$ the same qubit afterward. Prepare $\alpha\ket{0}_{A_\mathrm{in}}+\beta\ket{1}_{A_\mathrm{in}}$, with $|\alpha|^2+|\beta|^2=1$, and prepare the full chain in $\ket{g_\epsilon}_{R\bar{R}}$. Instead of selecting the pulse sign classically, apply the controlled unitary
\begin{equation}
 \mathrm{C}V
 \coloneqq
 \ket{0}\!\bra{0}_{A_\mathrm{in}}\otimes V_+
 +\ket{1}\!\bra{1}_{A_\mathrm{in}}\otimes V_- .
 \label{eq:qubit-controlled-pulse}
\end{equation}
Here $V_+$ and $V_-$ denote the pulses with signs $+1$ and $-1$, respectively, as defined in Eq.~\eqref{eq:pulse}. Evolving only the chain then gives
\begin{equation}
 \ket{\Psi_{\alpha,\beta}(t)}_{AR\bar{R}}
 = \alpha\ket{0}_A\ket{\psi_{+,\epsilon}(t)}_{R\bar{R}} 
 +\beta\ket{1}_A\ket{\psi_{-,\epsilon}(t)}_{R\bar{R}},
 \label{eq:qubit-encoded-state}
\end{equation}
where $\ket{\psi_{s,\epsilon}(t)}=U_\epsilon(t)V_s\ket{g_\epsilon}$ is the state in Eq.~\eqref{eq:global-branches}. The distinction is that the relative phase between the two alternatives is now part of the information to be preserved.

We let the receiver have joint access to $A$ and $R$, while $\bar R$ is inaccessible. The preparation defines an isometry $W_{\epsilon,t}$ by $W_{\epsilon,t}\ket{x}_{A_{\mathrm{in}}} =\ket{x}_A\ket{\psi_{(-1)^x,\epsilon}(t)}_{R\bar{R}}$, for $x\in\{0,1\}$, and hence a channel $\mathcal N_{R,t}\colon A_{\mathrm{in}}\to AR$, which acts on an input density operator $\vartheta$ as
\begin{equation}
 \mathcal N_{R,t}(\vartheta)
 \coloneqq
 \Tr_{\bar R}\!\left[
 W_{\epsilon,t}\vartheta W_{\epsilon,t}^{\dagger}\right].
 \label{eq:qubit-channel}
\end{equation}
We suppress the fixed model parameter $\epsilon$ in the channel notation. For any region $X$ of the chain, $\rho_s(X,t)$ denotes the reduced branch state in Eq.~\eqref{eq:branch-state} with $R$ replaced by $X$. Discarding $A$ gives the measure-and-prepare channel that outputs $\rho_+(R,t)$ or $\rho_-(R,t)$ of Eq.~\eqref{eq:branch-state} according to the computational basis input. Such a channel is entanglement breaking: it destroys entanglement between the input and an arbitrary reference, and has zero quantum capacity. The recovery task considered below requires access to the channel $A$ as part of the receiver's output.

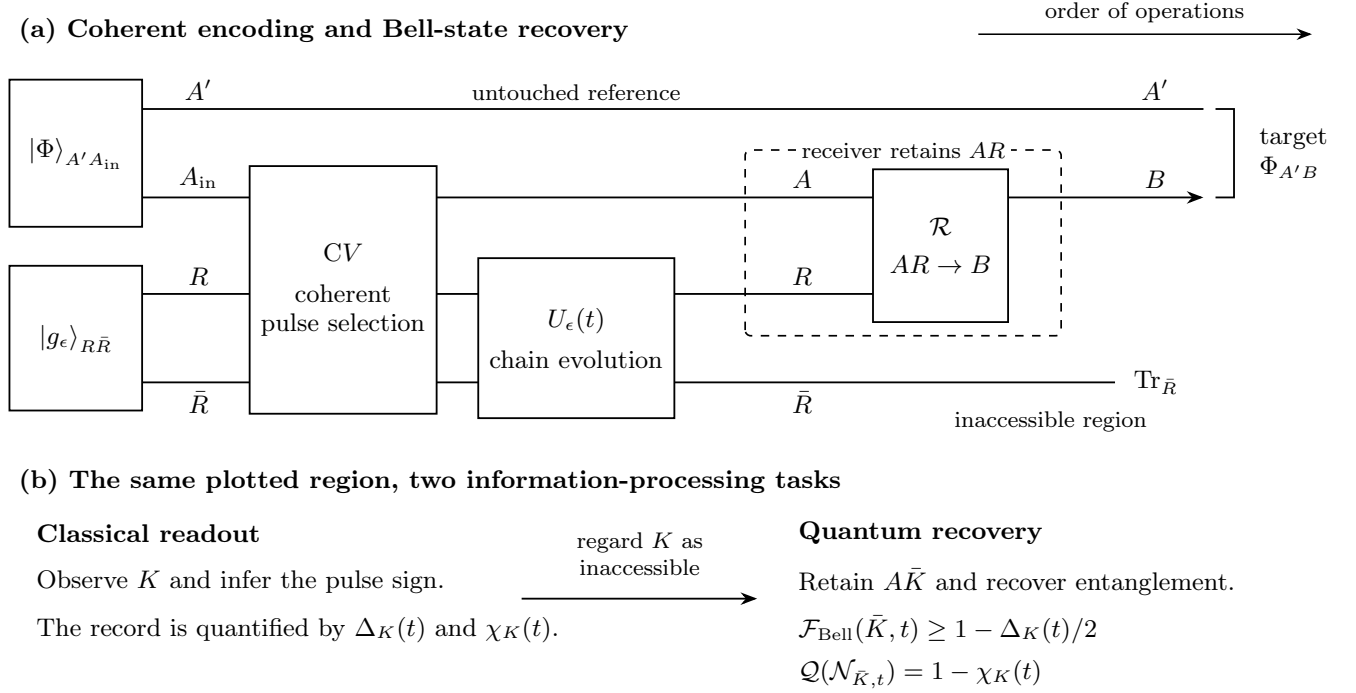
\begin{figure*}[t]
 \centering
 \resizebox{0.98\textwidth}{!}{%
 \begin{tikzpicture}[x=1cm,y=1cm,>=Stealth,
  every node/.style={font=\small},
  gate/.style={draw,line width=0.65pt,fill=white,align=center,
               inner sep=4pt},
  wire/.style={line width=0.65pt},
  heading/.style={font=\small\bfseries,anchor=west}]
 \node[heading] at (0,4.08) {(a) Coherent encoding and Bell-state recovery};
 \draw[->,line width=0.55pt] (12.05,4.03)--(16.20,4.03);
 \node[above,font=\footnotesize] at (14.12,4.06) {order of operations};
 \draw[wire] (1.65,3.10)--(14.85,3.10);
 \draw[wire] (1.65,2.00)--(10.75,2.00);
 \draw[wire] (1.65,0.80)--(10.75,0.80);
 \draw[wire] (1.65,-0.30)--(13.75,-0.30);
 \node[gate,minimum width=1.65cm,minimum height=1.83cm]
   at (0.825,2.55) {$\ket{\Phi}_{A'A_{\rm in}}$};
 \node[gate,minimum width=1.65cm,minimum height=1.80cm]
   at (0.825,0.25) {$\ket{g_\epsilon}_{R\bar R}$};
 \node[above] at (2.35,3.10) {$A'$};
 \node[above] at (2.35,2.00) {$A_{\rm in}$};
 \node[above] at (2.35,0.80) {$R$};
 \node[below] at (2.35,-0.30) {$\bar R$};
 \node[gate,minimum width=1.85cm,minimum height=3.08cm]
   at (4.15,0.85) {$\mathrm{C}V$\\[4pt] coherent\\pulse selection};
 \node[gate,minimum width=1.80cm,minimum height=1.99cm]
   at (7.05,0.25) {$U_\epsilon(t)$\\[4pt] chain evolution};
 \node[above,font=\footnotesize] at (7.05,3.10) {untouched reference};
 \draw[dashed,line width=0.6pt,rounded corners=2pt]
   (9.15,0.28) rectangle (13.08,2.59);
 \node[fill=white,inner sep=2pt,font=\footnotesize]
   at (11.12,2.59) {receiver retains $AR$};
 \node[above] at (9.86,2.00) {$A$};
 \node[above] at (9.86,0.80) {$R$};
 \node[below] at (9.86,-0.30) {$\bar R$};
 \node[gate,minimum width=1.68cm,minimum height=1.90cm]
   at (11.58,1.40) {$\mathcal R$\\[3pt] $AR\to B$};
 \draw[wire,->] (12.42,2.00)--(14.85,2.00);
 \node[above] at (14.25,3.10) {$A'$};
 \node[above] at (14.25,2.00) {$B$};
 \draw[wire] (15.00,3.10)--(15.23,3.10)--(15.23,2.00)--(15.00,2.00);
 \node[anchor=west,align=left] at (15.43,2.55) {target\\$\Phi_{A'B}$};
 \node[anchor=west] at (13.87,-0.30) {$\Tr_{\bar R}$};
 \node[below,font=\footnotesize] at (12.95,-0.54) {inaccessible region};
 \node[heading] at (0,-1.53) {(b) The same plotted region, two information-processing tasks};
 \node[anchor=west,font=\small\bfseries] at (0.22,-2.17) {Classical readout};
 \node[anchor=west] at (0.22,-2.78) {Observe $K$ and infer the pulse sign.};
 \node[anchor=west] at (0.22,-3.37) {The record is quantified by $\Delta_K(t)$ and $\chi_K(t)$.};
 \draw[->,line width=0.65pt] (6.37,-2.99)--(9.30,-2.99);
 \node[align=center,font=\footnotesize] at (7.84,-2.43)
   {regard $K$ as\\inaccessible};
 \node[anchor=west,font=\small\bfseries] at (9.70,-2.17) {Quantum recovery};
 \node[anchor=west] at (9.70,-2.78) {Retain $A\bar K$ and recover entanglement.};
 \node[anchor=west] at (9.70,-3.37)
   {$\mathcal{F}_{\mathrm{Bell}}(\bar K,t)\geq 1-\Delta_K(t)/2$};
 \node[anchor=west] at (9.70,-3.93)
   {$\mathcal{Q}(\mathcal N_{\bar K,t})=1-\chi_K(t)$};
\end{tikzpicture}%
 }
 \caption{\textbf{The source record as an obstruction to quantum recovery.} (a) One share of a Bell state coherently selects the pulse sign. The many-body evolution acts only on $R\bar R$. The receiver retains the control $A$ and the region $R$, applies an arbitrary joint decoding channel, and seeks a Bell state between the output $B$ and the untouched reference $A'$. The two lower input wires are prepared jointly in $\ket{g_\epsilon}$; they need not be in a product state. The block $\mathrm{C}V$ denotes both nontrivial pulses in Eq.~\eqref{eq:qubit-controlled-pulse}. (b) The arrow changes the interpretation of the plotted region. An existing plot for a region $K$ has a second interpretation: information about the pulse sign in $K$ obstructs recovery from $A\bar K$. The capacity equality uses the spin-flip symmetry of the present models; the fidelity bound holds for an arbitrary pure two-state preparation.}
 \label{fig:qubit-encoding}
\end{figure*}

For the entanglement-recovery experiment, initially prepare $\ket{\Phi}_{A'A_{\mathrm{in}}} =(\ket{00}+\ket{11})/\sqrt{2}$, with $A'$ an untouched reference qubit. After the controlled pulse and the evolution, the joint state is
\begin{align}
 \ket{\Omega(t)}_{A'AR \bar R}
 =\frac{1}{\sqrt{2}}\bigl(&
 \ket{0}_{A'}\ket{0}_A\ket{\psi_{+,\epsilon}(t)}_{R \bar R} \nonumber\\
 &+\ket{1}_{A'}\ket{1}_A\ket{\psi_{-,\epsilon}(t)}_{R \bar R}\bigr).
 \label{eq:qubit-bell-encoding}
\end{align}
The receiver and reference share the reduced state $\omega_{A'AR}\coloneqq\Tr_{\bar R}\ket{\Omega(t)}\!\bra{\Omega(t)}$. We use $\omega$ to distinguish this state, which includes the coherent control and reference, from the branch states $\rho_s(R,t)$  in Eq.~\eqref{eq:branch-state}. The receiver applies a decoding channel $\mathcal R\colon AR\to B$, where $B$ is an output qubit, with the goal of recovering $\Phi_{A'B}=\ket{\Phi}\!\bra{\Phi}_{A'B}$. Figure~\ref{fig:qubit-encoding} separates the coherent preparation, the inaccessible subsystem, and the recovery operation. No measurement of $A'$ or of the source sign is performed during this protocol.

\subsection{Single-copy Bell-state recovery from trace-distance data}

We define the fidelity of two states $\rho$ and $\sigma$ as
\begin{equation}
 \mathcal{F}(\rho,\sigma)\coloneqq\norm{\sqrt{\rho}\sqrt{\sigma}}_1^2.
 \label{eq:qubit-fidelity-convention}
\end{equation}
In particular, fidelity with a pure target is its projection probability. The optimal deterministic Bell-state fidelity is then given by
\begin{align}
 \mathcal{F}_{\mathrm{Bell}}(R,t)
 \coloneqq\max\bigl\{&\mathcal F\bigl(\Phi_{A'B},
 (\operatorname{id}_{A'}\otimes\mathcal R )(\omega_{A'AR})\bigr):
 \nonumber\\[-2pt]
 &\mathcal R\colon AR\to B\ \text{is a quantum channel}\bigr\}.
 \label{eq:qubit-bell-fidelity-definition}
\end{align}
Here and below, a quantum channel is completely positive and trace preserving. This is a single-copy, deterministic recovery task: no postselection is allowed, and the reference is not acted on. It is distinct from the asymptotic distillation task discussed below.

Consider the two conditional states $\rho_+(\bar R,t)$ and $\rho_-(\bar R,t)$ of the inaccessible region, obtained by tracing $R$ out of the corresponding dynamical states. They are also the conditional states obtained by measuring $A'$ and $A$ in the computational basis and obtaining $00$ or $11$, respectively. This hypothetical measurement identifies the relevant ensemble; it is not part of the recovery procedure. The exact optimum is
\begin{equation}
 \mathcal{F}_{\mathrm{Bell}}(R,t)
 =\frac{1+\sqrt{\mathcal{F}\bigl(\rho_+(\bar R,t),\rho_-(\bar R,t)\bigr)}}{2}.
 \label{eq:qubit-exact-bell-fidelity}
\end{equation}
The underlying controlled-unitary recovery is the one used in~\cite[Figure~4]{QubitHayden2013}. Supplemental Sec.~\ref{supp:qubit-proofs} derives its exact Bell-state fidelity and proves that no other decoding channel can do better. The argument uses Uhlmann's theorem~\cite{QubitUhlmann1976}, without requiring an efficient implementation of the optimizing unitary.

Applying the Fuchs--van de Graaf inequalities~\cite{QubitFuchs1999} gives bounds entirely in terms of the normalized trace distance in Eq.~\eqref{eq:Trace distance}, evaluated on $\bar R$:
\begin{equation}
 1-\frac{\Delta_{\bar R}(t)}{2}
 \ \leq\ \mathcal{F}_{\mathrm{Bell}}(R,t)
 \ \leq\ \frac{1+\sqrt{1-\Delta_{\bar R}(t)^2}}{2}.
 \label{eq:qubit-bell-trace-bounds}
\end{equation}
Both bounds are sharp when only the normalized trace distance is known, although a particular spin-chain family need not attain them. Thus, $\Delta_{\bar R}=0$ permits perfect recovery, whereas $\Delta_{\bar R}=1$ forces the optimum down to $1/2$. The physical distinction from classical readout is important: preserving an unknown superposition requires the inaccessible region to reveal little about the computational-basis value. The receiver can then align the two conditional purifications by a unitary controlled by $A$, without learning the source sign. If the environment states are identical, this alignment completely disentangles the recovered Bell pair from the chain. A measurement chosen to read the pulse sign is not, by itself, such a recovery operation.

\subsection{Coherent information and the qubit-transmission rate}

For the state in Eq.~\eqref{eq:qubit-bell-encoding}, define the coherent information by
\begin{equation}
 I(A'\rangle AR)_\omega
 \coloneqq S(AR)_\omega-S(A'AR)_\omega\,.
 \label{eq:qubit-coherent-information-definition}
\end{equation}
For comparison with the classical source ensemble, use the register $C$ from Eq.~\eqref{eq:cq-state}, and form the auxilary classical--quantum state on $C \bar R$ with the equiprobable signs, $\rho_{C\bar R} \coloneqq\frac12\sum_{s=\pm1}\ket{s}\!\bra{s}_C \otimes\rho_s(\bar R,t)$. The Holevo identity in Eq.~\eqref{eq:Holevo-mutual} then has the quantum counterpart
\begin{equation}
 I(A'\rangle AR)_\omega
 =1-I(C:\bar R)_{\rho_{C\bar R}}
 =1-\Holevo_{\bar R}(t).
 \label{eq:qubit-coherent-holevo}
\end{equation}
This quantity lies between zero and one. The control makes the computational-basis value perfectly available to the receiver, while $\Holevo_{\bar R}$ measures how much information about that value has also reached the inaccessible region.

To interpret Eq.~\eqref{eq:qubit-coherent-holevo} as a transmission rate, fix the model, evolution time, and partition. Each use of the channel starts with a newly prepared ground state, and coding is allowed across independent uses of the resulting channel. The quantum capacity is equal to the supremum of rates for which the entanglement-transmission fidelity tends to one as the number of uses grows, without assisting communication or shared entanglement. The quantum coding theorem then makes $1-\Holevo_{\bar R}(t)$ an achievable rate, in qubits per channel use~\cite{QubitLloyd1997,QubitDevetak2005}. This asymptotic statement is distinct from the fidelity of recovering one uncoded Bell pair.

In the present setting there is a further simplification. The channel is degradable, meaning that its complementary channel, which outputs the inaccessible system $\bar R$, can be simulated from $AR$ by measuring $A$ in the computational basis and preparing the corresponding state on $\bar R$. This simulation establishes a property of the channel. Degradability makes the quantum capacity equal to the optimized single-use coherent information \cite{QubitDevetakShor2005}. Moreover, the global spin-flip symmetry introduced in Sec.~\ref{sec:protocol} exchanges the two conditional states. Together with input dephasing, this symmetry makes the maximally mixed input optimal. Consequently,
\begin{equation}
 \mathcal{Q}(\mathcal N_{R,t})=1-\Holevo_{\bar R}(t),
 \label{eq:qubit-exact-capacity}
\end{equation}
where $\mathcal{Q}(\mathcal N_{R,t})$ denotes the quantum capacity of the channel $\mathcal N_{R,t}$ defined in \eqref{eq:qubit-channel}. The capacity is that of the complete channel $A_{\mathrm{in}}\to AR$, with no additional assisting communication. It does not assign a quantum capacity to the dynamics with access to $R$ alone. Supplemental Sec.~\ref{supp:qubit-proofs} proves the symmetry reduction and gives the prior-optimized formula when the exchange symmetry is absent.

There is a related but distinct statement for entanglement distillation. Suppose two parties share independent copies of $\omega_{A'AR}$, with one holding $A'$ and the other holding $AR$. They may now act locally on all of their copies and communicate classically.  Now, let $\mathcal{D}_{\to}(A':AR)_\omega$ denote the Bell-pair distillation rate with classical communication from $A'$ to $AR$, and $\mathcal{D}_{\leftrightarrow}(A':AR)_\omega$ with communication in both directions. Then,  their optimal rates, in Bell pairs per copy, satisfy
\begin{equation}
 \mathcal{D}_{\to}(A':AR)_\omega
 =\mathcal{D}_{\leftrightarrow}(A':AR)_\omega
 =1-\Holevo_{\bar R}(t).
 \label{eq:qubit-distillation-main}
\end{equation}
Unlike the channel-capacity optimization, this state identity requires no spin-flip symmetry. A hashing lower bound and a matching relative-entropy converse establish it \cite{QubitDevetakWinter2005,QubitVedralPlenio1998}; both are detailed in Supplemental Sec.~\ref{supp:qubit-proofs}.

\subsection{Classical capacity and the role of the control}
\label{sec:qubit-classical-capacity}

The coherent-control construction also strengthens the classical interpretation of the earlier data. Discarding the control defines the full qubit-input channel $\mathcal M_{R,t}\colon A_{\mathrm{in}}\to R$,
\begin{equation}
 \mathcal M_{R,t}(\vartheta)
 \coloneqq (\Tr_A\circ\mathcal N_{R,t})(\vartheta)
 =\sum_{x=0}^1\vartheta_{xx}\rho_{(-1)^x}(R,t),
 \label{eq:qubit-classical-channel}
\end{equation}
with $\vartheta_{xx}=\bra{x}\vartheta\ket{x}$. Discarding the control removes the input coherences, as anticipated after Eq.~\eqref{eq:qubit-channel}. Let $\mathcal{C}(\mathcal M)$ denote unassisted classical capacity, in bits per independent channel use, with arbitrary quantum input codewords, collective decoding, and vanishing decoding error in the limit of many uses. For the present spin-flip-symmetric model,
\begin{equation}
 \mathcal{C}(\mathcal M_{R,t})=\Holevo_R(t),
 \qquad \mathcal{C}(\mathcal N_{R,t})=1.
 \label{eq:qubit-classical-capacities}
\end{equation}
The first equality remains valid when the sender can use superpositions and entangled inputs across experiments. The second concerns a different receiver: the retained control in $AR$ preserves one perfectly readable classical bit. Supplemental Sec.~\ref{supp:qubit-proofs} proves these statements using the classical--quantum coding theorem \cite{QubitHolevo1998,QubitSchumacherWestmoreland1997} and an explicit converse for arbitrary input codewords. The first equality is consistent with entanglement-breaking-channel additivity~\cite{QubitShor2002}, not a claim of additivity for arbitrary qubit-input channels.

Combining Eqs.~\eqref{eq:qubit-exact-capacity} and \eqref{eq:qubit-classical-capacities} gives
\begin{equation}
 \mathcal{Q}(\mathcal N_{R,t})+\mathcal{C}(\mathcal M_{\bar R,t})=1.
 \label{eq:qubit-complementary-capacities}
\end{equation}
The same controlled encoding therefore relates two capacities: quantum communication to $AR$ and classical communication to the inaccessible region $\bar R$. This relation depends on which systems are delivered; it is not a simultaneous-transmission trade-off for one receiver. Likewise, $\mathcal{C}(\mathcal M_{R,t})=\Holevo_R(t)$ is a coding statement, not a claim that one measurement on a single output always attains its Holevo information.

\subsection{A quantum interpretation of the earlier plots}

The existing data can be reused directly, provided that the observed and inaccessible regions are distinguished. Let $K$ denote a contiguous region appearing in the earlier plots to distinguish the plotted region from the receiver region $R$ used above. Regard $K$ as inaccessible and give the receiver $A\bar K$. Equations~\eqref{eq:qubit-bell-trace-bounds} and \eqref{eq:qubit-exact-capacity} become
\begin{align}
 1-\frac{\Delta_K(t)}{2}
 &\leq \mathcal{F}_{\mathrm{Bell}}(\bar K,t)
 \leq\frac{1+\sqrt{1-\Delta_K(t)^2}}{2},
 \label{eq:qubit-plot-fidelity}\\
 \mathcal{Q}(\mathcal N_{\bar K,t})&=1-\Holevo_K(t).
 \label{eq:qubit-plot-capacity}
\end{align}
Thus, the trace-distance wakes in Fig.~\ref{fig:D-wakes} and the source-centered curves in Fig.~\ref{fig:D-source} supply pointwise bounds on Bell-state recovery after the plotted block becomes inaccessible. The Holevo wakes in Fig.~\ref{fig:holevo} already give the classical capacity $\mathcal{C}(\mathcal M_{K,t})$ by Eq.~\eqref{eq:qubit-classical-capacities}; they become exact quantum capacity maps under the transformation $\Holevo_K\mapsto1-\Holevo_K$. A stronger classical record in a lost block means a smaller quantum capacity for its complement and the retained control. Likewise, the increasing Holevo information with aperture in Fig.~\ref{fig:aperture-summary} describes a decreasing capacity when that aperture is interpreted as a growing inaccessible region. For a growing accessible region, both the optimal recovery fidelity and the capacity are nondecreasing.

For a single lost site $K=\{j\}$, the connection is exact even for the fidelity. The one-site symmetry analysis in Sec.~\ref{sec:encoding} gives $\rho_\pm(K,t)=(I\pm m_j(t)\sigma_j^z)/2$, with $\Delta_K(t)=|m_j(t)|$. Therefore,
\begin{align}
 \mathcal{F}_{\mathrm{Bell}}(\bar K,t)
 &=\frac{1+\sqrt{1-\Delta_K(t)^2}}{2},
 \label{eq:qubit-one-site-fidelity}\\
 \mathcal{Q}(\mathcal N_{\bar K,t})
 &=h_2\!\left(\frac{1+\Delta_K(t)}{2}\right).
 \label{eq:qubit-one-site-capacity}
\end{align}
Here $h_2(p)=-p\log_2p-(1-p)\log_2(1-p)$ is the binary entropy, with $0\log_2 0=0$. In the weak-record regime, the Bell infidelity is $\Delta_K^2/4+O(\Delta_K^4)$ and the capacity loss is $\Delta_K^2/(2\ln2)+O(\Delta_K^4)$. For $t>0$ at the source site, Eqs.~\eqref{eq:phase-cycle} and~\eqref{eq:retarded-green} give $\Delta_1=2|G^R_{zz}|$, so both losses are therefore quadratic in the magnitude of the retarded response. The corresponding coefficients are derived in Supplemental Sec.~\ref{supp:qubit-proofs}.

These reinterpretations do not require new many-body simulations. They do, however, change which subsystem is available to the receiver. Determining recovery from the original small region $AR$ instead requires the diagnostics on $\bar R$ and they are not generally determined by $\Delta_R$ and $\Holevo_R$ alone. Supplemental Sec.~\ref{supp:qubit-proofs} gives an alternative computational route using one additional reduced cross operator on $R$ and matrices of dimension at most $2^{k+1}$ when $R$ contains $k$ spins. The needed cross operator can be contracted from the coherently specified many-body states, rather than constructing a density matrix on all of $\bar R$. The resulting optima allow arbitrary joint recovery operations, while no restriction to the spin-readout families studied earlier is imposed.

The entanglement considered here is shared between an external reference and a coherently encoded qubit. We next return to the entanglement already present within the many-body states that carry the source record.

\section{Entanglement in the states carrying the record}
\label{sec:entanglement}

Entanglement already enters the dynamical structure factor problem in an important way. The dynamical susceptibility can be integrated to obtain a quantum Fisher information that witnesses multipartite entanglement and places a lower bound on the entanglement depth of the many-body state~\cite{hauke2016measuring,hyllus2012fisher}. This connection was used in the neutron scattering benchmarking protocol of Ref.~\cite{lee2026benchmarking}.

Here we have access to more than the response function. The two dynamical states are many-body states, and their reduced density matrices allow us to study their entanglement directly. We first look at the entanglement between an observed block and the rest of the chain, then at the entanglement among spins inside the block, and finally ask whether a specified local reduced state is genuinely multipartite entangled.

\begin{figure*}[t]
 \centering
 \includegraphics[width=0.98\textwidth]{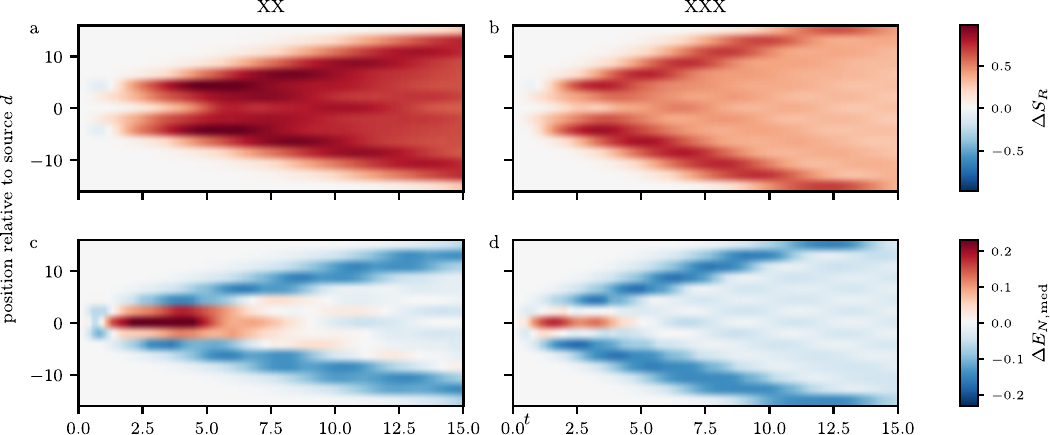}
\caption{\textbf{Entanglement between and within the observed block changes across the propagating wake.} Width-seven branch-averaged changes relative to the physical $t=0$ post-pulse state in the block--exterior entanglement entropy $\Delta S_R$ for (a) XX and (b) XXX, and in the median logarithmic negativity across internal contiguous bipartitions, $\Delta E_{N,\mathrm{med}}$, for (c) XX and (d) XXX. Positive $\Delta S_R$ indicates increased entanglement between the observed block and the rest of the chain, while negative $\Delta E_{N,\mathrm{med}}$ indicates a reduction of entanglement across typical cuts within the block. Each quantity uses a symmetric color scale shared by the two models.}
 \label{fig:reorganization}
\end{figure*}

\subsection{Entanglement between the observed block and its exterior}

For either pulse, the global state remains pure throughout the evolution. The entropy of a reduced region~$R$,
\begin{equation}
 S_R=-\Tr(\rho_R\log_2\rho_R),
 \label{eq:block-entropy}
\end{equation}
is therefore the entanglement entropy across the bipartition between $R$ and the rest of the chain. For each branch we measure the change relative to the physical $t=0$ state, i.e., the state immediately after the source pulse and before many-body evolution,
\begin{equation}
 \Delta S_R(t)=S_R(t)-S_R(0).
 \label{eq:block-entropy-change}
\end{equation}
The maps and summary statistics below report the mean of this branch-resolved change over the two pulse signs.

Figures~\ref{fig:reorganization}(a,b) show $\Delta S_R$ for width seven. In both models, the block becomes more entangled with its exterior over much of the region swept out by the perturbation. The real-space patterns are noticeably different. In XX, the entropy increase fills a broad region behind the propagating front. In XXX, the change remains more concentrated along moving ridges. Thus, the different structures seen in the response wakes also appear in the entanglement generated across the boundary of a finite observation region.

\subsection{Entanglement within the observed block}

The entropy $S_R$ does not indicate how the spins inside~$R$ are entangled with one another. To probe this feature, we divide the block across a contiguous bipartition $A|B$ and use the logarithmic negativity \cite{vidal2002computable,plenio2005logarithmic},
\begin{equation}
 E_N(A{:}B)
 \coloneqq 
 \log_2\!\norm{\rho_{AB}^{T_B}}_1,
 \label{eq:logneg}
\end{equation}
where $T_B$ denotes partial transpose on $B$. A positive value certifies negative partial transpose (NPT) entanglement across that cut.

For each block, we calculate $E_N$ across the inequivalent contiguous cuts and use their median to summarize the internal bipartite entanglement. We again use the physical post-pulse $t=0$ state as the baseline,
\begin{equation}
 \Delta E_{N,\mathrm{med}}(t)
 =
 E_{N,\mathrm{med}}(t)-E_{N,\mathrm{med}}(0).
 \label{eq:internal-neg-change}
\end{equation}
The plotted and quoted values are then averaged over the two pulse states, as for $\Delta S_R$. Figures~\ref{fig:reorganization}(c,d) show a predominantly negative wake in both models. Hence, while the block generally becomes more entangled with the rest of the chain, the entanglement across cuts lying entirely inside the block tends to decrease. A positive region remains close to the source at early times, after which the negative structure moves outward with the perturbation.

\begin{figure*}[t]
 \centering
 \includegraphics[width=0.98\textwidth]{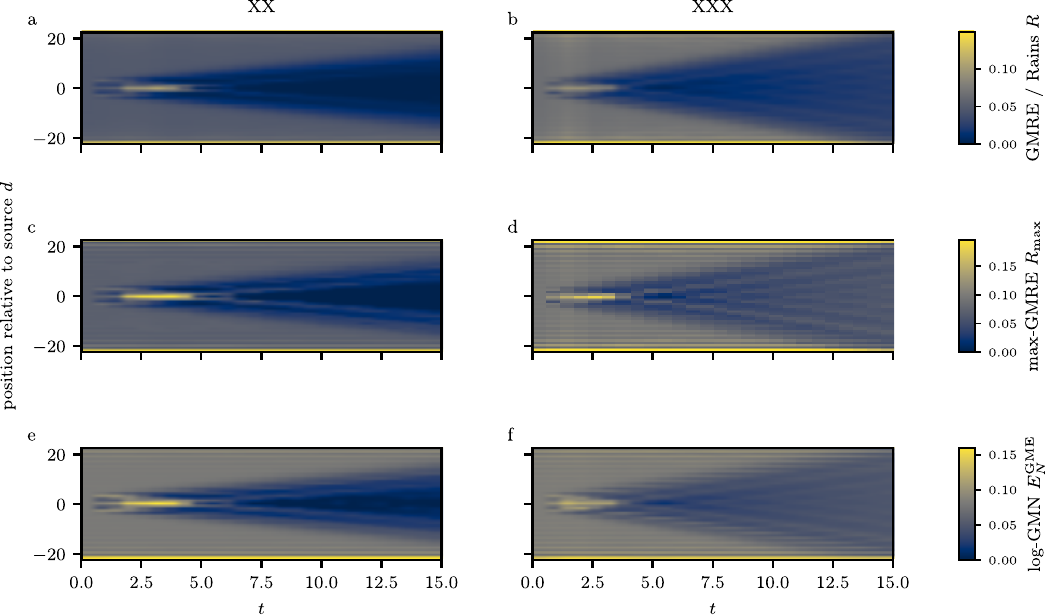}
\caption{\textbf{Genuine five-partite entanglement in the local reduced states.} Width-five multipartite Rains entanglement $R$ for (a) XX and (b) XXX, max-Rains entanglement $R_{\max}$ for (c) XX and (d) XXX, and genuine multipartite log-negativity $E_N^{\rm GME}$ for (e) XX and (f) XXX. Positive values of each quantity certify that the five-spin reduced state is not biseparable. For each measure, the XX and XXX panels use the same color scale, allowing the magnitude and space-time structure to be compared directly between the two models.}
 \label{fig:gme}
\end{figure*}

\subsection{Direct certification of genuine five-partite entanglement}

Next, let us note that entanglement across a particular cut is not the same as genuine multipartite entanglement. A multipartite state is biseparable if it can be written as a convex mixture of states that are separable across possibly different bipartitions. For example, a five-spin mixed state can be biseparable even if no single bipartition accounts for every state in the mixture. Genuine multipartite entanglement rules out all such decompositions.

We test this directly using the multipartite Rains entanglement introduced in Ref.~\cite{murray2026genuine}. Let $\mathcal B$ denote the set of inequivalent bipartitions of the spins, and let $T_m$ denote partial transpose with respect to bipartition~$m$. We define
\begin{equation}
 \mathcal T
 \coloneqq 
 \left\{
 \sigma=\sum_{m\in\mathcal B}\tau_m:
 \tau_m\geq0,\,
 \sum_{m\in\mathcal B}\norm{T_m(\tau_m)}_1\leq1
 \right\},  
 \label{eq:rains-set}
\end{equation}
where $\tau_m$ denotes the positive semidefinite operator associated with partition~$m$. The Rains entanglement is then defined as
\begin{equation}
 R(\rho)
 \coloneqq 
 \inf_{\sigma\in\mathcal T}
 D(\rho\Vert\sigma),
 \label{eq:gmre}
\end{equation}
with $D$ the quantum relative entropy defined in Sec.~\ref{sec:protocol}. All biseparable states belong to the positive partial transpose-mixture relaxation contained in $\mathcal T$ and have zero Rains entanglement. A positive value of $R$ therefore certifies genuine multipartite entanglement, although the converse is not guaranteed.

We also consider the max-Rains entanglement \cite{murray2026flood}, defined as
\begin{equation}
 R_{\max}(\rho)
 \coloneqq 
 \inf_{\sigma\in\mathcal T}
 D_{\max}(\rho\Vert\sigma),
 \label{eq:max-gmre}
\end{equation}
where the max-relative entropy is defined as~\cite{datta2009minmax}
\begin{equation}
 D_{\max}(\rho\Vert\sigma)
 \coloneqq 
 \log_2\inf\left\{
 \lambda\geq 0:\rho\leq\lambda\sigma
 \right\}.
 \label{eq:dmax}
\end{equation}
The feasible set is therefore the same as for $R$, but the relative entropy is replaced by the max-relative entropy. This form is useful because $R_{\max}$ can be evaluated directly as a semidefinite program.

For comparison, we also calculate the genuine multipartite negativity (GMN)~\cite{jungnitsch2011taming,hofmann2014analytical}. If
\begin{equation}
 N_m(\rho)
 \coloneqq 
 \frac{\norm{T_m(\rho)}_1-1}{2}
\end{equation}
is the ordinary negativity across bipartition $m$, the GMN is defined as
\begin{equation}
 N_{\rm GME}(\rho)
 \coloneqq 
 \min_{\rho=\sum_m p_m\rho_m}
 \sum_m p_m N_m(\rho_m).
 \label{eq:gmn}
\end{equation}
Here, $(p_m)_m$ is the probability distribution associated with bipartition $m$, such that $\sum_{m} p_m = 1$. We report it in logarithmic form,
\begin{equation}
 E_N^{\rm GME}
 \coloneqq 
 \log_2\!\left(2N_{\rm GME}+1\right).
 \label{eq:loggmn}
\end{equation}
The Rains, max-Rains, and log-GMN quantities are related by
\begin{equation}
 R(\rho)\leq R_{\max}(\rho)\leq E_N^{\rm GME}(\rho),
 \label{eq:gme-ordering}
\end{equation}
with the second inequality following from the closely related semidefinite programs for max-Rains and GMN~\cite{murray2026flood}.

Figure~\ref{fig:gme} shows these quantities for the width-five reduced states. The Rains entanglement is positive over almost the entire space-time region shown in both models, with median values $0.0164$ in XX and $0.0329$ in XXX. The max-Rains panels provide a second Rains-family view of the same local states, while the log-GMN gives the same qualitative certification, with medians $0.1138$ and $0.1542$, respectively. Thus, over most of the dynamics shown in the figure, the five neighboring spins are not merely entangled across one selected cut; their reduced state is certified to be genuinely five-partite entangled.

The magnitude of the genuine multipartite entanglement is nevertheless strongly reorganized by the perturbation. In both models, the Rains and log-GMN values are reduced over much of the region between the two propagating fronts, with the XX data developing a broad low-entanglement interior and the XXX data retaining a more structured pattern of ridges. The local state can therefore remain genuinely multipartite entangled while carrying less genuine multipartite entanglement than it did before the wake passed through the region. We make this comparison directly in the next subsection by relating the change in Rains entanglement to the change in block--exterior entropy.

To finish, we note that the width-five results are part of a broader aperture dependence. Supplemental Sec.~\ref{supp:gme-aperture} and Fig.~\ref{fig:gme-aperture} show the same three entanglement measures for $k=2,3,4,5$, with the width-five data repeated for completeness. For $k=3$ and $k=4$, genuine multipartite entanglement is again present over broad regions of the dynamics, while the propagating perturbation reorganizes its magnitude in a manner consistent with the width-five results.

\subsection{Reorganization of entanglement across spatial scales}

The spatial maps already show two opposite trends. In Fig.~\ref{fig:reorganization}(a,b), the entanglement between the observed block and the rest of the chain grows over much of the wake. At the same time, Fig.~\ref{fig:reorganization}(c,d) shows a predominantly negative change in the logarithmic negativity across cuts inside the block. For the same set of reduced states used in the Rains calculation, the median changes in block entropy are $0.179$ in XX and $0.126$ in XXX, while the corresponding median changes in internal log-negativity are $-0.0313$ and $-0.0302$.

\begin{figure}[t]
 \centering
 \includegraphics[width=0.98\columnwidth]
 {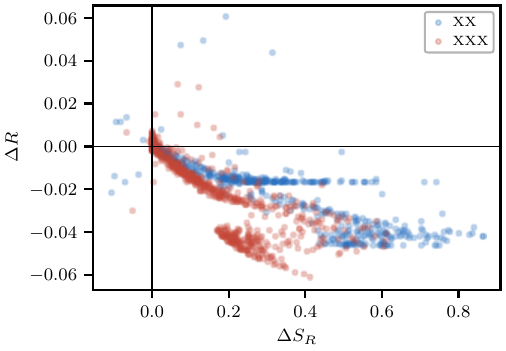}
 \caption{\textbf{Change in genuine multipartite Rains entanglement versus block--exterior entropy.} Change in the Rains entanglement $\Delta R$ versus the change in block entropy $\Delta S_R$ for the same reduced states in the XX and XXX models. Most observations have $\Delta S_R>0$ and $\Delta R<0$, showing that increased entanglement between the block and its exterior typically accompanies a reduction of genuine multipartite entanglement within the block. The distinct branches traced by XX and XXX show that this reorganization differs between the two dynamics.}
 \label{fig:rains-reorganization}
\end{figure}

The genuine multipartite entanglement follows the same overall trend. Its median change is $-0.0137$ in XX and $-0.0170$ in XXX. Figure~\ref{fig:rains-reorganization} compares $\Delta R$ directly with $\Delta S_R$ for the same reduced states. Most observations lie in the region with $\Delta S_R>0$ and $\Delta R<0$: as the block becomes more entangled with the rest of the chain, the genuine multipartite entanglement contained within that block typically decreases.

The relation is not the same in the two models. The XX data contain a pronounced branch in which $R$ changes only modestly while $S_R$ grows, followed by a second branch with a stronger decrease of $R$. In XXX, the decrease in $R$ is generally stronger at comparable changes in $S_R$, with a distinct branch reaching $\Delta R\simeq-0.06$. Thus, the redistribution of entanglement between the block and its exterior is not described by a single relation between the two quantities, and its detailed structure depends on the underlying many-body dynamics.

Complete block-entropy and internal-negativity maps for apertures $k=3,5,7$ are given in Supplemental Sec.~\ref{supp:full-entanglement} and Fig.~\ref{fig:s11}.

Finally, we note that the classical source label can itself affect how much entanglement is visible in the local state. If the pulse direction is known, the observer has access to the branch-resolved states $\rho_\pm$. However, if it is forgotten, the state becomes $\rbar=(\rho_++\rho_-)/2$. Since negativity is convex, mixing the two branches can reduce the entanglement visible across a fixed bipartition. Supplemental Sec.~\ref{supp:label-forgetting} and Fig.~\ref{fig:s9} quantify this effect. The loss grows strongly with aperture and is substantially larger in XX, and its increase relative to $t=0$ is positive in both models. Thus, the source sign is not only a classical record carried by the dynamical state, as retaining that record as side information can also reveal entanglement that is washed out when the two branches are mixed.

\section{Discussion and conclusions}

In this work we have explored how dynamical quantum simulations serve as laboratories for quantum information. Protocols designed to compute Green functions, dynamical structure factors, or other response functions necessarily prepare nontrivial many-body states before the final measurement is performed. Here we treated those states themselves as objects of study. Using local-pulse protocols in finite XX and XXX chains, opposite source rotations provide a natural classical label that can be followed through the many-body dynamics. This allowed us to study where the source record propagates, how much of it is available to observers of different sizes, and how its encoding changes from a simple one-site record to a noncommuting and jointly distributed multisite one.

A central result is that simple representation and simple readout need not coincide. The XX dynamics remain Gaussian, so covariance data completely determine the local branch states, whereas the same description becomes strongly incomplete in XXX. Nevertheless, the restricted spin measurements considered here recover a larger fraction of the available source information in XXX. The same ordering persists when the task is changed from discriminating two finite pulses to estimating a continuously varying source angle at matched local QFI. Thus, a compact description in one set of degrees of freedom need not imply an equally simple measurement in the physical degrees of freedom available to the experiment.

The reduced states also give direct access to the entanglement generated during the dynamics. We find that the propagating perturbation reorganizes entanglement across spatial scales: the observed block typically becomes more entangled with its exterior while bipartite and genuine multipartite entanglement contained within the block decrease. At the same time, five-site reduced states remain genuinely five-partite entangled over much of the dynamics, as certified directly by multipartite Rains and negativity-based quantities. This complements the indirect route available from the dynamical structure factor, where the dynamical susceptibility can be used to extract a QFI witness of entanglement depth \cite{hauke2016measuring,hyllus2012fisher,lee2026benchmarking}. The two approaches answer different questions, as the response-derived QFI bounds how many particles participate in the entanglement, while access to a specified reduced density matrix allows genuine multipartite entanglement of that local mixed state to be tested directly.

There are several natural extensions. The same questions can be asked in higher-dimensional systems, frustrated magnets, disordered systems, and models motivated more directly by quantum materials. An important practical question is how much of this information can be obtained without reconstructing complete reduced density matrices. Randomized measurements and classical shadows provide natural tools for this purpose, including measurement schemes adapted to the symmetries and representations of the system \cite{huang2020predicting,elben2022randomized,west2025real,west2026classical}. It will also be interesting to determine whether the representation--readout separation found here can be connected to more direct properties of the underlying many-body dynamics. More broadly, our results show that a dynamical quantum simulation need not be tied to the single observable for which it was designed, as the same prepared states can support a substantially broader set of questions at the interface of quantum simulation, many-body physics, and quantum information.

Another possible extension is to consider source ensembles in which the two branches are not prepared with equal probability. In that setting, the trace distance alone no longer determines the optimal discrimination probability, since the decision problem depends explicitly on the prior probabilities assigned to the two states. More generally, one may also consider situations in which the two possible decision errors carry different operational costs. Such settings connect naturally to asymmetric quantum hypothesis testing \cite{wang2012one,hiai1991proper}, where one error probability is constrained while the other is minimized. In the many-copy limit, the corresponding optimal error exponent is governed by the quantum relative entropy through quantum Stein's lemma \cite{ogawa2000strong}.

\section*{Artificial Intelligence Disclosure}
The authors acknowledge the use of Codex and ChatGPT 5.6 Sol for code development and for writing and reviewing the manuscript. In particular, most of the research was conducted as if we were working with a colleague. This meant an original discussion of the project followed by some initial tests, and then a back and forth between the authors and ChatGPT. Throughout, the model performed simulations and derivations, presented results for joint evaluation and  determination of the next steps to take. Details of the AI workflow are presented in Supplemental Sec.~\ref{sec:workflow}. All generated results were reviewed and validated by the authors.

\medskip

\section*{Acknowledgments}

SYC and MC were supported by the Laboratory Directed Research and Development (LDRD) program of Los Alamos National Laboratory (LANL) under project number 20260043DR. ML and MC acknowledge support from Los Alamos National Laboratory (LANL) ASC Beyond Moore’s Law project. This work was also supported by the Quantum Science Center (QSC), a National Quantum Information Science Research Center of the U.S. Department of Energy (DOE).

\bibliography{quantum,references}

\clearpage
\startsupplementtoc

\begingroup
\setcounter{section}{0}
\setcounter{subsection}{0}
\setcounter{equation}{0}
\setcounter{figure}{0}
\counterwithout{equation}{section}
\renewcommand{\thesection}{S\arabic{section}}
\renewcommand{\thesubsection}{\thesection.\arabic{subsection}}
\renewcommand{\theequation}{S\arabic{equation}}
\renewcommand{\thefigure}{S\arabic{figure}}
\renewcommand{\theHsection}{S\arabic{section}}
\renewcommand{\theHsubsection}{S\arabic{section}.\arabic{subsection}}
\renewcommand{\theHequation}{S\arabic{equation}}
\renewcommand{\theHfigure}{S\arabic{figure}}

\onecolumngrid

\begin{center}
 {\large\bfseries Supplemental Information for\\[0.35em]
 \textit{Dynamical quantum simulations as laboratories for quantum information}}
\end{center}

The Supplemental Information collects additional results that support the main text. We discuss how the experiments considered relate to dynamical structure factors, the dependence on observer geometry, the relation between trace distance and Holevo information, restricted spin readouts, alternative spatial partitions, covariance reconstruction, Fisher-information readout, label-sensitive entanglement, and the numerical domain and matched-model comparisons.

\begingroup
\hypersetup{linkcolor=magenta}
\supplementtoc
\endgroup

\section{Details for numerical simulations}\label{supp:details}

All many-body simulations were performed using matrix-product-state tensor networks with \textsf{iTensor}~\cite{fishman2022itensor}. For each Hamiltonian, the initial ground state in the total $S_z=0$ sector was obtained using the density-matrix renormalization group (DMRG) algorithm~\cite{schollwock2005density}. After applying the local source pulse, the state was evolved in time within the matrix-product-state representation.

Before fixing the production parameters, we repeated representative time evolutions with increasing maximum bond dimension and verified that the observables and reduced density matrices used in the analysis were unchanged within the numerical accuracy of the calculation. The results reported in this work were therefore obtained in a regime where the dynamics were not limited by the chosen bond-dimension cutoff.

For each time and source position, reduced density matrices on the contiguous observation regions were obtained directly from the tensor network using standard contractions of the matrix-product state. These reduced states form the numerical input for all of the information, readout, Fisher-information, and entanglement calculations discussed in the main text.

\section{From a local perturbation to the dynamical structure factor}
\label{supp:Green function-dsf}

Here we briefly review the connection between the local-pulse protocol considered in the main text, the retarded Green function, and the dynamical structure factor. This well-established connection has enabled recent comparisons between spectra computed on quantum computing devices and those obtained from inelastic neutron scattering experiments~\cite{baez2020dynamical,lee2026benchmarking}.

In an inelastic neutron scattering experiment, the incoming neutron exchanges energy and momentum with the magnetic degrees of freedom of the material. The corresponding scattering intensity is determined by the dynamical structure factor $S^{\alpha\beta}(q,\omega)$, which is the momentum- and frequency-resolved spin correlation function. Equivalently, it can be obtained from the imaginary part of the retarded spin Green function through the fluctuation--dissipation relation
\begin{equation}
 S^{\alpha\beta}(q,\omega)
 =
 -\frac{1}{\pi}
 \left[1+n_B(\omega)\right]
 \operatorname{Im}
 G^{R}_{\alpha\beta}(q,\omega),
 \label{eq:si_fdt}
\end{equation}
where $n_B(\omega)$ is the Bose distribution. At zero temperature and positive frequency, $n_B(\omega)=0$. The momentum-frequency Green function is obtained by Fourier transforming its real-space and real-time counterpart, which is proportional to the causal commutator
\begin{equation}
 G^{R}_{\alpha\beta}(i,j,t)
=
 -i\,\Theta(t)
 \left\langle
 \left[
 S_i^\alpha(t),S_j^\beta
 \right]
 \right\rangle .
 \label{eq:si_retarded_green}
\end{equation}
The precise prefactors depend on the convention used for spin operators. Throughout this work we take $S^\alpha=\sigma^\alpha/2$.

The quantum-computing protocol accesses this commutator by simulating the propagation of the perturbation in the material, i.e., without explicitly simulating the neutron. Starting from the ground state, one applies a local rotation generated by a Pauli operator $P$ at the source site,
\begin{equation}
 V_s=e^{-is\pi P/4}
 =\frac{I-isP}{\sqrt{2}},
 \qquad s=\pm1,
 \label{eq:si_pulse}
\end{equation}
and subsequently evolves the state under the spin Hamiltonian. For an observable $O$ measured after the evolution, the two pulse directions are implemented in separate circuit runs. The difference between the resulting expectation values yields
\begin{equation}
 \frac{
 \langle O\rangle_{+}
 -
 \langle O\rangle_{-}
 }{2}
 =
 -\frac{i}{2}
 \left\langle
 \left[
 O(t),P
 \right]
 \right\rangle .
 \label{eq:si_phase_cycle}
\end{equation}
Thus, opposite local rotations isolate the commutator that enters the retarded response. Repeating the measurement for different positions and evolution times gives the real-space and real-time Green function. A Fourier transform then produces $G^R(q,\omega)$ and, through Eq.~\eqref{eq:si_fdt}, the dynamical structure factor measured in neutron scattering.

\section{Quadratic scaling of the Holevo information}
\label{supp:holevo-quadratic}

Here we justify the quadratic behavior of the Holevo information when the two reduced states $\rho_+$ and $\rho_-$ become locally close. Writing
\begin{equation}
    \rho_{\pm}=\bar\rho\pm\xi,
    \qquad
    \Delta_R=\|\xi\|_1,
\end{equation}
the Holevo information for the equally likely source labels can be written as
\begin{equation}
    \Holevo_R=
    \frac{1}{2}D(\rho_+\Vert\bar\rho)
    +
    \frac{1}{2}D(\rho_-\Vert\bar\rho),
    \label{eq:si-holevo-relative}
\end{equation}
where the relative entropy is measured in bits.

For a full-rank state $\bar\rho$, fix $\epsilon > 0$ and a Hermitian operator $A$ to satisfy $\xi = \epsilon A$, with $\Tr(A)=0$ and $|\epsilon|\ll1$. The second-order expansion of the quantum relative entropy around $\bar\rho$ is governed by the Bogoliubov--Kubo--Mori metric~\cite{lesniewski1999monotone}. To see this, define
\begin{equation}
    f(\epsilon)
    \coloneqq 
    D(\bar\rho+\epsilon A\Vert\bar\rho).
\end{equation}
Since $\Tr(A)=0$, one has
\begin{equation}
    f(0)=f'(0)=0,
\end{equation}
while
\begin{equation}
    f''(0)
    =
    \frac{1}{\ln2}
    \Tr\!\left[
        A\,D\log(\bar\rho)[A]
    \right]
    =
    \frac{1}{\ln2}
    \Tr\!\left[
        A\,\mathcal{T}_{\bar\rho}(A)
    \right].
\end{equation}
Here, the completely positive map $\mathcal{T}_{\bar\rho}$ is defined as
\begin{equation}
    \mathcal{T}_{\bar\rho}(A)
    \coloneqq
    \int_0^\infty
    (\bar\rho+sI)^{-1}
    A
    (\bar\rho+sI)^{-1}\,ds
\end{equation}
and is equal to the Fr\'echet derivative of the matrix logarithm at $\bar\rho$. Taylor expansion therefore gives
\begin{equation}
    D(\bar\rho+\epsilon A\Vert\bar\rho)
    =
    \frac{\epsilon^2}{2\ln2}
    \Tr\!\left[
        A\,\mathcal{T}_{\bar\rho}(A)
    \right]
    +O(\epsilon^3).
\end{equation}

Since Eq.~\eqref{eq:si-holevo-relative} averages the $+\epsilon$ and $-\epsilon$ states, the odd orders cancel. Therefore
\begin{equation}
    \Holevo_R
    =
    \frac{\epsilon^2}{2\ln2}
    \Tr\!\left[
        A\,\mathcal{T}_{\bar\rho}(A)
    \right]
    +O(\epsilon^4).
\end{equation}
At the same time,
\begin{equation}
    \Delta_R=|\epsilon|\,\left\|A\right\|_1,
\end{equation}
and hence
\begin{equation}
    \Holevo_R=C_R \Delta_R^2+O(\Delta_R^4),
    \label{eq:si-holevo-quadratic}
\end{equation}
where $C_R$ depends on the local average state and on the direction in state space along which the two states separate. The quantum Pinsker inequality further implies
\begin{equation}
    \Holevo_R\geq\frac{\Delta_R^2}{2\ln2}.
\end{equation}

For the one-site states considered in the main text, $U(1)$ symmetry makes both states diagonal in the local $z$ basis, while global spin-flip symmetry exchanges the two states and fixes $\bar\rho_1=I/2$. Choosing the orientation of the local $z$ axis so that the $+$ state has positive magnetization therefore gives
\begin{equation}
    \rho_\pm=\frac{1}{2}\left(I\pm \Delta_1\sigma^z\right),
\end{equation}
so the relation can be obtained exactly,
\begin{equation}
    \Holevo_1
    =
    1-h_2\!\left(\frac{1+\Delta_1}{2}\right)
    =
    \frac{\Delta_1^2}{2\ln2}
    +O(\Delta_1^4).
\end{equation}

\begin{figure*}[t]
 \centering
 \includegraphics[width=0.98\textwidth]{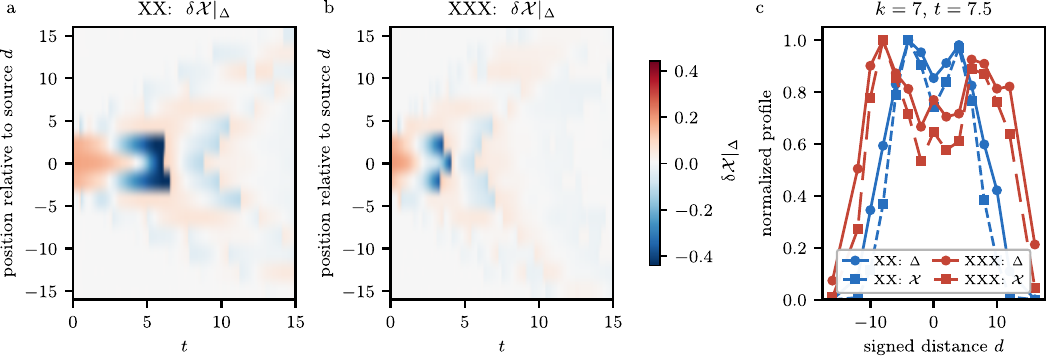}
\caption{\textbf{Holevo information is not fixed by the local binary distinguishability.} Panels (a) and (b) show the conditional residual $\delta\Holevo|_D$ for the width-seven XX and XXX wakes. Positive (negative) values indicate more (less) source information than the median value found at comparable $\Delta_R$ within the same model and geometry. Panel (c) compares independently normalized $\Delta_R$ and $\Holevo_R$ profiles at $k=7$ and $t=7.5$, showing that the two quantities emphasize somewhat different spatial regions of the same wake.}
 \label{fig:s2}
\end{figure*}

\section{Trace distance and Holevo information}
\label{supp:DX}

Away from the small-$\Delta_R$ regime analyzed in the main text, the trace distance does not uniquely determine the Holevo information. Fixing $\Delta_R$ fixes the optimal binary guessing probability, $P_{\rm succ}=(1+\Delta_R)/2$, but it does not fix how much information about the source label is contained in the local ensemble. Indeed, different pairs of states can have the same trace distance while differing in their average state $\bar\rho$ and therefore in their Holevo information.

To separate this effect from the overall strength of the source record, we compare observations with similar values of~$\Delta_R$. Within each model, aperture, and geometry, we group points with nearby trace distance and define $\Holevo_{\rm ref}(\Delta_R)$ as the median Holevo information in the corresponding group. We then introduce
\begin{equation}
 \delta\Holevo|_\Delta
 \coloneqq 
 \Holevo
 -
 \Holevo_{\rm ref}[\Delta].
 \label{eq:conditional-holevo}
\end{equation}
Thus, $\delta\Holevo|_\Delta>0$ identifies regions that contain more source information than other regions with comparable optimal guessing probability, whereas $\delta\Holevo|_D<0$ identifies regions that contain comparatively less.

Figure~\ref{fig:s2}(a,b) shows this quantity for the width-seven XX and XXX wakes. In both models the residual is clearly structured in space and time. In particular, regions near the source develop a pronounced negative contribution over part of the evolution, while positive residuals appear in neighboring portions of the wake. Hence, two regions with comparable optimal guessing probability can still differ in how much information about the source label is contained in their reduced states. The effect is therefore not captured by the overall magnitude of the response alone.

Panel (c) makes the same distinction directly at $t=7.5$. We normalize the $\Delta_R$ and $\Holevo_R$ profiles independently by their respective maxima and compare their spatial dependence. The Holevo profile is more concentrated than the corresponding trace-distance profile and redistributes weight across the wake. This shows that even beyond the perturbative regime, $\Delta_R$ and $\Holevo_R$ need not assign the same relative importance to different spatial regions of the evolving state.

\section{Changing the observer location and geometry}
\label{supp:numerical-domain}

In the main text, the apertures with $k=1,3,5,7$ are all centered on the same site. Thus, increasing $k$ asks what happens when an observer at a fixed position is given access to more of the surrounding many-body state. However, this leaves open a simple question. \textit{If the centered spin has a very small $\Delta_1$ while the surrounding seven-site region has a much larger $\Delta_7$, is the source record really spread over several spins, or did we simply place the one-site observer at a weak point of the propagating wake?}

We first illustrate this distinction in Fig.~\ref{fig:s1}(a). Here we select delayed observations away from the source for which the centered one-site signal is small, $\Delta_1<0.01$, while the seven-site region with the same center still has $\Delta_7\geq0.05$. There are many points for which the central spin is nearly blind even though its immediate neighborhood carries a sizeable source record. By itself, however, this comparison does not tell us where the additional distinguishability comes from. For instance, the chosen center could lie between two stronger portions of the wake, in which case moving the one-site observer by one or two sites could recover much of the signal.

\begin{figure*}[t]
 \centering
 \includegraphics[width=0.98\textwidth]{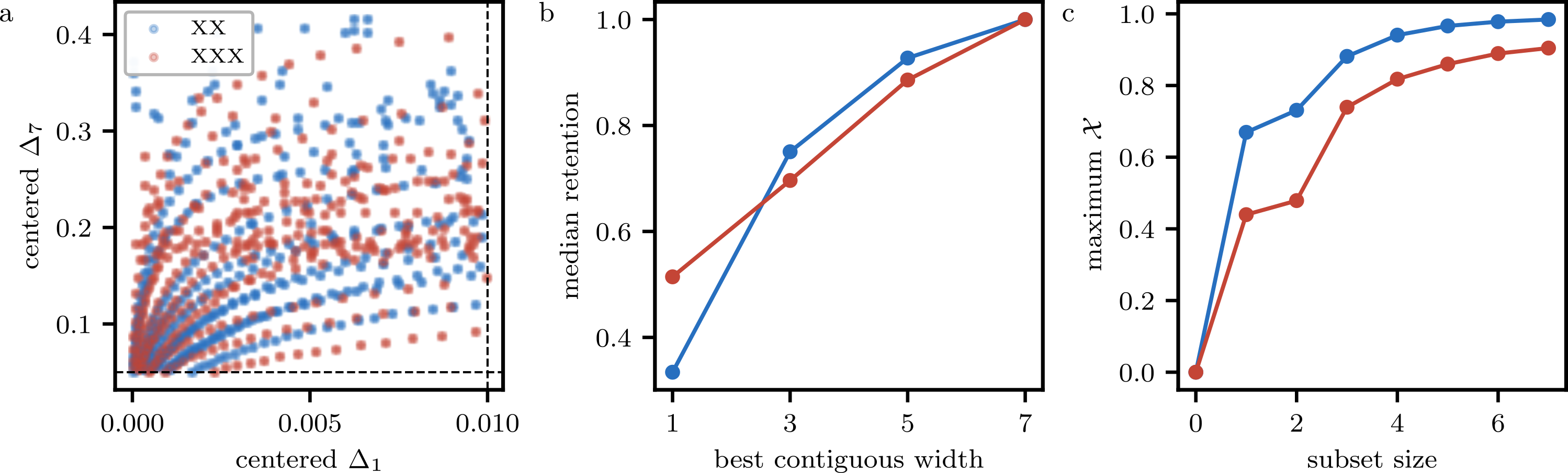}
\caption{\textbf{Dependence of the recovered source record on the location and geometry of the observed region.} (a) Trace distance of a centered one-site observer versus that of the seven-site region with the same center, restricted to delayed observations with $\Delta_1<0.01$ and $\Delta_7\geq0.05$. (b) Median fraction $\Delta_k^\star/\Delta_7$ recovered by the best contiguous $k$-site region contained within the same seven-site block. (c) Maximum Holevo information obtained when arbitrary subsets of a given size are retained. Allowing the smaller observer to move recovers much of the signal missed at a fixed center, although increasing the number of accessible spins continues to provide additional source information.}
 \label{fig:s1}
\end{figure*}

To test this possibility, we next allow the smaller observation window to move within the same seven-site region. Given a seven-site block $R_7$, we define
\begin{equation}
 \Delta_k^\star(R_7)
 \coloneqq
 \max_{\substack{A\subseteq R_7\\ |A|=k\\
                  A\ {\rm contiguous}}}
 \Delta_A ,
 \label{eq:best-contiguous-D}
\end{equation}
where $\Delta_k^\star$ is the largest trace distance obtained from any contiguous $k$-site region contained in $R_7$. In particular, $\Delta_1^\star$ is simply the largest one-site trace distance among the seven spins. Figure~\ref{fig:s1}(b) shows the median ratio $\Delta_k^\star/\Delta_7$.

Allowing the smaller observer to move changes the picture substantially. In the delayed-time data, we no longer find the  separation $\Delta_1^\star<0.01$ together with $\Delta_7\geq0.05$. Thus, much of the extreme one-site blindness seen in panel (a) can indeed be traced to the position of the chosen center relative to the moving wake. Nevertheless, moving the detector does not remove the aperture dependence. The best individual spin retains only about one third of the seven-site distinguishability in XX and about one half in XXX at the median, and the retained fraction increases as additional neighboring spins are included. Hence, even after choosing the best location, no single spin typically contains the full record available in the larger region.

We can go one step further and ask whether the requirement that the observed spins form a contiguous block is itself important. For every subset $A\subseteq R_7$, we trace out the remaining spins and compute the corresponding Holevo information $\Holevo_A$. At fixed subset size $m$, we define
\begin{equation}
 \Holevo_m^\star(R_7)
 =
 \max_{\substack{A\subseteq R_7\\ |A|=m}}
 \Holevo_A .
 \label{eq:best-subset-X}
\end{equation}
Figure~\ref{fig:s1}(c) shows the largest Holevo information obtained in this scan as the number of retained spins is increased. A small number of appropriately chosen spins can already carry a substantial amount of source information, with the XX data in particular approaching the information available to the full seven-site region rather quickly.

To make this comparison more precise, we also asked, for each seven-site region separately, how many spins are needed to retain $90\%$ or $95\%$ of $\Holevo_7$. We performed this optimization first over all subsets and then over contiguous subsets only. Across the active regions considered, allowing disconnected subsets did not reduce the minimum number of spins required at either threshold. Hence, although the position of the observation window matters, selecting separated spins does not provide an additional compression of the seven-site source record at these retention levels.

These checks sharpen the aperture results in the main text. A fixed one-site observer can miss the record simply because it lies at a weak point of the wake. Once its position is optimized, this extreme effect disappears, but a larger observation region still retains substantially more distinguishability than the best individual spin. The advantage of increasing the aperture is therefore not solely a consequence of where the observer was placed.

\section{Alternative partitions and separable flanks}
\label{supp:i3}

\begin{figure*}[t]
 \centering
 \includegraphics[width=0.98\textwidth]{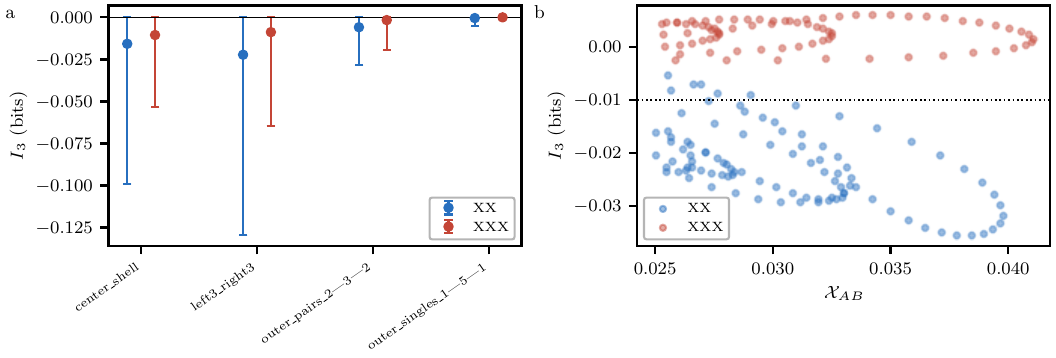}
\caption{\textbf{Joint source information for alternative spatial partitions and separable flanks.} (a) Quantile summaries of $I_3(C{:}A{:}B)$ for several partitions of the width-seven observer. Negative $I_3$ persists beyond the primary $3|1|3$ geometry, with a magnitude that depends on the size and separation of the two observed regions. (b) Width-three nonlocal observations for which the positive-pulse, negative-pulse, and sign-forgotten two-qubit flank states are all separable. Negative $I_3$ persists in this regime, showing that joint source information between the two flanks does not require entanglement between them.}
 \label{fig:s5}
\end{figure*}

The main text uses the two three-site flanks of a seven-site block, with the central spin omitted, to quantify how source information is shared between the two sides of the wake. Here we first check how much this result depends on that particular partition. Figure~\ref{fig:s5}(a) compares the main $3|1|3$ geometry with several alternatives that change both the size and separation of the regions $A$ and $B$. Negative $I_3$ is not specific to the partition used in the main text. It persists for several choices of the two regions, although its magnitude decreases as $A$ and $B$ are made smaller and the amount of source information contained in their union decreases. Thus, the strength of the joint contribution depends on which parts of the wake are collected, rather than being a fixed property of the full seven-site block.

We next ask whether this joint source information requires entanglement between the two observed regions. The width-three geometry provides a particularly simple test. Starting from a $1|1|1$ block, we trace out the central spin and take the two remaining spins as $A$ and $B$. The state on $AB$ is then a two-qubit mixed state, for which positivity under partial transpose is equivalent to separability.

Figure~\ref{fig:s5}(b) shows the resulting two-qubit flank data. The separation between the models is particularly clear. In XX, $I_3$ is predominantly negative and reaches values below $-0.03$ bits as $\Holevo_{AB}$ increases. The XXX observations instead remain clustered much closer to $I_3=0$. Across the $176$ nonlocal observations satisfying $\Holevo_{AB}\geq0.025$, all three flank states $\rho_+$, $\rho_-$, and $\rbar$ are separable, while $115$ observations have negative $I_3$. Of these, $99$ belong to XX and $16$ to XXX. Hence the strong XX--XXX difference seen here cannot be attributed simply to entanglement between the two observed flanks.

To finish, we note that the width-three signal is weaker than the seven-site signal considered in the main text. In particular, the separable-flank observations do not survive the stronger cut $\Holevo_{AB}\geq0.05$ used there. We therefore show the width-three data for $\Holevo_{AB}\geq0.025$, for which the negative-$I_3$ behavior remains well resolved. This lower threshold is used only for the separability comparison in Fig.~\ref{fig:s5}(b).

\section{Gaussian reconstruction of internal entanglement}
\label{supp:gaussian-ent}

The main text shows that, in the XXX chain, covariance information can reproduce the trace distance and Holevo information reasonably well even when it fails to reconstruct the full signed-response operator. Here we ask whether the same Gaussian surrogate also reproduces another property of the local branch states, namely the entanglement between spins inside the observed block.

For each physical branch $\rho_s$, with $s=\pm1$, we construct the Gaussian state $\rho_s^G$ with the same fermionic covariance matrix, as defined in Sec.~\ref{sec:inversion}. We then partition the $k$-site observation region into two contiguous pieces,
\begin{equation}
    R=A\cup B,
    \qquad
    A\cap B=\varnothing,
\end{equation}
and compare the logarithmic negativity across the cut $A|B$. For a state $\rho_{AB}$, the logarithmic negativity~\cite{vidal2002computable} is
\begin{equation}
    E_N(A{:}B)
    \coloneqq 
    \log_2 \norm{\rho_{AB}^{T_B}}_1,
\end{equation}
where $T_B$ denotes the partial transpose on subsystem $B$. For each branch and cut, we define the Gaussian reconstruction error
\begin{equation}
 a_N^G(s,A{:}B)
 \coloneqq 
 \abs{
 E_N(\rho_s,A{:}B)
 -
 E_N(\rho_s^G,A{:}B)
 }.
 \label{eq:gaussian-neg-error}
\end{equation}
Thus, $a_N^G=0$ means that the covariance-matched Gaussian state reproduces the physical entanglement across that cut exactly.

For XX this error vanishes in exact arithmetic, since every branch state is itself Gaussian. We therefore focus on the nontrivial XXX case. Figure~\ref{fig:s6}(a) shows the median $a_N^G$, taken over the two source branches, all contiguous cuts of the block, and the resolved space-time observations at each aperture. The error increases steadily with $k$, from roughly $0.15$ at $k=3$ to more than $0.4$ at $k=7$. Thus, as the observed region grows, covariance data become progressively less accurate at reproducing the entanglement contained within the physical XXX branch states.

Figure~\ref{fig:s6}(b) compares this entanglement error with the operator-level covariance error
\begin{equation}
    \eG
    =
    \frac{\norm{\xi-\xi^G}_1}{\norm{\xi}_1}
\end{equation}
introduced in the main text. Each point corresponds to one of the XXX aperture and spatial-region strata used in the analysis. The two errors increase together. Regions for which covariance information gives a poor reconstruction of the signed-response operator also show a larger error in the reconstructed logarithmic negativity. Importantly, we note that the near agreement of $\Delta^G$ and $\Holevo^G$ with their physical values should therefore not be interpreted as a generally accurate reconstruction of the local many-body state.

\begin{figure*}[t]
 \centering
 \includegraphics[width=0.86\textwidth]{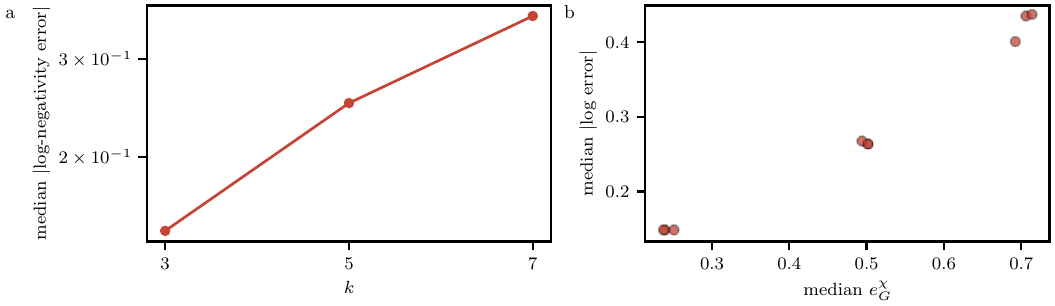}
\caption{\textbf{Covariance information does not reconstruct the internal entanglement of the XXX branch states.} (a) Median absolute error in the logarithmic negativity between the physical branch $\rho_s$ and its covariance-matched Gaussian surrogate $\rho_s^G$, shown as a function of aperture $k$. The median is taken over the two source branches, contiguous bipartitions of the block, and resolved space-time observations. (b) Median logarithmic-negativity error versus the corresponding signed-response reconstruction error $\eG$ for the XXX aperture and spatial-region strata. Larger errors in the reconstructed response are accompanied by larger errors in the internal entanglement.}
 \label{fig:s6}
\end{figure*}

\section{Representation and commuting readout}
\label{supp:inversion-controls}

The main text establishes the representation--readout inversion using measurements in the physical Pauli basis. Here we compare the covariance reconstruction with the different readout restriction introduced in Sec.~\ref{sec:encoding}, where the decision observable is required to commute with the sign-forgotten state $\rbar$.

Recall that
\begin{equation}
    g_{\rm coh}
    =
    1-\frac{\Delta_{\rm comm}}{\Delta}
\end{equation}
is the fraction of the optimal decision bias lost under this restriction. Thus, larger $g_{\rm coh}$ means that more of the Helstrom advantage depends on coherences between eigenspaces of $\rbar$.

Figure~\ref{fig:s7} compares XX and XXX after matching observations at similar source-record strength, as in the main-text readout analysis. The two quantities have opposite model orderings. The covariance error is larger in XXX,
\begin{equation}
    e_{G,\rm XXX}^{\xi}-e_{G,\rm XX}^{\xi}>0,
\end{equation}
while the commuting-readout loss is smaller,
\begin{equation}
    g_{{\rm coh},\rm XXX}-g_{{\rm coh},\rm XX}<0.
\end{equation}
Thus, although covariance information provides a poorer representation of the XXX response, a larger fraction of the optimal decision remains available when the readout is restricted to commute with $\rbar$. The reversed ordering found for Pauli measurements therefore also appears under a different restriction on the physical readout.

\begin{figure*}[t]
 \centering
 \includegraphics[width=0.98\textwidth]{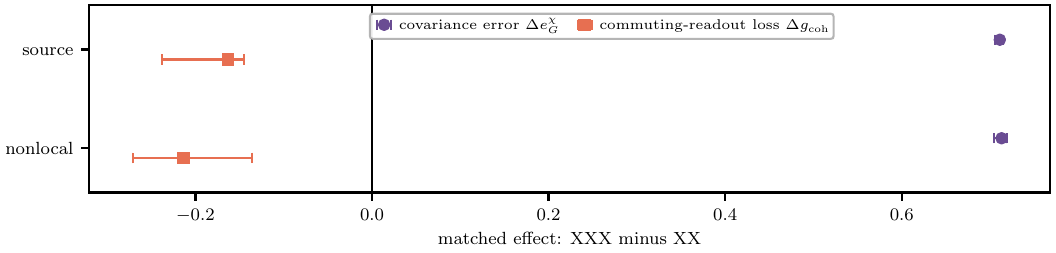}
\caption{\textbf{Covariance reconstruction and commuting readout have opposite model orderings.} Matched XXX-minus-XX differences at width seven for source and nonlocal regions. The covariance reconstruction error $\Delta e_G^\xi=e_{G,\mathrm{XXX}}^\xi-e_{G,\mathrm{XX}}^\xi$ is positive, showing that covariance data reconstruct the XXX signed response less accurately. In contrast, the commuting-readout loss $\Delta g_{\rm coh}=g_{\rm coh}^{\mathrm{XXX}}-g_{\rm coh}^{\mathrm{XX}}$ is negative, showing that a smaller fraction of the optimal decision bias is lost in XXX when the readout is restricted to commute with $\rbar$. Error bars show the corresponding confidence intervals.}
 \label{fig:s7}
\end{figure*}

\section{Pauli structure of the optimal decision}
\label{supp:pauli-hierarchy}

The main text compares the optimal binary decision with several restricted measurements in the physical spin basis. Here we look at the same problem from the opposite direction and ask what the optimal Helstrom observable itself looks like when expanded in spin operators.

For the two equally likely branches, the optimal binary decision is implemented by the Helstrom sign observable
\begin{equation}
    H=\operatorname{sgn}(\xi).
\end{equation}
We expand $H$ in the $k$-spin Pauli basis,
\begin{equation}
 H=\sum_{P\in\mathcal P_k}h_P P,\qquad
 h_P=2^{-k}\Tr(PH),
 \label{eq:helstrom-pauli}
\end{equation}
where $\mathcal P_k=\{I,X,Y,Z\}^{\otimes k}$. The Pauli weight $\operatorname{wt}(P)$ is the number of spins on which $P$ acts nontrivially. Since the Pauli strings form an orthogonal operator basis, the squared coefficients $|h_P|^2$ tell us how the Hilbert--Schmidt weight of the Helstrom observable is distributed over one-spin, two-spin, and higher-weight terms.

We collect this information through
\begin{equation}
 W_m^{(H)}
 =
 \frac{\displaystyle
 \sum_{\operatorname{wt}(P)=m}\abs{h_P}^2}
 {\displaystyle
 \sum_{P\neq I}\abs{h_P}^2}.
 \label{eq:helstrom-weight}
\end{equation}
Figure~\ref{fig:s4}(a) shows the median $W_m^{(H)}$ for the width-seven data. In both models, the Helstrom observable has substantial support beyond weight one. The distributions are also not identical: XXX has less weight-two and more weight-three contribution than XX, while the higher-weight tails become similar. Thus, the optimal decision on a seven-spin block is not represented by a single local spin observable or by one preferred Pauli weight.

Figure~\ref{fig:s4}(b) asks the corresponding operational question. We compare the best single Pauli string with feasible decision observables constructed from Pauli strings of weight at most two or three, and report the fraction of the full Helstrom bias that they recover. At $k=3$, allowing combinations of low-weight strings gives a substantial improvement and the weight-three construction reaches the full decision bias. The gain becomes smaller as the aperture grows. By $k=5$ and $k=7$, the low-weight constructions recover only about $60$--$65\%$ of the Helstrom bias, indicating that the optimal decision on the larger blocks cannot be captured well by these simple low-weight descriptions.

The weight-at-most-two and weight-at-most-three results in panel~(b) are feasible lower bounds obtained from the corresponding numerical constructions, rather than certified global optima over the full low-weight operator spaces. Their values therefore need not be monotonically ordered.

Finally, these restrictions should not be confused with the product-Pauli measurements used in Sec.~\ref{sec:inversion}. A single Pauli string retains only one binary correlator. A product-Pauli measurement instead measures every spin in a chosen local Pauli basis and retains the full string of local outcomes. The latter can therefore extract information that is not contained in the expectation value of any single Pauli string, even though both measurements are built from the same physical spin operators.

\begin{figure*}[t]
 \centering
 \includegraphics[width=0.88\textwidth]{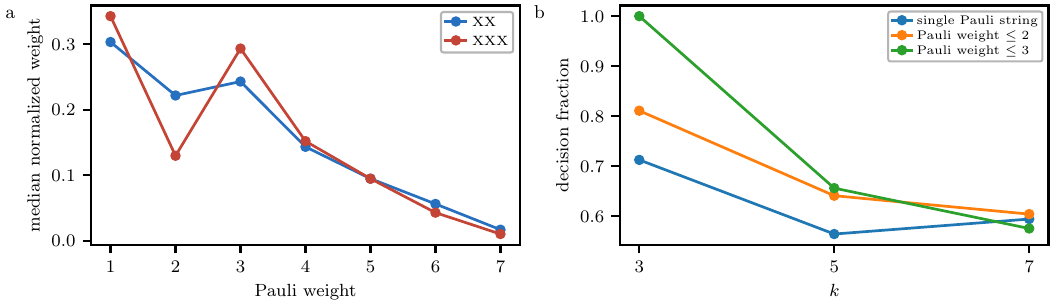}
\caption{\textbf{Pauli structure and low-weight approximations to the optimal binary decision.} (a) Median distribution $W_m^{(H)}$ of the Helstrom observable over Pauli weight $m$ for width-seven XX and XXX blocks. In both models, substantial weight extends beyond one-spin terms. (b) Fraction of the optimal Helstrom decision bias recovered by the best single Pauli string and by feasible observables constructed from Pauli strings of weight at most two or three. Low-weight observables recover most of the decision bias for the smallest aperture, while a larger fraction remains inaccessible as the observed block grows.}
 \label{fig:s4}
\end{figure*}
\section{Transfer of finite-pulse readout bases}
\label{supp:fisher-controls}

The finite-pulse and source-angle tasks optimize different objective functions, but they are implemented on the same local spin degrees of freedom. This raises a practical question: \textit{if a product-Pauli basis has already been optimized for the finite-pulse experiment, how much is lost by reusing that basis for source-angle estimation?}

For each observation, we first optimize the product-Pauli basis for either the finite-pulse decision bias or the finite-pulse Holevo information. We then keep that basis fixed and evaluate its classical Fisher information at $\theta=0$. To compare with the best product-Pauli measurement for the Fisher task, we define the transfer ratio
\begin{equation}
 T_F^{(\mu)}
 =
 \frac{
 F_C\!\left[\mathcal B_{\mu}^{\star}\right]
 }{
 F_C^{\rm prodP}
 },
 \qquad
 \mu\in\{D,\Holevo\},
\end{equation}
where $\mathcal B_{\mu}^{\star}$ is the product basis optimized for the finite-pulse task $\mu$. A value of one means that the finite-pulse basis is also optimal within the product-Pauli family for estimating the source angle.

Figure~\ref{fig:s8} shows that the basis optimized for finite-pulse Holevo information transfers extremely well. Its median transfer ratio is close to one in every model and geometry considered. The decision-optimal basis is less robust. The largest loss occurs for the XX width-five source region, where the median transfer ratio is approximately $0.889$, with a substantially broader distribution. The corresponding XXX bases transfer much more efficiently.

Thus, although the finite-pulse and infinitesimal tasks have different optimality criteria, optimizing the product basis for source-label information gives a measurement that can usually be reused for the Fisher task with little loss. Optimizing only for binary discrimination does not provide the same guarantee.

\begin{figure*}[t]
 \centering
 \includegraphics[width=0.98\textwidth]{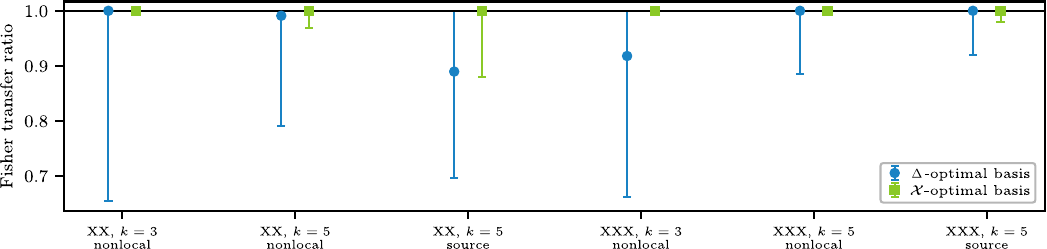}
\caption{\textbf{Finite-pulse product bases can be reused for source-angle estimation.} Fisher-information transfer ratio obtained by applying, without reoptimization, product-Pauli bases optimized for finite-pulse discrimination ($\Delta$) or finite-pulse Holevo information ($\Holevo$). The ratio is taken with respect to the Fisher-optimal product-Pauli measurement at the same space-time point. Markers show medians and error bars the 10th--90th percentile range. Bases optimized for $\Holevo$ retain nearly all of the available product-Pauli Fisher information, while decision-optimal bases can incur a larger loss, particularly in the XX width-five source region.}
 \label{fig:s8}
\end{figure*}

\section{Aperture dependence of local multipartite entanglement}
\label{supp:gme-aperture}

The main text focuses on $k=5$, where the reduced state can be tested directly for genuine five-partite entanglement. Here we collect the corresponding results for apertures $k=2,3,4,5$ in a single figure. We show the multipartite Rains entanglement $R$, the max-Rains entanglement $R_{\max}$, and the genuine multipartite log-negativity $E_N^{\rm GME}$ introduced in Sec.~\ref{sec:entanglement}. The width-five data are repeated here for completeness. For $k=2$ there is only one bipartition, so the multipartite constructions reduce to the corresponding bipartite quantities. For $k\geq3$, positive values certify genuine $k$-partite entanglement of the local reduced state.

\begin{figure*}[t]
 \centering
 \includegraphics[width=0.98\textwidth]
 {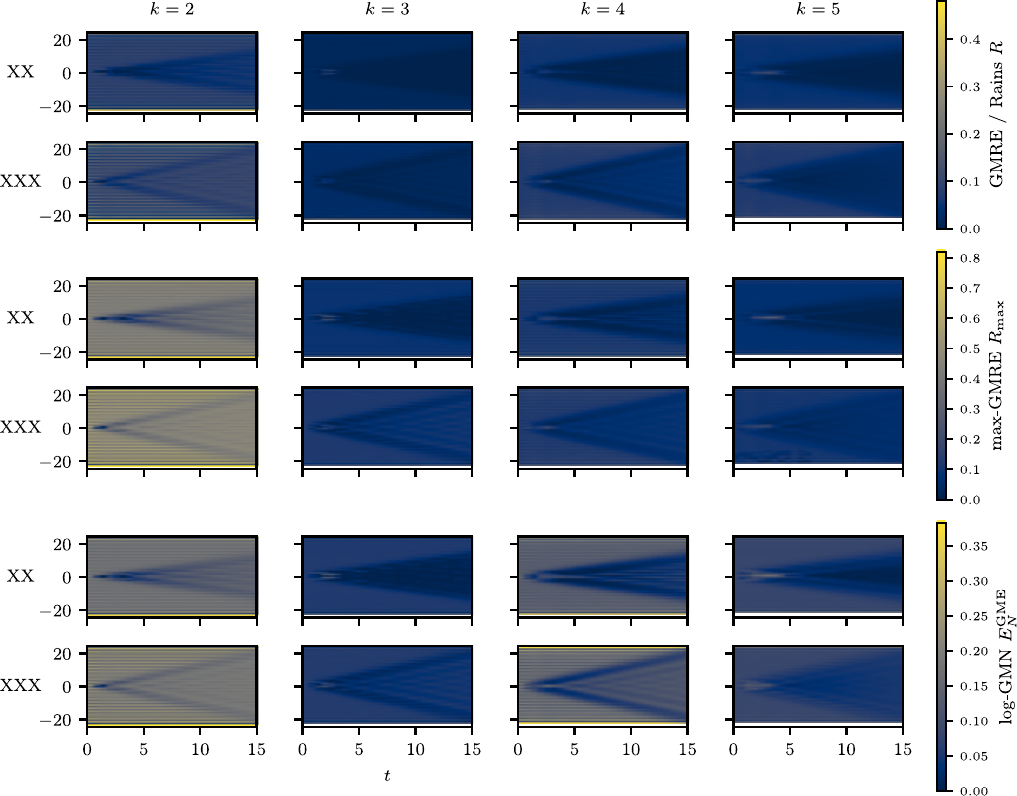}
 \caption{\textbf{Aperture dependence of the local entanglement structure.} Multipartite Rains entanglement $R$ (top four rows), max-Rains entanglement $R_{\max}$ (middle four rows), and genuine multipartite log-negativity $E_N^{\rm GME}$ (bottom four rows) for apertures $k=2,3,4,5$. Within each measure, the upper and lower rows show XX and XXX, respectively, while columns increase the aperture from left to right. For $k=2$ the quantities reduce to their bipartite counterparts; for $k\geq3$, positive values certify genuine multipartite entanglement of the observed block. The width-five data from the main text are repeated for comparison.}
 \label{fig:gme-aperture}
\end{figure*}

Figure~\ref{fig:gme-aperture} shows that the qualitative structure found at width five is already present at smaller apertures. For $k=3$ and $k=4$, the local states remain genuinely multipartite entangled over broad regions of the dynamics, while all three measures develop a clear space-time structure associated with the propagating perturbation. As the aperture grows, the reduction of local multipartite entanglement inside the wake becomes more pronounced. This is particularly visible in XX, where a broad low-entanglement interior develops behind the moving fronts, whereas XXX retains a more structured pattern of propagating ridges.

The absolute values of the three measures should not be compared directly across apertures, since both the number of spins and the entanglement measure itself change. The useful comparison here is the space-time structure at fixed $k$ and the way that structure develops as additional spins are included in the observed region.

\begin{figure*}[t]
 \centering
 \includegraphics[width=0.98\textwidth]{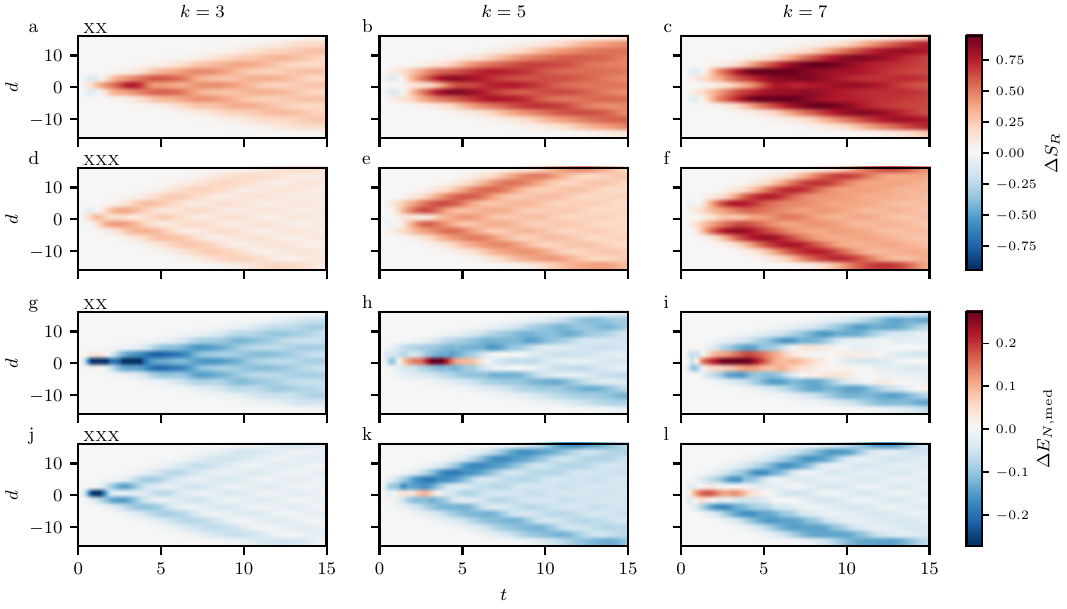}
\caption{\textbf{Aperture dependence of block--exterior and internal entanglement.} Changes in block--exterior entropy $\Delta S_R$ for XX (a)--(c) and XXX (d)--(f), and changes in the median logarithmic negativity across internal contiguous cuts, $\Delta E_{N,\mathrm{med}}$, for XX (g)--(i) and XXX (j)--(l), at apertures $k=3,5,7$. Increasing the aperture broadens the region of enhanced block--exterior entanglement, while much of the same propagating region shows reduced entanglement across cuts within the block.}
 \label{fig:s11}
\end{figure*}

\section{Complete entropy and internal-negativity fields}
\label{supp:full-entanglement}

Figure~\ref{fig:s11} extends the entanglement maps in the main text to apertures $k=3,5,7$. Panels~(a)--(f) show the change in the block--exterior entropy. In both models the increase becomes larger and fills a broader region as the aperture grows. The difference between the two dynamics is also clear across all three widths. XX develops a broad region of positive $\Delta S_R$ behind the propagating front, whereas XXX retains sharper moving ridges and a less uniform interior.

Panels~(g)--(l) show the corresponding change in the median logarithmic negativity across contiguous cuts inside the block. Over much of the same space-time region where $\Delta S_R$ is positive, $\Delta E_{N,\mathrm{med}}$ is negative. The wider blocks also develop a short-lived positive region close to the source before the decrease in internal entanglement spreads outward. Thus, the aperture dependence does not simply increase the amount of entanglement seen by the observer, as it also makes clear how the perturbation changes which spatial partitions carry that entanglement.

The propagating structure is visible in both quantities. In particular, the sharper ridges of the XXX wake appear in the block--exterior entropy and in the reduction of internal negativity, while XX shows the broader interior structure discussed in the main text.

\section{Entanglement hidden by forgetting the source sign}
\label{supp:label-forgetting}

If the source sign is retained, the observer knows whether the local state is $\rho_+$ or $\rho_-$. If that classical information is discarded, the state becomes
\begin{equation}
    \rbar=\frac{\rho_++\rho_-}{2}.
\end{equation}
Because entanglement is nonlinear in the state, these two descriptions need not contain the same visible entanglement.

For a fixed bipartition, let $N(\rho)$ denote the ordinary negativity. We call the contiguous cut closest to the middle of the block the primary cut; the remaining inequivalent contiguous cuts are used as controls. We compare the negativity available when the branch is known with that of the sign-forgotten mixture through
\begin{equation}
 L_N
 =
 \frac12\left[N(\rho_+)+N(\rho_-)\right]
 -
 N(\rbar).
 \label{eq:label-gap}
\end{equation}
Convexity of the negativity guarantees $L_N\geq0$. Thus, $L_N$ is the amount of branch-averaged negativity lost solely by forgetting which pulse was applied. Since $\rho_\pm=\rbar\pm\xi$, this loss is tied directly to the branch-dependent part of the dynamical state studied throughout the paper.

Figure~\ref{fig:s9}(a) shows the median $L_N$ as the aperture is increased. The gap grows strongly with $k$ in both models and is much larger in XX for the wider blocks. Hence, as the observer gains access to more spins, an increasing amount of the internal entanglement depends on knowing which dynamical branch was prepared.

Then, Figure~\ref{fig:s9}(b) compares this gap with its value in the physical post-pulse state at $t=0$,
\begin{equation}
 \Delta L_N(t)=L_N(t)-L_N(0).
\end{equation}
Although convexity constrains $L_N$ itself rather than $\Delta L_N$, the median change is positive for the apertures shown in both models. The dynamics therefore increase the amount of entanglement that is visible in the branch-resolved states but lost after the classical source sign is discarded.

\begin{figure*}[t]
 \centering
 \includegraphics[width=0.82\textwidth]{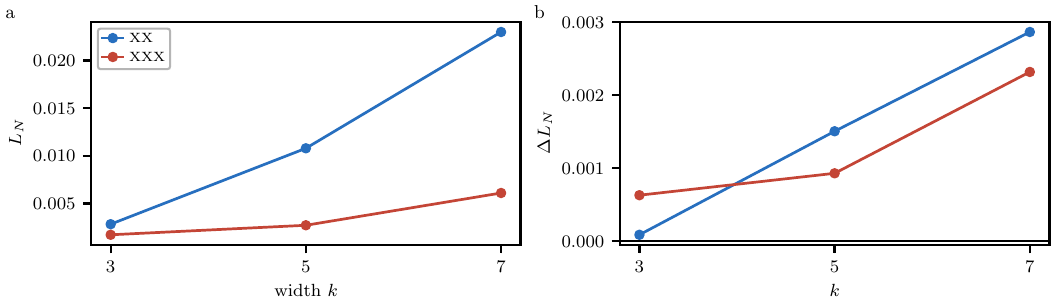}
\caption{\textbf{Entanglement hidden by forgetting the source sign.} (a) Median label-forgetting gap $L_N$ across the primary contiguous cut as a function of aperture $k$ for XX and XXX. (b) Median change relative to the physical post-pulse state at $t=0$, $\Delta L_N(t)=L_N(t)-L_N(0)$.}
Convexity guarantees $L_N\geq0$, while $\Delta L_N$ can have either
sign. The positive medians in panel (b) show that the dynamics increase
the negativity lost when the two source branches are mixed.
 \label{fig:s9}
\end{figure*}

\section{Numerical simulations and matched-model comparisons}
\label{supp:matching-domain}

The finite-pulse reduced states are evaluated on the stored grid $t=0,0.075,\ldots,15$ for the four apertures $k=1,3,5,7$. We call a block a source observation when $|d|<6$ and a nonlocal observation when $|d|\geq6$. The primary activity masks used for normalized finite-pulse readout quantities are $\Delta\geq10^{-3}$ for decision-based fractions and $\Holevo\geq10^{-6}$ for information-based fractions. Matched XX--XXX readout effects additionally use the material activity floor $\Delta\geq0.05$ or $\Holevo\geq0.05$, according to the task.

For the finite-pulse controlled comparison, matching is performed separately within each aperture and source/nonlocal stratum. The matching coordinate is the pooled empirical rank of the relevant state-level budget, $\Delta$ for decision readouts and $\Holevo$ for information readouts. Pairs are chosen without replacement by deterministic minimum-cost bipartite matching, with a primary rank caliper of $0.10$. Uncertainty intervals are obtained from $2000$ moving-time-block bootstrap replicates using blocks of five consecutive time samples; whole spatial profiles are resampled together so neighboring times are not treated as independent observations.

For the source-angle comparison, the Fisher quantities used for matching are evaluated at $\theta=0$. XX and XXX are matched separately within each aperture and source/nonlocal stratum by matching in $\log_{10}F_Q$, with a primary caliper of $0.05$ decades. The stored numerical floor is used only to regularize the logarithm near vanishing QFI. In each moving-time-block replicate, the two model time axes are resampled and the deterministic matching is rerun, so the reported intervals include both temporal dependence and uncertainty in which observations are paired.

\section{Coherent encoding, Bell-state recovery, and qubit transmission}
\label{supp:qubit-proofs}

This supplement proves the statements in Sec.~\ref{sec:qubit-encoding}. The recovery and entropy identities apply to an arbitrary pair of pure many-body states, which need not be orthogonal. The proof has three main steps: identify the information lost to the environment, align the two conditional purifications, and optimize the relevant communication rate. The capacity formula at the equiprobable input additionally uses spin-flip symmetry; the one-site specialization uses this symmetry together with conservation of total magnetization. The final subsections distinguish the two classical-capacity questions and show precisely which earlier data can be reused. Throughout, the initial chain state is pure and the evolution is unitary, as in the model in Sec.~\ref{sec:protocol}.

\subsection{The encoded state and the complementary channel}
\label{supp:qubit-channel-structure}

Fix $\epsilon$, $t$, and $R$. To simplify the derivations, write $E=\bar R$, $\ket{\psi_0}_{RE}=\ket{\psi_{+,\epsilon}(t)}$, and $\ket{\psi_1}_{RE}=\ket{\psi_{-,\epsilon}(t)}$. Define
\begin{equation}
 \rho_{x,R}\coloneqq\Tr_E\ket{\psi_x}\!\bra{\psi_x},
 \qquad
 \rho_{x,E}\coloneqq\Tr_R\ket{\psi_x}\!\bra{\psi_x},
 \qquad x\in\{0,1\},
 \label{eq:qubit-app-conditionals}
\end{equation}
and the reduced cross operator
\begin{equation}
 T_R\coloneqq\Tr_E\ket{\psi_0}\!\bra{\psi_1}.
 \label{eq:qubit-cross-operator}
\end{equation}
Here $x$ labels the computational-basis input, rather than the pre-pulse ground-state density operator. No orthogonality assumption on $\ket{\psi_0}$ and $\ket{\psi_1}$ is needed for the isometry $W\ket{x}=\ket{x}_A\ket{\psi_x}_{RE}$, because the states of $A$ are orthogonal.

For an input density operator $\vartheta$, write $\vartheta_{xy}\coloneqq\bra{x}\vartheta\ket{y}$. Expanding $W\vartheta W^\dagger$ gives the block form
\begin{equation}
 \mathcal N(\vartheta)
 =\sum_{x,y=0}^1\vartheta_{xy}\ket{x}\!\bra{y}_A
   \otimes\Tr_E\ket{\psi_x}\!\bra{\psi_y},
 \qquad
 \mathcal N^c(\vartheta)=\sum_{x=0}^1\vartheta_{xx}\rho_{x,E}.
 \label{eq:qubit-app-channel-blocks}
\end{equation}
The complementary channel loses all input off-diagonal terms when $A$ is traced out. Similarly, discarding $A$ from the receiver's output gives
\begin{equation}
 (\Tr_A\circ\mathcal N)(\vartheta)
 =\sum_{x=0}^1\bra{x}\vartheta\ket{x}\,\rho_{x,R}.
 \label{eq:qubit-without-control}
\end{equation}
This is a measurement followed by a state preparation. For an arbitrary input correlated with a reference, its output is a mixture of product states across reference versus $R$, so it is entanglement breaking. Tensor powers have the same property, and an entanglement-breaking channel cannot transmit entanglement or an unknown qubit with vanishing error. More explicitly, every separable state has fidelity at most $1/d$ with a maximally entangled target of Schmidt rank $d$; the product-state calculation establishing this fact is given in Supplemental Sec.~\ref{supp:qubit-distillation-rate}.

For the Bell input, the retained state is
\begin{equation}
 \omega_{A'AR}=\frac12\bigl(
 \ket{00}\!\bra{00}_{A'A}\otimes\rho_{0,R}
 +\ket{11}\!\bra{11}_{A'A}\otimes\rho_{1,R}+\ket{00}\!\bra{11}_{A'A}\otimes T_R
 +\ket{11}\!\bra{00}_{A'A}\otimes T_R^\dagger\bigr).
 \label{eq:qubit-app-bell-blocks}    
\end{equation}
The first two terms encode the basis value, while the last two terms carry its coherence with the reference. Measuring $A'$ or $A$ in that basis removes the last two terms and therefore does not implement the quantum recovery task.

\subsection{Exact optimal Bell-state recovery}
\label{supp:qubit-optimal-recovery}

The receiver must preserve the computational-basis value while restoring its relative phase. We first construct a controlled unitary that aligns the two purifications as well as possible, and then bound the Bell fidelity of an arbitrary decoding channel. Define
\begin{equation}
 f\coloneqq\sqrt{\mathcal{F}(\rho_{0,E},\rho_{1,E})}.
 \label{eq:qubit-root-fidelity}
\end{equation}
Uhlmann's theorem and the variational characterization of the trace norm give~\cite{QubitUhlmann1976}
\begin{equation}
 f
 =\max_{U_R}\left|
 \bra{\psi_1}(U_R\otimes I_E)\ket{\psi_0}\right|
 =\max_{U_R}|\Tr(U_RT_R)|
 =\norm{T_R}_1,
 \label{eq:qubit-uhlmann-cross}
\end{equation}
where both maximizations are over unitaries on $R$.

Let $T_R=L_R|T_R|$ be a polar decomposition, where $|T_R|\coloneqq\sqrt{T_R^\dagger T_R}$. Extend the polar partial isometry $L_R$ to a unitary on $R$; this is possible because its initial and final subspaces have equal dimension. The same notation $L_R$ denotes this extension, including when $T_R$ has a kernel. Apply the controlled unitary
\begin{equation}
 C_L\coloneqq\ket{0}\!\bra{0}_A\otimes I_R
              +\ket{1}\!\bra{1}_A\otimes L_R,
 \qquad
 \mathcal D_*(\zeta_{AR})\coloneqq\Tr_R(C_L\zeta_{AR}C_L^\dagger),
 \label{eq:qubit-optimal-decoder}
\end{equation}
and relabel the remaining qubit as $B$. The diagonal terms in Eq.~\eqref{eq:qubit-app-bell-blocks} are unchanged after tracing out $R$, while their off-diagonal coefficient becomes $\Tr(T_RL_R^\dagger)=\Tr|T_R|=f$. Hence the recovered state is
\begin{align}
 \omega^{\mathrm{rec}}_{A'B}
 &=\frac12\bigl(
 \ket{00}\!\bra{00}+\ket{11}\!\bra{11}
 +f\ket{00}\!\bra{11}+f\ket{11}\!\bra{00}\bigr)\nonumber\\
 &=\frac{1+f}{2}\Phi_{A'B}
   +\frac{1-f}{2}\Phi^-_{A'B},
 \label{eq:qubit-recovered-state}
\end{align}
where $\ket{\Phi^-}=(\ket{00}-\ket{11})/\sqrt2$ and $\Phi^-=\ket{\Phi^-}\!\bra{\Phi^-}$. Its Bell-state fidelity is $(1+f)/2$.

For the converse, let $\mathcal R\colon AR\to B$ be an arbitrary decoding channel and set
\begin{equation}
 \theta_x\coloneqq\mathcal R(\ket{x}\!\bra{x}_A\otimes\rho_{x,R}),
 \qquad
 z\coloneqq\Tr\!\left[
 \ket{1}\!\bra{0}_B\,
 \mathcal R(\ket{0}\!\bra{1}_A\otimes T_R)\right].
 \label{eq:qubit-converse-terms}
\end{equation}
Expanding the Bell-state projection gives
\begin{equation}
 \mathcal{F}\bigl(\Phi_{A'B},(\operatorname{id}_{A'}\otimes\mathcal R)(\omega)\bigr)
 =\frac14\left(
 \bra0\theta_0\ket0+\bra1\theta_1\ket1+2\operatorname{Re}z\right).
 \label{eq:qubit-converse-expansion}
\end{equation}
Each $\theta_x$ is a normalized state, so the two diagonal terms are probabilities and each is at most one. The adjoint of a channel is a contraction in the operator norm: a Stinespring isometry gives $\mathcal R^\dagger(Y)=V^\dagger(Y\otimes I)V$ and therefore $\norm{\mathcal R^\dagger(Y)}_\infty\leq\norm{Y}_\infty$. Trace-norm/operator-norm duality consequently implies
\begin{align}
 |z|
 &\leq\norm{\mathcal R^\dagger(\ket1\!\bra0)}_\infty
       \norm{\ket0\!\bra1\otimes T_R}_1\nonumber\\
 &\leq\norm{T_R}_1=f.
 \label{eq:qubit-converse-cross-bound}
\end{align}
Thus Eq.~\eqref{eq:qubit-converse-expansion} is at most $(1+f)/2$, which proves Eq.~\eqref{eq:qubit-exact-bell-fidelity} and the optimality of $\mathcal R_*$.

For an arbitrary qubit input, this decoder produces the dephasing channel
\begin{equation}
 (\mathcal R_*\circ\mathcal N)(\vartheta)
 =\frac{1+f}{2}\vartheta+\frac{1-f}{2}Z\vartheta Z,
 \qquad Z\coloneqq\ket0\!\bra0-\ket1\!\bra1.
 \label{eq:qubit-decoded-channel}
\end{equation}
Equation~\eqref{eq:qubit-decoded-channel} describes the chosen single-copy decoder, not an equivalence of the original channel with a qubit dephasing channel. Discarding $R$ separately after each use can lose information useful for a collective transmission code.

The construction is the controlled-unitary argument of Ref.~\cite{QubitHayden2013}, with the recovery objective optimized explicitly. In that argument, one compares the global decoded state with a Bell state tensored with a prescribed conditional purification. With the present choice of $L_R$, the fidelity with the particular product target $\Phi\otimes\ket{\psi_0}\!\bra{\psi_0}$ is $[(1+f)/2]^2$. The Bell-state marginal instead has fidelity $(1+f)/2$, as Eq.~\eqref{eq:qubit-recovered-state} shows. This avoids imposing an unnecessary condition on the residual many-body state. The converse above establishes optimality for this Bell-state marginal, rather than only an achievable continuity estimate.

\subsection{Sharp bounds in terms of the normalized trace distance}
\label{supp:qubit-trace-bounds}

Write $d=\frac12\norm{\rho_{0,E}-\rho_{1,E}}_1$. The Fuchs--van de Graaf inequalities in the fidelity convention of Eq.~\eqref{eq:qubit-fidelity-convention} are~\cite{QubitFuchs1999}
\begin{equation}
 1-f\leq d\leq\sqrt{1-f^2}.
 \label{eq:qubit-fvdg}
\end{equation}
Equivalently, $1-d\leq f\leq\sqrt{1-d^2}$. Substituting these inequalities into $F_{\mathrm{Bell}}=(1+f)/2$ proves Eq.~\eqref{eq:qubit-bell-trace-bounds}.

For completeness, both bounds are attainable for every $d\in[0,1]$ in the class of arbitrary pure two-state preparations. To attain the lower bound, choose three orthonormal environment vectors $\ket c,\ket0,\ket1$ and the states
\begin{equation}
 \eta_0=(1-d)\ket c\!\bra c+d\ket0\!\bra0,
 \qquad
 \eta_1=(1-d)\ket c\!\bra c+d\ket1\!\bra1.
 \label{eq:qubit-sharp-lower-example}
\end{equation}
Their normalized trace distance is $d$ and their root fidelity is $1-d$. To attain the upper bound, take
\begin{equation}
 \eta_0=\ket0\!\bra0,
 \qquad
 \eta_1=\ket{v_d}\!\bra{v_d},
 \qquad
 \ket{v_d}=\sqrt{1-d^2}\ket0+d\ket1.
 \label{eq:qubit-sharp-upper-example}
\end{equation}
The distance is $d$ and the root fidelity is $\sqrt{1-d^2}$. Each pair can be purified into $RE$. If orthogonal global conditional vectors are desired, append to $R$ orthogonal flags identifying the two purifications. This does not change their environment states. Thus no universal improvement depending only on $d$ is possible; additional structure of a specific Hamiltonian or state family can still give stronger conclusions.

\subsection{The coherent-information identity}
\label{supp:qubit-coherent-information}

Tracing $A'$ out of Eq.~\eqref{eq:qubit-app-bell-blocks} gives
\begin{equation}
 \omega_{AR}=\frac12\sum_{x=0}^1
 \ket{x}\!\bra{x}_A\otimes\rho_{x,R},
 \qquad
 S(AR)_\omega=1+\frac12\sum_{x=0}^1S(\rho_{x,R}).
 \label{eq:qubit-output-entropy}
\end{equation}
The global encoded state is pure, so
\begin{equation}
 S(A'AR)_\omega
 =S(\omega_E)
 =S\!\left(\frac{\rho_{0,E}+\rho_{1,E}}{2}\right).
 \label{eq:qubit-environment-entropy}
\end{equation}
Each conditional state $\ket{\psi_x}_{RE}$ is also pure, implying $S(\rho_{x,R})=S(\rho_{x,E})$. Subtracting Eq.~\eqref{eq:qubit-environment-entropy} from Eq.~\eqref{eq:qubit-output-entropy} therefore yields
\begin{align}
 I(A'\rangle AR)_\omega
 &=1+\frac12\sum_x S(\rho_{x,E})
   -S\!\left(\frac{\rho_{0,E}+\rho_{1,E}}2\right)\nonumber\\
 &=1-\Holevo_E.
 \label{eq:qubit-app-coherent-holevo}
\end{align}
The state of $A'E$ is already $\frac12\sum_x\ket{x}\!\bra{x}_{A'}\otimes\rho_{x,E}$. Relabeling its classical basis by the source register $C$ identifies $\Holevo_E=I(C;E)$ and proves Eq.~\eqref{eq:qubit-coherent-holevo}. In particular, $I(A'\rangle AR)_\omega=S(C|E)$ for this classical--quantum state, so it is nonnegative and no larger than one.

With a freshly prepared chain on every use, the quantum coding theorem applies to the memoryless channel $\mathcal N$. Its coherent information at the maximally mixed input is therefore an achievable rate \cite{QubitLloyd1997,QubitDevetak2005}. Different times in a single trajectory do not by themselves constitute independent channel uses.

\subsection{Degradability and the exact capacity under spin-flip symmetry}
\label{supp:qubit-capacity}

The complementary channel in Eq.~\eqref{eq:qubit-app-channel-blocks} is obtained from the receiver's channel by
\begin{equation}
 \mathcal G(\rho_{AR})
 \coloneqq\sum_{x=0}^1
 \Tr\!\left[(\ket{x}\!\bra{x}_A\otimes I_R)\rho_{AR}\right]
 \rho_{x,E},
 \qquad
 \mathcal G\circ\mathcal N=\mathcal N^c.
 \label{eq:qubit-degrading-channel}
\end{equation}
The channel is consequently degradable, and its quantum capacity is its optimized single-use coherent information \cite{QubitDevetakShor2005}. The measurement in $\mathcal G$ establishes this structural property; it is not a prescription for the quantum receiver to measure its control.

One can see the relevant additivity directly. For an arbitrary joint input to two independent uses, write $B_i=A_iR_i$ and $E_i$ for the two receiver and environment outputs. Entropy identities give
\begin{align}
 S(B_1B_2)-S(E_1E_2)
 ={}&[S(B_1)-S(E_1)]+[S(B_2)-S(E_2)]\nonumber\\
 &-I(B_1;B_2)+I(E_1;E_2)\nonumber\\
 \leq{}&[S(B_1)-S(E_1)]+[S(B_2)-S(E_2)].
 \label{eq:qubit-coherent-subadditivity}
\end{align}
The inequality is data processing under $\mathcal G\otimes\mathcal G$. Maximization, induction, and product inputs establish additivity of the optimized coherent information; the regularized quantum coding formula then reduces to one use. Here $I_{\mathrm c}(\vartheta,\mathcal N) \coloneqq S(\mathcal N(\vartheta))-S(\mathcal N^c(\vartheta))$ denotes the coherent information for input $\vartheta$.

It remains to optimize that use. Define the completely dephasing channel by $\overline{\Delta}(\vartheta)\coloneqq(\vartheta+Z\vartheta Z)/2$. Equation~\eqref{eq:qubit-app-channel-blocks} implies
\begin{equation}
 \mathcal N^c(\overline{\Delta}(\vartheta))=\mathcal N^c(\vartheta),
 \qquad
 \mathcal N(\overline{\Delta}(\vartheta))
 =\tfrac12\bigl[\mathcal N(\vartheta)
 +(Z_A\otimes I_R)\mathcal N(\vartheta)(Z_A\otimes I_R)\bigr].
 \label{eq:qubit-input-dephasing}
\end{equation}
Entropy concavity shows that the receiver's output entropy cannot decrease, while the environment entropy is unchanged. Thus $I_{\mathrm c}(\overline{\Delta}(\vartheta),\mathcal N) \geq I_{\mathrm c}(\vartheta,\mathcal N)$. This is a mathematical reduction of the input-state optimization, not a prescription to measure an unknown qubit before sending it. An optimal input density operator can therefore be taken to be $\vartheta_p=p\ket0\!\bra0+(1-p)\ket1\!\bra1$. For this input,
\begin{align}
 I_{\mathrm c}(\vartheta_p,\mathcal N)
 &=h_2(p)-\Holevo_p(E),\nonumber\\
 \Holevo_p(E)
 &\coloneqq S\bigl(p\rho_{0,E}+(1-p)\rho_{1,E}\bigr)
           -pS(\rho_{0,E})-(1-p)S(\rho_{1,E}).
 \label{eq:qubit-biased-coherent-information}
\end{align}
Consequently, without an exchange symmetry, the exact general formula is
\begin{equation}
 \mathcal Q(\mathcal N)
 =\max\bigl\{h_2(p)-\Holevo_p(E):0\leq p\leq1\bigr\}.
 \label{eq:qubit-general-capacity}
\end{equation}
The uniform-input identity remains valid even when $p=1/2$ is not optimal.

For the XX and XXX setting in the manuscript, define $F_S=\prod_j\sigma_j^x=F_R\otimes F_E$. The Hamiltonian commutes with $F_S$, the pulse generator changes sign under it, and the ground state has a definite parity, $F_S\ket{g_\epsilon}=\eta\ket{g_\epsilon}$, with $\eta\in\{+1,-1\}$. It follows that
\begin{equation}
 F_S\ket{\psi_{s,\epsilon}(t)}
 =\eta\ket{\psi_{-s,\epsilon}(t)},
 \qquad
 \rho_{1,E}=F_E\rho_{0,E}F_E^\dagger,
 \qquad F_E^2=I_E.
 \label{eq:qubit-spin-flip-symmetry}
\end{equation}
The function $q(p)\coloneqq h_2(p)-\Holevo_p(E)$ is therefore symmetric under $p\leftrightarrow1-p$. It is also concave: it is the conditional entropy $S(C|E)$ of the state $p\ket0\!\bra0_C\otimes\rho_{0,E} +(1-p)\ket1\!\bra1_C\otimes\rho_{1,E}$, and quantum conditional entropy is concave by strong subadditivity. Thus
\begin{equation}
 q(1/2)\geq\tfrac12[q(p)+q(1-p)]=q(p)
 \quad\text{for every }p\in[0,1].
 \label{eq:qubit-uniform-optimality}
\end{equation}
Combining Eqs.~\eqref{eq:qubit-general-capacity} and \eqref{eq:qubit-uniform-optimality} proves $\mathcal Q(\mathcal N_{R,t})=1-\Holevo_{\bar R}(t)$.

\subsection{The exact asymptotic Bell-pair distillation rate}
\label{supp:qubit-distillation-rate}

There is also an exact statement about independently prepared copies of $\omega_{A'AR}$. Let $\mathcal D_{\to}(A';AR)_\omega$ denote the Bell-pair distillation rate with classical communication from $A'$ to $AR$, and let $\mathcal D_{\leftrightarrow}(A';AR)_\omega$ allow communication in both directions. Rates are measured in Bell pairs per copy. We claim that
\begin{equation}
 \mathcal D_{\to}(A';AR)_\omega
 =\mathcal D_{\leftrightarrow}(A';AR)_\omega
 =I(A'\rangle AR)_\omega
 =1-\Holevo_E.
 \label{eq:qubit-exact-distillation-rate}
\end{equation}
This is the main-text statement in Eq.~\eqref{eq:qubit-distillation-main}. We prove it by matching an achievable rate with an upper bound. Unlike uniform-input optimality for the channel, this state identity does not require spin-flip symmetry.

The hashing inequality gives $\mathcal D_{\to}(A';AR)_\omega\geq I(A'\rangle AR)_\omega$ \cite{QubitDevetakWinter2005}. For a matching converse, dephase $A'$ to obtain the separable state
\begin{equation}
 \tau_{A'AR}
 =\frac12\sum_{x=0}^1
 \ket{x}\!\bra{x}_{A'}\otimes\ket{x}\!\bra{x}_A\otimes\rho_{x,R}.
 \label{eq:qubit-separable-comparison}
\end{equation}
For states with $\operatorname{supp}\rho\subseteq \operatorname{supp}\sigma$, the quantum relative entropy is $D(\rho\Vert\sigma)\coloneqq \Tr[\rho(\log_2\rho-\log_2\sigma)]$; it is $+\infty$ otherwise. Here $\omega\leq2\tau$, because $2\tau=\omega+(Z_{A'}\otimes I_{AR})\omega(Z_{A'}\otimes I_{AR})$, so the support condition holds. The logarithm of $\tau$ is block diagonal in the dephasing basis, giving $\Tr(\omega\log_2\tau)=\Tr(\tau\log_2\tau)$. Moreover, $S(\tau)=S(\omega_{AR})$. It follows that
\begin{equation}
 D(\omega\Vert\tau)
 =S(\tau)-S(\omega)
 =I(A'\rangle AR)_\omega.
 \label{eq:qubit-relative-entropy-converse}
\end{equation}
This is the relative-entropy method for bounding distillation \cite{QubitVedralPlenio1998}; a direct argument completes the proof.

Suppose a local-operations-and-classical-communication protocol converts $n$ copies of $\omega$ to $m$ Bell pairs with fidelity at least $1-\varepsilon$, where $\varepsilon\to0$. Applied to $\tau^{\otimes n}$, the same protocol produces a separable state. Its probability of passing the target maximally entangled projection, of Schmidt rank $2^m$, is at most $2^{-m}$: for normalized product vectors $\ket a\ket b$,
\begin{equation}
 \left|\bra{\Phi^{(d)}}\ket a\ket b\right|^2
 =\frac1d\left|\sum_{j=1}^d a_jb_j\right|^2
 \leq\frac1d,
 \qquad d=2^m.
 \label{eq:qubit-separable-bell-overlap}
\end{equation}
Here $\ket{\Phi^{(d)}}=d^{-1/2}\sum_{j=1}^d\ket{j}\ket{j}$, and $a_j$ and $b_j$ are the components of the two normalized vectors. The inequality is Cauchy--Schwarz, and the extension to separable states follows by convexity. Data processing of relative entropy, first through the protocol and then through this binary test, gives, for $0\leq\varepsilon\leq1/2$,
\begin{equation}
 nD(\omega\Vert\tau)
 \geq(1-\varepsilon)m-h_2(\varepsilon).
 \label{eq:qubit-distillation-finite-bound}
\end{equation}
Indeed, for test probabilities $p\geq1-\varepsilon$ and $r\leq2^{-m}$, binary relative entropy is at least $p\log_2(1/r)-h_2(p)\geq(1-\varepsilon)m-h_2(\varepsilon)$. Dividing by $n$ and taking the distillation limit proves the converse in Eq.~\eqref{eq:qubit-exact-distillation-rate}.

\subsection{Classical capacity with and without the control}
\label{supp:qubit-classical-capacity}

We distinguish the channel $\mathcal M\coloneqq\Tr_A\circ\mathcal N$ from the channel $\mathcal N$ itself. For $\mathcal M$, the receiver gets only $R$, so input off-diagonal terms vanish. Nevertheless, the capacity optimization is over arbitrary quantum encodings, including entangled inputs to many uses. We first show why this larger encoding class cannot improve the rate.

\paragraph{Reduction of arbitrary quantum codewords.}
Let $M$ be a uniformly distributed message with $L$ possible values. An arbitrary encoder assigns an $n$-qubit state $\vartheta_m$ to message $m$. Equation~\eqref{eq:qubit-without-control} gives
\begin{equation}
 \mathcal M^{\otimes n}(\vartheta_m)
 =\sum_{x^n\in\{0,1\}^n}
 \underbrace{\bra{x^n}\vartheta_m\ket{x^n}}_{p(x^n|m)}
 \rho_{x_1,R}\otimes\cdots\otimes\rho_{x_n,R}.
 \label{eq:qubit-classical-codeword-reduction}
\end{equation}
Thus every quantum codeword is equivalent at the output to a classical randomized choice of a pulse-sign string. For a fixed decoding measurement, the success probability for message $m$ is linear in $p(x^n|m)$. Some deterministic string therefore performs at least as well for that message. Replacing each codeword by such a computational- basis string cannot increase either the average error or the maximal error, even at finite blocklength. The encoder has no input constraints beyond the prescribed qubit domain.

It consequently suffices to consider deterministic strings $x^n(m)$. Let $p_i$ denote the fraction of messages for which $x_i(m)=0$. Subadditivity of entropy and additivity of entropy for each conditional product output give
\begin{align}
 I(M:R^n)
 &=S\!\left(\frac1L\sum_m
           \bigotimes_{i=1}^n\rho_{x_i(m),R}\right)
   -\frac1L\sum_m\sum_{i=1}^nS(\rho_{x_i(m),R})\nonumber\\
 &\leq\sum_{i=1}^n
 \left[S\!\left(p_i\rho_{0,R}+(1-p_i)\rho_{1,R}\right)
       -p_iS(\rho_{0,R})-(1-p_i)S(\rho_{1,R})\right]\nonumber\\
 &=\sum_{i=1}^n\Holevo_{p_i}(R)
 \leq n\max_{0\leq p\leq1}\Holevo_p(R).
 \label{eq:qubit-classical-converse-information}
\end{align}
For a decoding estimate $\widehat M$ with average error probability $P_{\mathrm e}$, Fano's inequality and data processing imply
\begin{equation}
 (1-P_{\mathrm e})\log_2 L
 \leq I(M;\widehat M)+h_2(P_{\mathrm e})
 \leq n\max_{0\leq p\leq1}\Holevo_p(R)+h_2(P_{\mathrm e}).
 \label{eq:qubit-classical-fano}
\end{equation}
Vanishing error therefore requires a rate no larger than the optimized single-use Holevo information. The matching achievability is the classical--quantum coding theorem \cite{QubitHolevo1998,QubitSchumacherWestmoreland1997}. Hence, without any symmetry assumption,
\begin{equation}
 \mathcal C(\mathcal M)
 =\max\bigl\{\Holevo_p(R):0\leq p\leq1\bigr\}.
 \label{eq:qubit-classical-general-capacity}
\end{equation}
This direct proof is consistent with the more general additivity result for entanglement-breaking channels~\cite{QubitShor2002}.

\paragraph{Uniform-input optimality and the retained control.}
Spin-flip symmetry gives $\rho_{1,R}=F_R\rho_{0,R}F_R^\dagger$ and $F_R^2=I_R$, by the same partial-trace argument as in Eq.~\eqref{eq:qubit-spin-flip-symmetry}. Consequently $\Holevo_p(R)=\Holevo_{1-p}(R)$. Since $\Holevo_p(R)$ is concave in $p$, averaging $p$ and $1-p$ yields $\Holevo_{1/2}(R)\geq\Holevo_p(R)$. This proves $\mathcal C(\mathcal M_{R,t})=\Holevo_R(t)$.

For $\mathcal N$, the computational-basis outputs $\ket0\!\bra0_A\otimes\rho_{0,R}$ and $\ket1\!\bra1_A\otimes\rho_{1,R}$ are orthogonal, so one bit per use is achievable with zero error. Conversely, every ensemble of $n$-qubit input states has Holevo information at most $n$. Data processing through $\mathcal N^{\otimes n}$ and the decoder, followed by Fano's inequality, bounds the classical communication rate by one. Thus $\mathcal C(\mathcal N_{R,t})=1$, without any exchange-symmetry assumption. These arguments prove Eq.~\eqref{eq:qubit-classical-capacities}; applying the first equality to $\bar R$ and using Eq.~\eqref{eq:qubit-exact-capacity} proves Eq.~\eqref{eq:qubit-complementary-capacities}.

For comparison, an isometry sending $\ket{x}$ directly to $\ket{\psi_x}_{RE}$, when those vectors are orthogonal, would retain cross terms at the output $R$. It is not implemented by discarding $A$ from the controlled encoding. None of the classical-capacity equalities above assumes that different physical encoding.

\subsection{Spatial reinterpretation, limiting cases, and the one-site formulas}
\label{supp:qubit-plot-interpretation}

Equations~\eqref{eq:qubit-plot-fidelity} and \eqref{eq:qubit-plot-capacity} follow by setting $R=\bar K$ and $E=K$ in the proved formulas. The region in the original plot is now inaccessible, whereas its complement and the same control qubit are retained. This explains the reversal of the aperture interpretation. If the accessible region is enlarged, the receiver can always discard the additional spins and use its former decoder. Thus its optimal single-copy fidelity cannot decrease. The same simulation argument applies to transmission codes; equivalently, in the symmetric setting, $\mathcal{Q}=1-\Holevo_E$ increases because tracing out part of $E$ cannot increase $\Holevo_E$.

There are useful boundary checks. With the entire chain available, the receiver can reverse the controlled encoding and recover perfectly. With no chain spins available, the environment is the full chain. The two conditional global states for the quarter turns in Eq.~\eqref{eq:pulse} are orthogonal: with $P=\sigma^z_{j_0}$, $\braket{\psi_0|\psi_1} =\bra{g_\epsilon}e^{i\pi P/2}\ket{g_\epsilon} =i\bra{g_\epsilon}P\ket{g_\epsilon}=0$, where the final equality follows from spin-flip parity. Consequently, $\mathcal F_{\mathrm{Bell}}=1/2$ and $Q=0$. At $t=0$, if $R$ contains the source site, the pulses act only in $R$ and leave the same reduced state in $E$. Then $\Delta_E=\Holevo_E=0$ and both the fidelity and capacity are one. The inverse controlled pulse is an explicit decoder in this case. None of these statements identifies classical readability in $R$ with recoverability of the coherent qubit.

For a single inaccessible site $K=\{j\}$, conservation of total magnetization makes both reduced states diagonal in the local $z$ basis. Equation~\eqref{eq:qubit-spin-flip-symmetry} reverses their magnetization. With $d=\Delta_K=|m_j|$, their average is $I/2$ and
\begin{equation}
 \sqrt{\mathcal F(\rho_+(K,t),\rho_-(K,t))}=\sqrt{1-d^2},
 \qquad
 \Holevo_K=1-h_2\!\left(\frac{1+d}{2}\right).
 \label{eq:qubit-one-site-ingredients}
\end{equation}
For example, the root fidelity is the sum of the two square roots of products of the corresponding probabilities $(1\pm m_j)/2$. Substitution proves Eqs.~\eqref{eq:qubit-one-site-fidelity} and \eqref{eq:qubit-one-site-capacity}. Their small-$d$ expansions are
\begin{align}
 1-\mathcal F_{\mathrm{Bell}}(\bar K,t)
 &=\frac{d^2}{4}+O(d^4),\nonumber\\
 1-\mathcal Q(\mathcal N_{\bar K,t})
 &=\frac{d^2}{2\ln2}+O(d^4).`
 \label{eq:qubit-one-site-expansions}
\end{align}
For $t>0$ at the source site, Eqs.~\eqref{eq:phase-cycle} and \eqref{eq:retarded-green} give $d=2|G^R_{zz}(j_0,j_0;t)|$. Substitution yields
\begin{align}
 1-\mathcal F_{\mathrm{Bell}}(\overline{\{j_0\}},t)
 &=|G^R_{zz}(j_0,j_0;t)|^2+O(|G^R_{zz}|^4),\nonumber\\
 1-\mathcal Q(\mathcal N_{\overline{\{j_0\}},t})
 &=\frac{2}{\ln2}|G^R_{zz}(j_0,j_0;t)|^2+O(|G^R_{zz}|^4).
 \label{eq:qubit-green-function-losses}
\end{align}
Here $G^R$ denotes the retarded Green function, whereas $\overline{\{j_0\}}$ denotes the chain with the source site removed. The fidelity expansion is special to the one-site structure; a small trace distance alone gives only the universal linear guarantee in Eq.~\eqref{eq:qubit-bell-trace-bounds}.

\subsection{Evaluating local recovery with small matrices}
\label{supp:qubit-small-matrices}

The complementary region of a small accessible block can contain most of the chain. Nevertheless, its large density matrix need not be formed explicitly. In addition to $\rho_{0,R}$ and $\rho_{1,R}$, compute the cross operator $T_R$ in Eq.~\eqref{eq:qubit-cross-operator} and form
\begin{equation}
 \Gamma_{AR}\coloneqq\frac12
 \begin{pmatrix}
  \rho_{0,R}&T_R\\
  T_R^\dagger&\rho_{1,R}
 \end{pmatrix}.
 \label{eq:qubit-small-state}
\end{equation}
This is the $AR$ marginal of the pure state $\ket{\Upsilon}_{ARE} =(\ket0_A\ket{\psi_0}_{RE}+\ket1_A\ket{\psi_1}_{RE})/\sqrt2$. In particular, it is a positive semidefinite operator with unit trace. It is not the block-diagonal marginal $\omega_{AR}$ of the Bell-input experiment. The $E$ marginal of $\ket{\Upsilon}$ is $(\rho_{0,E}+\rho_{1,E})/2$, so purity and the previous identities give
\begin{align}
 \mathcal F_{\mathrm{Bell}}(R,t)
 &=\frac{1+\norm{T_R}_1}{2},\nonumber\\
 I(A'\rangle AR)_\omega
 &=1+\frac{S(\rho_{0,R})+S(\rho_{1,R})}{2}-S(\Gamma_{AR}).
 \label{eq:qubit-small-evaluation}
\end{align}
The second expression is the exact capacity in the spin-flip-symmetric setting and an achievable rate without that symmetry.

Even the complementary trace distance has a small-matrix expression:
\begin{equation}
 \Delta_E
 =\norm{\sqrt{\Gamma_{AR}}(Z_A\otimes I_R)
                       \sqrt{\Gamma_{AR}}}_1.
 \label{eq:qubit-small-complement-distance}
\end{equation}
To prove this, observe first that
\begin{equation}
 \frac{\rho_{0,E}-\rho_{1,E}}2
 =\Tr_{AR}\!\left[(Z_A\otimes I_{RE})
              \ket{\Upsilon}\!\bra{\Upsilon}\right].
 \label{eq:qubit-signed-marginal}
\end{equation}
Write a Schmidt decomposition $\ket{\Upsilon}=\sum_j\sqrt{\lambda_j}\ket{u_j}_{AR}\ket{v_j}_E$, where $\Gamma_{AR}=\sum_j\lambda_j\ket{u_j}\!\bra{u_j}$ on its support. In the $\ket{v_j}$ basis, the right side of Eq.~\eqref{eq:qubit-signed-marginal} has matrix entries $\sqrt{\lambda_j\lambda_k}\bra{u_k}(Z_A\otimes I_R)\ket{u_j}$. It is therefore the transpose, up to an isometry on its support, of $\sqrt{\Gamma_{AR}}(Z_A\otimes I_R)\sqrt{\Gamma_{AR}}$. Transposition and this isometric embedding preserve the trace norm, proving Eq.~\eqref{eq:qubit-small-complement-distance}. All matrices in Eqs.~\eqref{eq:qubit-small-evaluation} and \eqref{eq:qubit-small-complement-distance} have dimension at most $2\dim R=2^{k+1}$ for a $k$-spin block. The tensor-network contraction cost remains dependent on the representation of the many-body states; this dimension statement concerns the reduced matrices, not the complexity of simulating the dynamics.

The cross operator contains information not available from the two conditional reduced density matrices alone. For example, the pairs
\begin{equation}
 (\ket0_R\ket+_E,\ \ket1_R\ket+_E)
 \quad\text{and}\quad
 (\ket0_R\ket0_E,\ \ket1_R\ket1_E)
 \label{eq:qubit-diagonal-data-counterexample}
\end{equation}
have identical conditional states on $R$ and satisfy $\Delta_R=\Holevo_R=1$. In the first pair the environment states are identical, giving $\mathcal F_{\mathrm{Bell}}=Q=1$; in the second they are orthogonal, giving $\mathcal F_{\mathrm{Bell}}=1/2$ and $Q=0$. Thus even complete knowledge of those two reduced states does not in general determine local quantum recovery. These are illustrative state pairs, not additional simulated XX or XXX data. When both dynamical state vectors are retained as tensor networks, $T_R$ is obtained by a cross contraction. No additional time evolution is needed, although new contractions beyond the earlier scalar plots are required.

Finally, all complement identifications above refer to the ideal pure preparation and unitary evolution. For a mixed initial chain state or additional dynamical noise, the purifying systems and noise environments must also be included in the inaccessible system. The chain exterior alone need not then contain the complete record relevant to recovery.

\section{Codex- and ChatGPT-assisted research workflow}\label{sec:workflow}

Codex and ChatGPT were used throughout this work as research tools. The workflow had two stages. In the first, we used an agent-based manager--worker structure to explore relatively broad questions and to identify promising directions. Once the main physical picture had started to emerge, we moved to a second stage consisting of narrower studies designed by the authors to test specific claims and numerical observations.

All agent-based studies used the same underlying model, GPT-5.6 Sol, accessed through institutional OpenAI API access with \texttt{reasoning.effort=xhigh}. Codex provided the environment for orchestrating the agents and carrying out code-based investigations. The labels ``manager,'' ``theory,'' ``numerics,'' and ``referee'' used below therefore denote different roles and prompts, rather than different underlying models. All research plans, scientific conclusions, and final manuscript text were reviewed by the authors.

\subsection{Manager--worker exploration}

The first stage was used both for broad exploration of the project and for preliminary investigations of individual questions. An author provided the scientific problem, relevant references, and any known constraints. A root-manager agent then organized the problem into a research plan containing the principal claims to be tested and the calculations or arguments needed to test them. Research proceeded only after this plan was reviewed by an author.

The root manager then assigned separate tasks to literature, theory, and numerical agents as needed. The literature agent checked prior work and the use of references. The theory agent examined analytic arguments, assumptions, limiting cases, and possible counterexamples. The numerical agent implemented calculations and tested the resulting claims computationally. These tasks were carried out as separate agent runs and their outputs were returned to the root manager rather than exchanged directly between agents.

The root manager combined the results into a common scientific account, after which a separate referee agent was asked to challenge the conclusions. In particular, the referee searched for unsupported assumptions, numerical weaknesses, missing controls, and alternative interpretations. When an objection was substantive, the manager opened a targeted follow-up calculation or revision. The final output of each investigation distinguished results that were supported by the available evidence from those that required additional assumptions, failed a check, or remained unresolved. The purpose of this organization was to separate the generation of an argument from its criticism before the result was incorporated into the developing picture of the project.
\begin{figure*}[t]
\centering

\begin{tikzpicture}[
    >=Latex,
    font=\small,
    flow/.style={
        ->,
        line width=0.8pt
    },
    line/.style={
        line width=0.8pt
    },
    feedback/.style={
        ->,
        dashed,
        line width=0.8pt
    },
    author/.style={
        draw,
        rounded corners=2pt,
        fill=green!8,
        minimum height=9mm,
        text width=28mm,
        align=center
    },
    manager/.style={
        draw,
        rounded corners=2pt,
        fill=blue!8,
        minimum height=9mm,
        text width=30mm,
        align=center
    },
    worker/.style={
        draw,
        rounded corners=2pt,
        fill=black!4,
        minimum height=10mm,
        text width=28mm,
        align=center
    },
    referee/.style={
        draw,
        rounded corners=2pt,
        fill=red!7,
        minimum height=10mm,
        text width=29mm,
        align=center
    },
    result/.style={
        draw,
        rounded corners=2pt,
        fill=black!2,
        minimum height=9mm,
        text width=28mm,
        align=center
    }
]


\node[
    anchor=west,
    font=\bfseries
] at (0,0.55)
{(a) Exploratory manager--worker studies};


\node[author] (input) at (1.4,-0.8)
{Author question\\and references};

\node[manager] (plan) at (5.3,-0.8)
{Root manager\\research plan};

\node[author] (approval) at (9.3,-0.8)
{Author review\\and approval};

\draw[flow] (input.east) -- (plan.west);
\draw[flow] (plan.east) -- (approval.west);


\coordinate (dispatchC) at (9.3,-1.95);
\coordinate (dispatchL) at (3.4,-1.95);
\coordinate (dispatchR) at (11.4,-1.95);

\draw[flow] (approval.south) -- (dispatchC);

\draw[line] (dispatchL) -- (dispatchR);


\node[worker] (lit) at (3.4,-2.95)
{Literature\\prior work\\and sources};

\node[worker] (theory) at (7.4,-2.95)
{Theory\\arguments\\and limits};

\node[worker] (num) at (11.4,-2.95)
{Numerics\\calculations\\and tests};

\draw[flow] (dispatchL) -- (lit.north);
\draw[flow] (7.4,-1.95) -- (theory.north);
\draw[flow] (dispatchR) -- (num.north);


\coordinate (collectL) at (3.4,-4.35);
\coordinate (collectC) at (7.4,-4.35);
\coordinate (collectR) at (11.4,-4.35);

\draw[line] (lit.south) -- (collectL);
\draw[line] (theory.south) -- (collectC);
\draw[line] (num.south) -- (collectR);

\draw[line] (collectL) -- (collectR);


\node[manager] (integration) at (7.4,-5.35)
{Root-manager\\integration};

\draw[flow] (collectC) -- (integration.north);


\node[referee] (referee) at (11.7,-5.35)
{Referee\\adversarial review};

\draw[flow]
    (integration.east)
    -- (referee.west);

\coordinate (refdown) at (11.7,-6.30);
\coordinate (feedleft) at (8.50,-6.30);

\draw[feedback]
    (referee.south)
    -- (refdown)
    -- node[
        below,
        midway,
        font=\scriptsize
    ] {targeted revisions}
    (feedleft)
    -- (integration.south east);


\node[result] (result) at (7.4,-7.35)
{Adjudicated\\study result};

\draw[flow]
    (integration.south)
    -- (result.north);


\node[
    anchor=west,
    font=\bfseries
] at (0,-8.65)
{(b) Targeted numerical studies};

\tikzset{
    target/.style={
        draw,
        rounded corners=2pt,
        minimum height=8mm,
        text width=18mm,
        align=center,
        font=\footnotesize
    },
    targetauthor/.style={
        target,
        fill=green!8
    },
    targetplain/.style={
        target,
        fill=black!4
    },
    targetreview/.style={
        target,
        fill=red!7
    },
    targetresult/.style={
        target,
        fill=blue!8
    }
}

\node[targetauthor] (q) at (0.9,-9.75)
{Targeted\\question};

\node[targetplain] (spec) at (3.25,-9.75)
{Fixed\\specification};

\node[targetplain] (prod) at (5.60,-9.75)
{Production\\calculation};

\node[targetplain] (val) at (7.95,-9.75)
{Separate\\validation};

\node[targetreview] (review) at (10.30,-9.75)
{Adversarial\\review};

\node[targetauthor] (auth) at (12.65,-9.75)
{Author\\review};

\node[targetresult] (accepted) at (15.00,-9.75)
{Result used in\\manuscript};

\draw[flow] (q.east) -- (spec.west);
\draw[flow] (spec.east) -- (prod.west);
\draw[flow] (prod.east) -- (val.west);
\draw[flow] (val.east) -- (review.west);
\draw[flow] (review.east) -- (auth.west);
\draw[flow] (auth.east) -- (accepted.west);

\end{tikzpicture}

\caption{\textbf{AI-assisted research workflow used in this work.} (a) Exploratory studies were organized by a root-manager agent, which assigned separate literature, theory, and numerical tasks and integrated their results before adversarial review. (b) Targeted studies began from an author-defined question and fixed numerical specification, followed by separate production, validation, and review steps. Author review was required before a result entered the manuscript.}
\label{fig:codex-workflow}

\end{figure*}
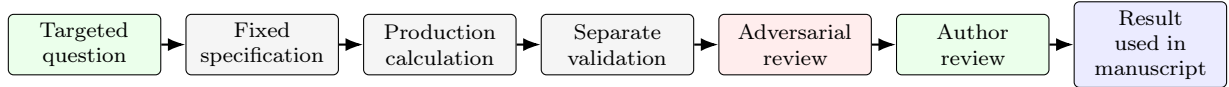

\subsection{Targeted studies}

Once the central questions of the paper had become clearer, we replaced open-ended exploration by a sequence of targeted studies. Each study began with a specific question chosen by the authors. Before evaluating the quantity of interest, we fixed the data to be used, the relevant observables, numerical tolerances, controls, and criteria for accepting or rejecting the proposed conclusion. This was particularly important for analyses involving cuts on the data or comparisons between the XX and XXX models, since it prevented those choices from being adjusted after inspecting the final result.

Production and validation were then performed separately. The production calculation generated the numerical data and figures for the specified question. A separate agent checked analytic limits, matrix properties, partial traces, optimization feasibility, sensitivity to numerical cutoffs, and reproducibility where appropriate. For studies using previously generated reduced data, the input files, coordinate conventions, and basic sanity tests were performed before proceeding.

Finally, a referee agent reviewed the numerical result together with its interpretation. Failed checks, missing data, negative results, and restricted domains were retained rather than removed from the research record. Only after this review, and subsequent author inspection, was a result used in the manuscript. This second stage produced most of the targeted comparisons reported in the main text and Supplemental Information.

\endgroup

\end{document}